\documentclass[preprint,showpacs,preprintnumbers,amsmath,amssymb,prd,floatfix,superscriptaddress]{revtex4}
\usepackage{graphicx}
\usepackage{dcolumn}
\usepackage{bm}
\def\ra    {\rightarrow}
\def\hra   {\hookrightarrow}
\def\Bb    {\ensuremath{\bar{B}}}
\def\Bd    {\ensuremath{\bar{B}^0}}
\def\Bu    {\ensuremath{B^-}}
\def\Bs    {\ensuremath{\bar{B}_s^0}}
\def\Lb    {\ensuremath{\Lambda_b^0}}
\def\Dz    {\ensuremath{D^0}}
\def\Dp    {\ensuremath{D^+}}
\def\Dst   {\ensuremath{D^{*+}}}
\def\Ds    {\ensuremath{D^+_s}}
\def\Lc    {\ensuremath{\Lambda_c^+}}
\def\lD    {\ensuremath{\ell^-D}}
\def\lDp   {\ensuremath{\ell^-\Dp}}
\def\lDz   {\ensuremath{\ell^-\Dz}}
\def\lDst  {\ensuremath{\ell^-\Dst}}
\def\lDs   {\ensuremath{\ell^-\Ds}}
\def\lLc   {\ensuremath{\ell^-\Lc}}
\def\BDln  {\ensuremath{\Bb\ra\ell^-\bar{\nu}_{\ell}D}}
\def\BdDpln {\ensuremath{\Bd\ra\ell^-\bar{\nu}_{\ell}\Dp}}
\def\BuDzln {\ensuremath{\Bu\ra\ell^-\bar{\nu}_{\ell}\Dz}}
\def\BsDsln {\ensuremath{\Bs\ra\ell^-\bar{\nu}_{\ell}\Ds}}
\def\LbLcln {\ensuremath{\Lb\ra\ell^-\bar{\nu}_{\ell}\Lc}}
\def\DzKpi   {\ensuremath{\Dz\ra K^-\pi^+}}
\def\DstDzpi {\ensuremath{\Dst\ra\Dz\pi^+}}
\def\DpKpipi {\ensuremath{\Dp\ra K^-\pi^+\pi^+}}
\def\Dsphipi {\ensuremath{\Ds\ra\phi\pi^+}}
\def\DsKKpi  {\ensuremath{\Ds\ra K^+K^-\pi^+}}
\def\LcpKpi  {\ensuremath{\Lc\ra pK^-\pi^+}}
\def\fu    {\ensuremath{f_u}}
\def\fd    {\ensuremath{f_d}}
\def\fs    {\ensuremath{f_s}}
\def\fb    {\ensuremath{f_{\Lambda_b}}}
\def\fsoud {\ensuremath{\fs/(\fu+\fd)}}
\def\fboud {\ensuremath{\fb/(\fu+\fd)}}
\def\fufd  {\ensuremath{\fu/\fd}}
\def\piZg  {\ensuremath{\pi^0/\gamma}}
\def\mevc  {\ifmmode {\rm MeV}/c \else MeV$/c$\fi}
\def\mevcc {\ifmmode {\rm MeV}/c^2 \else MeV$/c^2$\fi}
\def\gevc  {\ifmmode {\rm GeV}/c \else GeV$/c$\fi}
\def\gevcc {\ifmmode {\rm GeV}/c^2 \else GeV$/c^2$\fi}
\def\lxy   {\ifmmode L_{\rm xy} \else $L_{\rm xy}$\fi}
\def\dedx  {\ifmmode {\rm d}E/{\rm d}x \else d$E/$d$x$\fi}

\newcommand{\hsps}{\hspace*{0.6cm}}

\begin{document}



\title{\boldmath Measurement of Ratios of Fragmentation Fractions for 
Bottom Hadrons in $p\bar p$ Collisions at $\sqrt{s}=1.96$~TeV}


\affiliation{Institute of Physics, Academia Sinica, Taipei, Taiwan 11529, Republic of China} 
\affiliation{Argonne National Laboratory, Argonne, Illinois 60439} 
\affiliation{Institut de Fisica d'Altes Energies, Universitat Autonoma de Barcelona, E-08193, Bellaterra (Barcelona), Spain} 
\affiliation{Baylor University, Waco, Texas  76798} 
\affiliation{Istituto Nazionale di Fisica Nucleare, University of Bologna, I-40127 Bologna, Italy} 
\affiliation{Brandeis University, Waltham, Massachusetts 02254} 
\affiliation{University of California, Davis, Davis, California  95616} 
\affiliation{University of California, Los Angeles, Los Angeles, California  90024} 
\affiliation{University of California, San Diego, La Jolla, California  92093} 
\affiliation{University of California, Santa Barbara, Santa Barbara, California 93106} 
\affiliation{Instituto de Fisica de Cantabria, CSIC-University of Cantabria, 39005 Santander, Spain} 
\affiliation{Carnegie Mellon University, Pittsburgh, PA  15213} 
\affiliation{Enrico Fermi Institute, University of Chicago, Chicago, Illinois 60637} 
\affiliation{Comenius University, 842 48 Bratislava, Slovakia; Institute of Experimental Physics, 040 01 Kosice, Slovakia} 
\affiliation{Joint Institute for Nuclear Research, RU-141980 Dubna, Russia} 
\affiliation{Duke University, Durham, North Carolina  27708} 
\affiliation{Fermi National Accelerator Laboratory, Batavia, Illinois 60510} 
\affiliation{University of Florida, Gainesville, Florida  32611} 
\affiliation{Laboratori Nazionali di Frascati, Istituto Nazionale di Fisica Nucleare, I-00044 Frascati, Italy} 
\affiliation{University of Geneva, CH-1211 Geneva 4, Switzerland} 
\affiliation{Glasgow University, Glasgow G12 8QQ, United Kingdom} 
\affiliation{Harvard University, Cambridge, Massachusetts 02138} 
\affiliation{Division of High Energy Physics, Department of Physics, University of Helsinki and Helsinki Institute of Physics, FIN-00014, Helsinki, Finland} 
\affiliation{University of Illinois, Urbana, Illinois 61801} 
\affiliation{The Johns Hopkins University, Baltimore, Maryland 21218} 
\affiliation{Institut f\"{u}r Experimentelle Kernphysik, Universit\"{a}t Karlsruhe, 76128 Karlsruhe, Germany} 
\affiliation{Center for High Energy Physics: Kyungpook National University, Daegu 702-701, Korea; Seoul National University, Seoul 151-742, Korea; Sungkyunkwan University, Suwon 440-746, Korea; Korea Institute of Science and Technology Information, Daejeon, 305-806, Korea; Chonnam National University, Gwangju, 500-757, Korea} 
\affiliation{Ernest Orlando Lawrence Berkeley National Laboratory, Berkeley, California 94720} 
\affiliation{University of Liverpool, Liverpool L69 7ZE, United Kingdom} 
\affiliation{University College London, London WC1E 6BT, United Kingdom} 
\affiliation{Centro de Investigaciones Energeticas Medioambientales y Tecnologicas, E-28040 Madrid, Spain} 
\affiliation{Massachusetts Institute of Technology, Cambridge, Massachusetts  02139} 
\affiliation{Institute of Particle Physics: McGill University, Montr\'{e}al, Canada H3A~2T8; and University of Toronto, Toronto, Canada M5S~1A7} 
\affiliation{University of Michigan, Ann Arbor, Michigan 48109} 
\affiliation{Michigan State University, East Lansing, Michigan  48824} 
\affiliation{University of New Mexico, Albuquerque, New Mexico 87131} 
\affiliation{Northwestern University, Evanston, Illinois  60208} 
\affiliation{The Ohio State University, Columbus, Ohio  43210} 
\affiliation{Okayama University, Okayama 700-8530, Japan} 
\affiliation{Osaka City University, Osaka 588, Japan} 
\affiliation{University of Oxford, Oxford OX1 3RH, United Kingdom} 
\affiliation{University of Padova, Istituto Nazionale di Fisica Nucleare, Sezione di Padova-Trento, I-35131 Padova, Italy} 
\affiliation{LPNHE, Universite Pierre et Marie Curie/IN2P3-CNRS, UMR7585, Paris, F-75252 France} 
\affiliation{University of Pennsylvania, Philadelphia, Pennsylvania 19104} 
\affiliation{Istituto Nazionale di Fisica Nucleare Pisa, Universities of Pisa, Siena and Scuola Normale Superiore, I-56127 Pisa, Italy} 
\affiliation{University of Pittsburgh, Pittsburgh, Pennsylvania 15260} 
\affiliation{Purdue University, West Lafayette, Indiana 47907} 
\affiliation{University of Rochester, Rochester, New York 14627} 
\affiliation{The Rockefeller University, New York, New York 10021} 
\affiliation{Istituto Nazionale di Fisica Nucleare, Sezione di Roma 1, University of Rome ``La Sapienza," I-00185 Roma, Italy} 
\affiliation{Rutgers University, Piscataway, New Jersey 08855} 
\affiliation{Texas A\&M University, College Station, Texas 77843} 
\affiliation{Istituto Nazionale di Fisica Nucleare, University of Trieste/\ Udine, Italy} 
\affiliation{University of Tsukuba, Tsukuba, Ibaraki 305, Japan} 
\affiliation{Tufts University, Medford, Massachusetts 02155} 
\affiliation{Waseda University, Tokyo 169, Japan} 
\affiliation{Wayne State University, Detroit, Michigan  48201} 
\affiliation{University of Wisconsin, Madison, Wisconsin 53706} 
\affiliation{Yale University, New Haven, Connecticut 06520} 
\author{T.~Aaltonen}
\affiliation{Division of High Energy Physics, Department of Physics, University of Helsinki and Helsinki Institute of Physics, FIN-00014, Helsinki, Finland}
\author{J.~Adelman}
\affiliation{Enrico Fermi Institute, University of Chicago, Chicago, Illinois 60637}
\author{T.~Akimoto}
\affiliation{University of Tsukuba, Tsukuba, Ibaraki 305, Japan}
\author{M.G.~Albrow}
\affiliation{Fermi National Accelerator Laboratory, Batavia, Illinois 60510}
\author{B.~\'{A}lvarez~Gonz\'{a}lez}
\affiliation{Instituto de Fisica de Cantabria, CSIC-University of Cantabria, 39005 Santander, Spain}
\author{S.~Amerio}
\affiliation{University of Padova, Istituto Nazionale di Fisica Nucleare, Sezione di Padova-Trento, I-35131 Padova, Italy}
\author{D.~Amidei}
\affiliation{University of Michigan, Ann Arbor, Michigan 48109}
\author{A.~Anastassov}
\affiliation{Rutgers University, Piscataway, New Jersey 08855}
\author{A.~Annovi}
\affiliation{Laboratori Nazionali di Frascati, Istituto Nazionale di Fisica Nucleare, I-00044 Frascati, Italy}
\author{J.~Antos}
\affiliation{Comenius University, 842 48 Bratislava, Slovakia; Institute of Experimental Physics, 040 01 Kosice, Slovakia}
% ...
\collaboration{CDF Collaboration\footnote{With visitors from $^a$University of Athens, 15784 Athens, Greece, 
$^b$Chinese Academy of Sciences, Beijing 100864, China, 
$^c$University of Bristol, Bristol BS8 1TL, United Kingdom, 
$^d$University Libre de Bruxelles, B-1050 Brussels, Belgium, 
$^e$University of California Irvine, Irvine, CA  92697, 
$^f$University of California Santa Cruz, Santa Cruz, CA  95064, 
$^g$Cornell University, Ithaca, NY  14853, 
$^h$University of Cyprus, Nicosia CY-1678, Cyprus, 
$^i$University College Dublin, Dublin 4, Ireland, 
$^j$University of Edinburgh, Edinburgh EH9 3JZ, United Kingdom, 
$^k$University of Heidelberg, D-69120 Heidelberg, Germany, 
$^l$Universidad Iberoamericana, Mexico D.F., Mexico, 
$^m$University of Manchester, Manchester M13 9PL, England, 
$^n$Nagasaki Institute of Applied Science, Nagasaki, Japan, 
$^o$University de Oviedo, E-33007 Oviedo, Spain, 
$^p$Queen Mary, University of London, London, E1 4NS, England, 
$^q$Texas Tech University, Lubbock, TX  79409, 
$^r$IFIC(CSIC-Universitat de Valencia), 46071 Valencia, Spain, 
}}
\noaffiliation

\date{\today}

\begin{abstract}
This paper describes the first measurement of $b$-quark fragmentation
fractions into bottom hadrons in Run\,II of the Tevatron Collider at
Fermilab.  The result is based on a 360~pb$^{-1}$ sample of data
collected with the CDF\,II detector in $p\bar p$~collisions at
$\sqrt{s}=1.96$~TeV. Semileptonic decays of \Bd, \Bu, and \Bs~mesons,
as well as \Lb~baryons, are reconstructed.  For an effective bottom
hadron $p_T$~threshold of 7~\gevc, the fragmentation fractions are
measured to be $\fufd=1.054\pm0.018\,{\rm (stat)}\,
^{+0.025}_{-0.045}\,{\rm (sys)}\,\pm 0.058\,({\cal B})$,
$\fsoud=0.160\pm0.005\,{\rm (stat)}\, ^{+0.011}_{-0.010}\,{\rm
(sys)}\,^{+0.057}_{-0.034}\,({\cal B})$, and
$\fboud=0.281\pm0.012\,{\rm (stat)}\, ^{+0.058}_{-0.056}\,{\rm
(sys)}\,^{+0.128}_{-0.086}\,({\cal B})$, where the uncertainty ${\cal
B}$ is due to uncertainties on measured branching ratios.  The value
of $\fsoud$ agrees within one standard deviation with previous CDF
measurements and the world average of this quantity, which is
dominated by LEP measurements.  However, the ratio $\fboud$ is
approximately twice the value previously measured at LEP.  The
approximately 2\,$\sigma$ discrepancy is examined in terms of
kinematic differences between the two production environments.
\end{abstract}

                             
                              
\maketitle



\section{\label{sec:Intro}
Introduction}

Bottom quarks, $b$, produced in $p\bar p$ collisions combine with
anti-quarks or di-quarks to form bottom hadrons. In this process,
called fragmentation, the color force field creates quark-antiquark
pairs $q\bar q$ that combine with the bottom quark to create a
\Bb~meson $|b\bar q\rangle$ or $b$ baryon $|b q_1 q_2\rangle$. Since
the fragmentation process, which is governed by the strong force,
cannot be reliably calculated by perturbative
QCD~\cite{Ref:Politzer,Ref:Gross,Ref:Feynman_Field}, the fragmentation
properties of $b$~quarks must be determined empirically.  This paper
describes a measurement of the species dependence of the $b$-quark
fragmentation rates into bottom hadrons produced in $p\bar
p$~collisions at center of mass energy $\sqrt{s}=1.96$~TeV during
Run\,II of the Tevatron collider at Fermilab.

The probabilities that the fragmentation of a $b$~quark will result in
a \Bu~$|b\bar u\rangle$, \Bd~$|b\bar d\rangle$, or \Bs~$|b\bar
s\rangle$~meson or a \Lb~$|bdu\rangle$~baryon are denoted by \fu,
\fd, \fs, and \fb, respectively.  In this paper, $f_q$ indicates the
fragmentation fraction integrated above the momentum threshold of
sensitivity in the data: $f_q\equiv f_q(p_T(\Bb) >
p_T^{min})$~\cite{Ref:pT}. In the case that the fragmentation
fractions are momentum dependent, the measured fragmentation fractions
are proportional to the relative yields of the bottom hadrons
integrated above the effective $p_T^{min}$.  The contributions from
the production of excited bottom hadrons that decay into final states
containing a \Bu, \Bd, \Bs~meson or \Lb~baryon are implicitly included
in this definition of the fragmentation fractions, $f_q\equiv {\cal
B}(b\ra B_q X)$.  Throughout the paper, unless otherwise noted,
references to a specific charge state are meant to imply the charge
conjugate state as well.

In Run\,I of the Fermilab Tevatron, which collected data from
1992\,-\,1996, the fraction of \Bs~mesons produced relative to the
number of \Bd~mesons was measured $\approx$\,2\,$\sigma$ higher at
CDF~\cite{Ref:Simon1,Ref:Simon2,Ref:Simon3} than at the LEP
experiments~\cite{Ref:LEP_fs1,Ref:LEP_fs2,Ref:LEP_fs3}.
Interestingly, the time-integrated flavor averaged mixing parameter,
$\bar{\chi} = \fd\chi_d + \fs\chi_s$, where $\chi_d$ and $\chi_s$ are
the time-integrated mixing parameters of $\Bd$ and $\Bs$ mesons
respectively, was also measured $\approx$\,2\,$\sigma$ higher in
Run\,I~\cite{Ref:Paolo1,Ref:Paolo2} than the LEP averages of the same
quantity~\cite{Ref:chib1,Ref:chib2,Ref:chib3,Ref:chib4,Ref:chib5,Ref:chib6}.
This second discrepancy led to speculations about possible sources of
the enhanced average mixing rate at a hadron collider relative to
electron-positron collisions, including suggestions that new physics
may be the source of the disagreement~\cite{Ref:Berger}. Since the
average momentum of $b$~quarks produced at LEP,
$\langle\,p(b)\,\rangle$\,$\sim$\,40~\gevc, is significantly higher
than at the Tevatron, $\langle\,p(b)\,\rangle$\,$\sim$\,10~\gevc,
it is also possible that the fragmentation process depends on the
$b$-quark momentum.  Another possible explanation is that \fs~is
higher at the Tevatron than at LEP due to the different initial
mechanism of $b$-quark production.  Of course, a more mundane
possibility is that the Run\,I results relating to $\fs$ are simply
statistical fluctuations.  To shed light on the question of whether
$b$-quark fragmentation is different in a hadron environment than in
$e^+e^-$~collisions, the fragmentation fractions are measured in CDF
Run\,II with high statistical precision and an updated treatment of
the lepton-charm sample composition.

The analysis strategy is as follows. Semileptonic decays of bottom hadrons,
$\BDln X$, where $\ell^-$ stands for electron or muon, and $D$ represents a
charm meson or baryon, in case of semileptonic bottom baryon decays, unless
otherwise specified, provide large samples for studying the
fragmentation properties of $b$~quarks.  This measurement determines
the $b$-quark fragmentation fractions by reconstructing five
semileptonic signatures, \lDp, \lDz, \lDst, \lDs, and \lLc.  The
selection requirements are kept similar among the five lepton-charm
channels in order to cancel as many systematic uncertainties as
possible.  The final signal requirements, though similar, have been
selected to maintain good acceptance for the individual decays, which
have different kinematic features. The reconstructed \lD~signal
yields, originating from the various $\BDln X$ semileptonic decays,
are then related to the numbers of bottom hadrons (\Bu, \Bd, \Bs, or
\Lb) produced in the $b$-quark fragmentation process. Since the
neutrino from the semileptonic bottom hadron decay is not
reconstructed, the missing energy in the decay allows semileptonic
bottom hadron decays to excited charm states to contribute to the five
final state decay signatures.  This results in ``cross-talk'' between
the bottom hadron~species, particularly between the
\Bb~mesons.  The observed semileptonic \lD~decay signatures are
related to their corresponding parent bottom hadrons through a
procedure used to extract the sample composition, as described later
in the text.  In order to reduce systematic uncertainties in trigger
and tracking efficiencies, the $b$-quark fragmentation fractions are
measured relative to \fd. This means that the relative fragmentation
fractions \fufd, \fsoud\ and \fboud\ are extracted from the five
lepton-charm yields, taking the sample composition into account. Since
the fragmentation of $b$~quarks into $b$~baryons other than the \Lb\
are ignored, a constraint requiring the fragmentation fractions \fu,
\fd, \fs, and \fb\ to sum to unity is not applied.

This paper is organized as follows.  The semileptonic signal
reconstruction is discussed in Section~\ref{sec:data}, while the
sample composition procedure used to relate the lepton-charm
signatures to the parent bottom hadron is described in
Section~\ref{sec:sample_comp}.  The efficiencies needed to extract the
sample composition are determined in Section~\ref{sec:eff}.  The fit
to the fragmentation fractions is detailed in Section~\ref{sec:fit}.
Finally, the systematic uncertainties assigned to the measurement are
described in Section~\ref{sec:sys} and the final results are discussed
in Section~\ref{sec:results}.

\section{\label{sec:data} 
Data Reconstruction}

\subsection{\label{sec:cdfexp} Experimental Apparatus}

The data used in this measurement represent an integrated luminosity
of approximately 360~pb$^{-1}$ collected with the CDF\,II detector
between February 2002, and August 2004. The CDF\,detector employs a
cylindrical geometry around the $p\bar{p}$ interaction region with the
proton direction defining the positive $z$-direction.  Most of the
quantities used for candidate selection are measured in the plane
transverse to the $z$-axis.  In the CDF coordinate system, $\varphi$
is the azimuthal angle, $\theta$ is the polar angle measured from the
proton direction, and $r$ is the radius perpendicular to the beam
axis. The pseudorapidity $\eta$ is defined as
$\eta=-\ln[\,\tan(\theta/2)\,]$.  The transverse momentum,~$p_T$, is
the component of the track momentum,~$p$, transverse to the $z$-axis
($p_T = p\cdot \sin\theta$), while $E_T = E\cdot\sin\theta$, with $E$
being the energy measured in the calorimeter.

The CDF\,II detector features excellent lepton identification and charged
particle tracking and is described in detail
elsewhere~\cite{Ref:CDFdet,Ref:CDF_TDR}.  The parts of the detector
relevant to the reconstruction of semileptonic bottom hadron decays
used in this measurement are briefly summarized below.  The detector
nearest to the $p\bar{p}$ interaction region is a silicon vertex
detector (SVX\,II)~\cite{Ref:SVXII}, which consists of five concentric
layers of double-sided sensors located at radii between 2.5 and
10.6~cm. An additional single layer of silicon (L00)~\cite{Ref:L00} is
mounted on the beam pipe at radius $r$\,$\sim$\,1.5~cm, but the
information from this detector is not used in this measurement.  In
addition, two forward layers plus one central layer of double sided
silicon located outside the SVX at radii of 20-29~cm make up the
intermediate silicon layers (ISL)~\cite{Ref:ISL}. Together with the
SVX\,II, the ISL detector extends the sensitive region of the CDF\,II
tracking detector to $|\eta| \le 2.0$. CDF's silicon system provides
three-dimensional track reconstruction and is used to identify
displaced vertices associated with bottom hadron decays.  The
measurement of the momentum of charged particles in the silicon
detector is significantly improved with the central outer tracker
(COT)~\cite{Ref:COT}, an open-cell drift chamber with 30,200 sense
wires arranged in 96 layers combined into four axial and four stereo
super-layers (SL). It provides tracking from a radius of $\sim$\,40~cm
out to a radius of 132~cm covering $|z|<155$~cm.  The track
reconstruction efficiency of the COT is found to be
$(99.6^{+0.4}_{-0.9})\%$ for charged particles with $p_T >
1.5~\gevc$~\cite{Ref:BsmumuPRL} and $\gtrsim 94\%$~\cite{Ref:charmPRL}
for charged particles with $p_T = 0.4~\gevc$.  For high-momentum
charged particles, the $p_T$ resolution is found to be
$\sigma(p_T)/p_T = 0.0015~p_T/\gevc$.  The COT also provides specific
energy loss, \dedx, information for charged particle identification
with a separation between pions and kaons of approximately
1.4\,$\sigma$~\cite{Ref:CDF_dEdx}.  The central tracking system is
immersed in a superconducting solenoid that provides a 1.4~T axial
magnetic field.

Electromagnetic (CEM)~\cite{Ref:CEM} and hadronic (CHA)~\cite{Ref:CHA}
calorimeters are located outside the COT and the solenoid, where they
are arranged in a projective-tower geometry.  The electromagnetic and
hadronic calorimeters are lead-scintillator and iron-scintillator
sampling devices, respectively.  The energy resolution for the CDF
central calorimeter is $\sigma(E_T) / E_T = [(13.5\% / \sqrt{E_T})^2 +
(1.5\%)^2]^{1/2}$ for electromagnetic
showers~\cite{Ref:CEM,Ref:CEM2} and $\sigma(E_T) / E_T = [(75\% /
\sqrt{E_T})^2 + (3\%)^2]^{1/2}$ for hadrons~\cite{Ref:CDF_TDR,Ref:CHA},
where $E_T$ is measured in GeV.  A layer of proportional chambers
(CES), with wire and strip readout, is located six radiation lengths
deep in the CEM calorimeters, near the electromagnetic shower maximum.
The CES provides a measurement of electromagnetic shower
profiles in both the $\varphi$- and $z$-directions for use in electron
identification.  Muon candidates are identified with two sets of
multi-layer drift chambers and scintillator
counters~\cite{Ref:CMU,Ref:mu_upgrade}, one located outside the
calorimeters (CMU) and the other (CMP) behind an additional 60~cm of
iron shielding, equivalent to approximately 3 pion interaction
lengths.  The CMU provides coverage for particles with $|\eta|<0.6$
and $p_T>1.4~\gevc$.  The CMP covers the same pseudorapidity region,
but identifies muons with $p_T>2.0~\gevc$ with higher purity than
muons reconstructed~in~the~CMU~only.

\subsection{\label{sec:trig_req} Trigger Requirements} 

CDF uses a three-level trigger system~\cite{Ref:CDF_TDR}, where each
level provides a rate reduction sufficient to allow for processing at
the next level with minimal dead-time.  At level\,1, data from every
beam crossing are stored in a pipeline memory capable of buffering
data for $\approx 5.5\mu\mbox{s}$.  The level\,1 trigger either
rejects an event or copies the data into one of four level\,2
buffers. At level\,2, a substantial fraction of the event data is
available for analysis by the dedicated trigger processors. Events
that pass the level\,1 and level\,2 trigger selection criteria are
then sent to the level\,3 trigger~\cite{Ref:L3,Ref:EVB}, a cluster of
computers running a speed-optimized reconstruction code.  Events
selected by level\,3 are written to permanent mass storage.

\begin{figure}
\centerline{
\includegraphics[width=0.6\hsize]{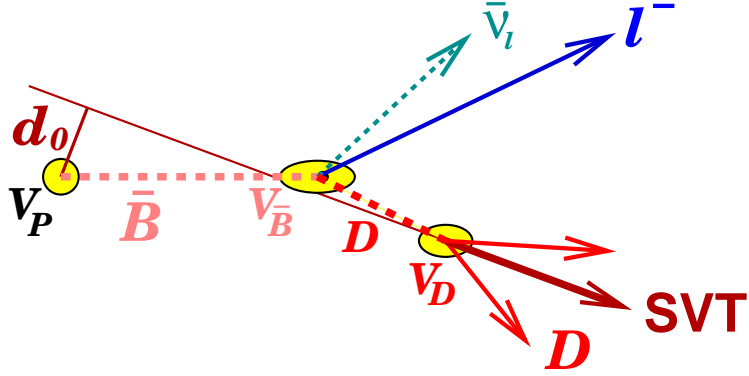}
}
\caption{\label{fig:Bdecay_sketch}
Sketch of semileptonic $B$-decay topology in the transverse plane, where
$V_P$ is the primary vertex, $V_B$ is the decay vertex of the bottom
hadron, $V_D$ is the decay vertex of the charm hadron, and $d_0$ is defined
in the text. ``SVT'' indicates the track selected by the displaced track
SVT trigger, which is also defined in the text.}
\end{figure}

Tracking plays a significant role in the triggers utilized for this
analysis. Semileptonic $\BDln X$~decays are recorded using a trigger
that requires a lepton and a track displaced from the interaction
point and identified with the silicon vertex trigger
(SVT)~\cite{Ref:SVT}. The decay topology of semileptonic $B$~decays is
sketched in Fig.~\ref{fig:Bdecay_sketch}.  Tracks are reconstructed at
level\,1 with the extremely fast tracker (XFT)~\cite{Ref:XFT} by
examining COT hits from the four axial super-layers. The XFT provides
$r$-$\varphi$ tracking information and can identify tracks with
$p_T>1.5$~\gevc\ with high efficiency ($>90\%$) and good transverse
momentum resolution, $\sigma(p_T)/p_T=0.016~p_T/[\gevc]$.  XFT tracks
can be matched with either calorimeter clusters to identify electron
candidates or with track segments in the muon detectors to identify
muon candidates.  The XFT tracks are extrapolated into the silicon
detector system, where the SVT uses the SVX\,II measurements of charge
deposits from charged particles to form simplified tracks.  In
addition, the SVT determines the distance of closest approach in the
transverse plane,~$d_0$, with respect to the $p\bar p$ beam line,
which is determined from a time-dependent line fit to the locus of
primary interaction vertices determined from all tracks available at
trigger level (see Fig.~\ref{fig:Bdecay_sketch}).  The impact
parameter resolution of the SVT is approximately
$50~\mu$m~\cite{Ref:SVT, Ref:SVT2}, which includes a contribution of
$35~\mu$m from the width of the $p\bar{p}$ interaction
region~\cite{Ref:Beam}.

The primary trigger used in this measurement requires that the lepton
and the displaced track (SVT track) must have transverse momentum
values greater than 4~\gevc\ and 2~\gevc, respectively.  The displaced
track's impact parameter,~$d_0$, must exceed 120~$\mu$m and be less
than 1~mm to reject decay products of long-lived hadrons decays such
as $K^0_S$ or $\Lambda^0$.  The opening angle, $\Delta\phi$, between
the lepton and SVT track is required to satisfy $2^{\circ} \leq
\Delta\phi(\ell^-,\mbox{SVT track}) \leq 90^{\circ}$ to increase the
probability that the two tracks originate from the same $\Bb$ hadron.
Additionally, the invariant mass between the trigger lepton and SVT
triggered track must be less than the nominal bottom~hadron mass,
$m(\ell^-,\mbox{SVT
track})\equiv\sqrt{(p^{\mu}(\ell^-)+p'^{\mu}(\mbox{SVT}))^2} <
5~\gevcc$, where the SVT track is assumed to have the pion mass.  The
trigger lepton requirements are described in conjunction with their
analysis selections in Section~\ref{sec:trig_lep}.  Events that pass
these trigger requirements are recorded to the lepton plus SVT trigger
data stream for further analysis.  In this measurement both the muon
and electron plus SVT trigger data ($e$+SVT and $\mu$+SVT) are used.
An additional trigger utilized for selecting
\Bb~events is the two-track trigger (TTT), which requires two
displaced tracks.  Large semileptonic \Bb~samples are also available with
this trigger~\cite{Ref:Likelihood,Ref:BsMixing}, although the false lepton
background is much larger as well.  Semileptonic events from the TTT are
used in this analysis for a study of the systematic uncertainty arising
from false leptons. 

\subsection{\label{sec:selec_req} Data Selection and Reconstruction} 

Events from the lepton plus SVT trigger data stream are used to
reconstruct semileptonic bottom hadron decays in this analysis. First,
trigger leptons are identified by re-confirming the trigger decision
with offline quantities after event reconstruction. Charm candidates are
then reconstructed, with the SVT track required to match one of the
daughter tracks from the charm decay.  The selections on the
lepton-charm signals obtained are optimized to reduce combinatoric
background and improve signal significance.  Non-combinatoric
backgrounds in the charm signals are handled separately.

\subsubsection{\label{sec:trig_lep} Trigger Lepton Identification} 

The data analysis begins by identifying the trigger leptons from the
$e$+SVT and $\mu$+SVT trigger streams.  The electron candidates are
identified by requiring the following selection criteria.  The
longitudinal shower profile must be consistent with that of an
electron shower, with a leakage energy from the CEM into the CHA of
less than 12.5\%, in order to suppress hadron contamination.  The
lateral shower profile of the CEM cluster is required to be consistent
with a profile obtained from test beam electrons after appropriate
corrections.  The association of
a single track with the calorimeter shower is made based on the
position matching at the CES plane, with both $|\Delta z\cdot\sin
\theta| < 5$~cm and $r|\Delta\varphi| < 3$~cm conditions required. To
achieve good agreement between data and Monte Carlo (MC) simulation
(see Sec.~\ref{sec:mc}), an isolation requirement is applied to the
trigger electron candidates by requiring that exactly only one track
is found that projects to the CEM towers used to define the electron
energy.  To reconfirm electron trigger cuts, the offline reconstructed
$E_T$ and $p_T$ of the electron candidate are required to be greater
than 4~GeV and 4~\gevc, respectively.  Additionally, electron
candidates from photon conversions in the detector material are
removed by rejecting those electron candidates that have a small
opening angle with oppositely charged particles in the event.

Trigger muon candidates are reconstructed by extrapolating tracks
measured in the COT to the muon system, where they are matched to
track segments (stubs) reconstructed in the muon chambers. A CMU or
CMP stub is required to have hits in at least three out of the four
layers of planar drift chambers. Trigger muons are required to have
hits in both the CMU and CMP muon chambers. The separation between a
track segment reconstructed in the muon chamber and the extrapolated
COT track is computed.  The uncertainty in this quantity is dominated
by multiple scattering in the traversed detector material. For good
track to stub matching, this separation is required to be less than
15~cm and 20~cm in the $r\varphi$-view for CMU and CMP, respectively.
The transverse momentum of a muon candidate reconstructed offline is
required to be greater than 4~\gevc.

\subsubsection{\label{sec:sig_sel} Charm Candidate Selection} 

The SVT track is required to match one of the final state tracks in
the five reconstructed charm signals: \DzKpi, 
$\Dst\ra\Dz\,(\ra K^-\pi^+)\,\pi^+$, \DpKpipi,
$\Ds\ra\phi\,(\ra K^+K^-)\,\pi^+$, and \LcpKpi.  Only
well-reconstructed tracks with $p_T \geq 0.4~\gevc$ and at least three
silicon $r$-$\varphi$ hits are retained for offline analysis.  To
ensure good track quality, all charm daughter tracks, except for the
soft pion from the \Dst~decay, are required to have at least five hits
in at least two axial and two stereo COT super-layers.  There are no
COT requirements on the $\Dst$~soft pion.  During data reconstruction
the track parameters are corrected for the ionization energy loss
appropriate to the mass hypothesis under consideration.  In addition,
tracks are required to be fiducial in the COT, so that only tracks
which are well-described by the simulation (see Sec.~\ref{sec:mc}) are
used for further analysis.  In particular, tracks that fall within
$|z|\leq$~1.5 cm of the COT mid-plane, where no track information is
recorded, and tracks that originate outside of the COT volume at
$|z|\geq$~155~cm are excluded from the analysis. In addition, all
tracks must at least pass through the axial SL\,6 before exiting the
COT. This means the exit radius of the track must be greater than the
radius of the sixth super-layer $r_{\rm SL\,6}$ = 106~cm. This
requirement is tightened for the SVT trigger track and the trigger
lepton. Both tracks must pass through SL\,8 of the COT ($r_{\rm
SL\,8}$ = 131~cm) as required in the trigger.  The invariant mass of
the $\DzKpi$~and $\DpKpipi$ is reconstructed within
$[1.40,2.00]$~\gevcc and $[1.70,2.00]$~\gevcc, respectively.  The
reconstructed $\Ds\ra\phi\pi^+$ mass is required to be within $[1.75,
2.2]$~\gevcc, while the $\LcpKpi$ is reconstructed within $[2.15,
2.40]$~\gevcc.  Finally, the reconstructed charm signals are combined
with the triggered lepton in a three-dimensional kinematic fit
constraining all tracks to a common vertex (see
Fig.~\ref{fig:Bdecay_sketch}) to establish signals that can be related
to semileptonic \Bu, \Bd, \Bs, and \Lb~decays.  The $\phi\ra K^+K^-$
vertex reconstruction does not use a constraint to the known 
$\phi$~mass~\cite{Ref:PDG_2004}, although $|m(\phi)-1.019|$~[\gevcc] is
required 
in order to select a pure sample of $\phi$ candidates.

\subsubsection{\label{sec:lD_bkgs} Backgrounds to Lepton-Charm Signals} 

Several backgrounds affect the semileptonic \Bb~signals.  Some of
these can be reduced by judicious signal selection, while some must be
included in the modeling of the signal or treated as sources of
systematic uncertainties.  The simplest of these backgrounds to
understand are those events arising from combinatoric sources, which
are generally estimated from the sidebands of the charm signal. In
these backgrounds, random tracks are combined to form a charm signal
which passes all charm selection requirements.  This combinatoric
background can most easily be reduced by selection requirements and
modeled by the sideband events, which are expected to exhibit the same
shape underneath the signal.  A related, but more subtle type of
background is that arising from the mis-identification of tracks in
one charm decay arising from incorrect assignment of particle
identifications in a real charm decay, resulting in "reflection"
backgrounds.  These backgrounds are often flat beneath the signal of
interest, but occasionally they exhibit particular shapes that can
affect the signal distribution non-uniformly.  Some reflection
backgrounds can be effectively reduced with particle identification
selections, such as the specific ionization of particles, \dedx\ (see
Section~\ref{sec:sig_opt}.)  Other reflection backgrounds, which have
non-uniform distribution in mass beneath the charm signal are included
in the fit to the signal (see Section~\ref{sec:reflec}.)  MC simulated
data is used to determine the shape of these reflection backgrounds.

The third type of background to the semileptonic signals arises from
physical processes that produce a real lepton and charm hadron, but
not through a decay directly to \lD.  This includes processes which
originate from the same \Bb, such as $\Bb\ra D\bar{D}$, where
$\bar{D}\ra\ell^-X$, and $\Bb\ra\tau^-\bar{\nu}_{\tau}D$, where
$\tau^-\ra\ell^-X$.  These ``physics backgrounds'' are included in the
fit to the sample composition (see Section~\ref{sec:sample_comp}).
Other backgrounds include processes in which the lepton and charm
hadron originate from separate $b\bar{b}$ and $c\bar{c}$ quark pairs,
{\it i.e.} $b\ra DX$, $\bar{b}\ra\ell^+X$, or $c\ra DX$,
$\bar{c}\ra\ell^-X$.  The $b\bar{b}$ background gives a wrong sign
(WS) lepton-charm combination, in which the charm and lepton have the
same charge, while the $c\bar{c}$ background gives right sign (RS)
lepton-charm combinations, in which the charm and lepton have opposite
charge.  All of these processes are also possible with a real charm
hadron and a false lepton.  In the case of false leptons, both right
sign and wrong sign lepton-charm are expected to be present.
Backgrounds which do not originate from the same \Bb~hadron are
treated as a source of systematic uncertainty and described by the
wrong sign lepton-charm events, which primarily describe false leptons
(see Section~\ref{sec:ws}.)  The $c\bar{c}$ background is assumed to
be small for a charm decaying to a lepton with $p_T >
4~\gevc$~\cite{Ref:Karenthesis} and is 
ignored, while the $b\bar{b}$ background is implicitly included in the
false lepton systematic uncertainty.

\subsubsection{\label{sec:sig_opt} Signal Optimization} 

Requirements to further enhance the lepton-charm signal include $p_T$ cuts on
the $p$, $K$, and $\pi$ charm daughter tracks, and cuts on the
invariant mass of the lepton-charm system,
$m(\lD)\equiv\sqrt{(p^{\mu}(\ell^-)+p'^{\mu}(D))^2}$, to limit
feed-down from excited charm and lepton-charm combinations which do
not originate from direct semileptonic bottom hadron decays.
Requirements are also made on the probability of the charm and
lepton-charm vertex fits.

\begin{figure}
\centerline{
\includegraphics[width=0.5\hsize]{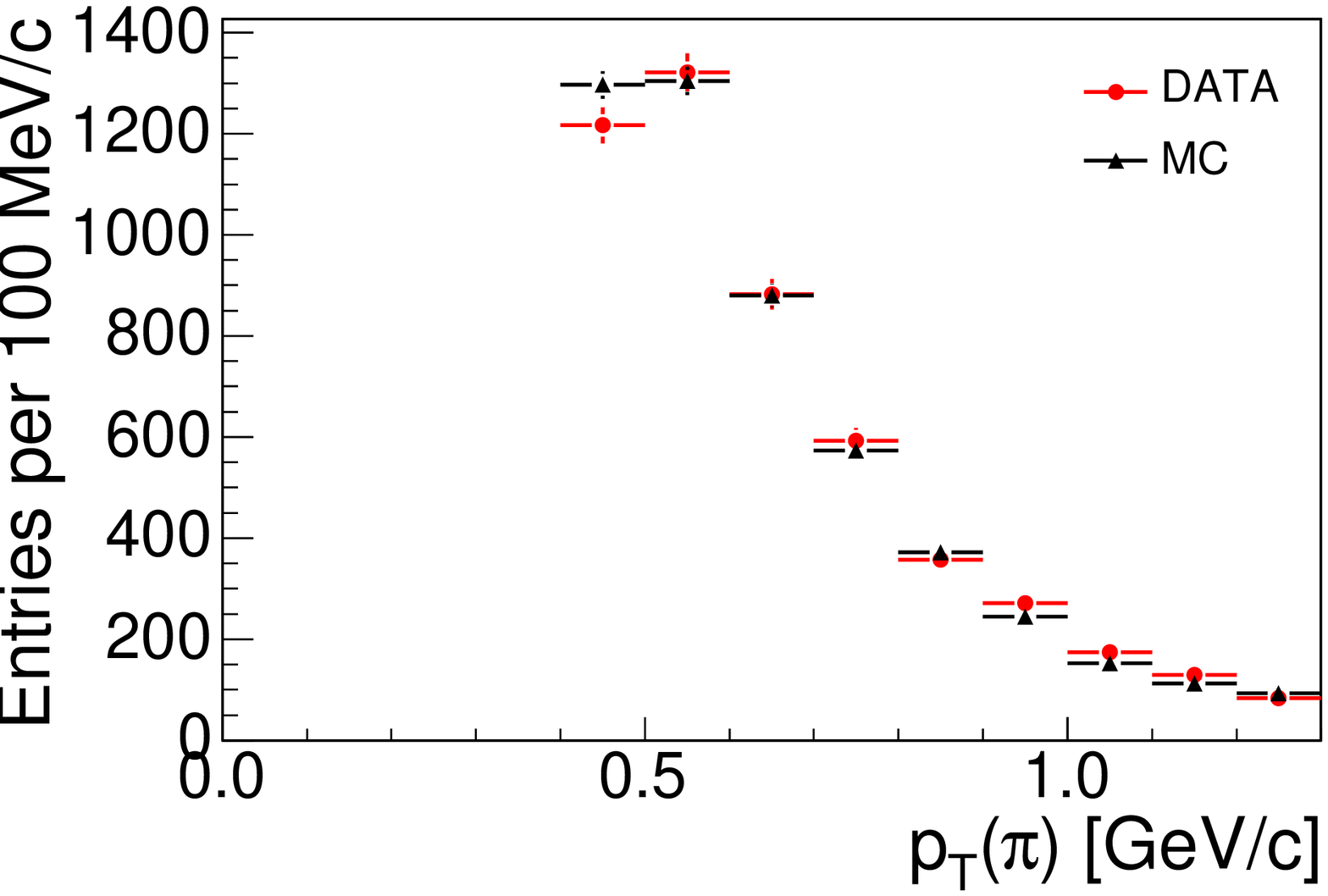}
\includegraphics[width=0.5\hsize]{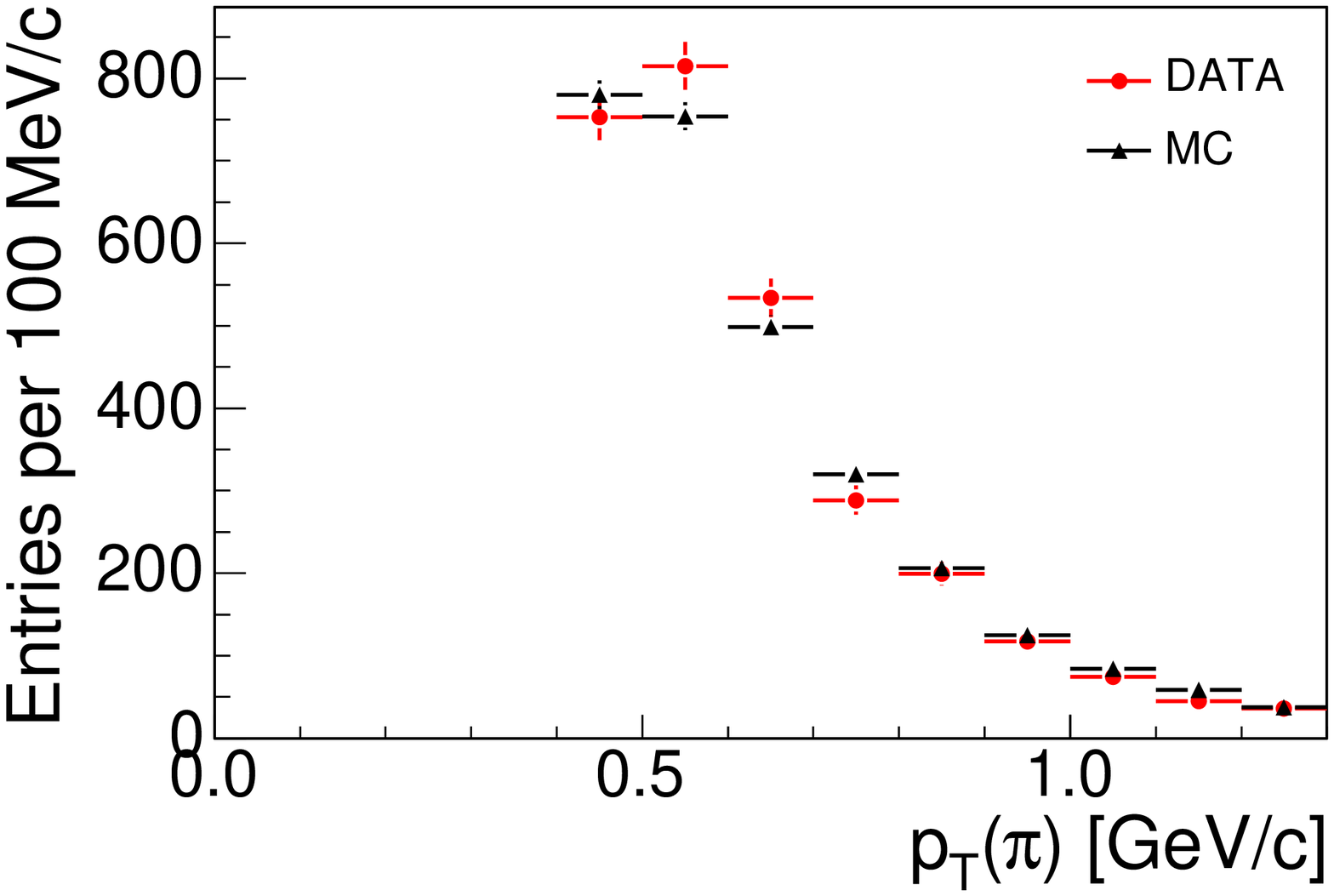}
\put(-420,125){\large\bf (a)}
\put(-180,125){\large\bf (b)}
}
\caption
{\label{fig:pi_star} Comparisons between data and simulation of
$p_T(\pi_*^+)$, the soft pion from the \Dst~decay, for the (a) $\mu^-\Dst$
and (b) $e^-\Dst$ mode.}
\end{figure}

Since bottom hadrons are longer-lived, a powerful discriminant against
these backgrounds is a cut on the proper time of \lD~candidate. The
decay distance of the \Bb~hadron is determined by defining a quantity,
$\lxy(\mbox{PV}\ra\lD)$, which is the transverse decay distance of the
lepton-charm combination from the primary interaction vertex (PV),
projected on the \lD~momentum direction. The missing neutrino produced
in the semileptonic decay prevents precise knowledge of $p_T(\Bb)$ and
thus of the proper decay time of the \Bb~candidate. Instead, a pseudo
proper decay time is constructed as:
\begin{eqnarray}
ct^*(\lD)\equiv \lxy(\mbox{PV}\ra\lD)\times\frac{m(\Bb)}{p_T(\lD)}.
\end{eqnarray}
A $ct^*(\lD) > 200~\mu\mbox{m}$ cut is applied to guarantee a signal
from long-lived bottom hadrons and to reduce signal contamination from
false leptons and other processes that can contribute a lepton and a
charm hadron from uncorrelated sources (see also
Section~\ref{sec:lD_bkgs}.) This requirement also drastically reduces
the combinatoric background of charm candidates with real leptons.  A
cut on the significance of the transverse decay distance of the charm
meson, $\lxy(\mbox{PV}\ra D)/\sigma_{\lxy({\rm PV}\ra D)}$, also
reduces the light flavored hadron contamination in the signal. A cut
on $p_T(D^0) > 5~\gevc$ is applied to improve agreement between the
\lDz~data and Monte Carlo simulation used in determining the
efficiencies (see Sec.~\ref{sec:eff}).  The selected
\Dst~candidates are a subset of the \Dz~candidates.  Instead of performing
a vertex fit on the soft pion,~$\pi^+_*$, from the $\DstDzpi_*$~decay,
A tight $\Delta m(\Dst,\Dz)\equiv m(\Dz\pi^+)-m(\Dz)\in [0.1440,
0.1475]~\gevcc$ cut is used to select a very clean \lDst~sample.  This
reduces the systematic uncertainty in the selection of the
\lDst~combination relative to a \lDz~pair, since no additional vertex
fit is performed.  Consequently, the efficiency to detect the soft
pion is better described by the simulation.  Since the data agrees
well with the simulation for tracks with $p_T$ greater than 400~\mevc,
as can be seen in Fig.~\ref{fig:pi_star}, the soft pion efficiency is
determined from the simulation.  A tight $\Delta m(\Dst,\Dz)\in
[0.1440, 0.1475]~\gevcc$ cut is used to select a very clean
\lDst~sample.

In order to determine the final analysis selection, kinematic
selection criteria are optimized with respect to the combinatoric
background for each lepton-charm channel, with additional cuts
designed to limit non-combinatoric background, such as the $ct^*(\lD)$
and $p_T(D)$ cuts, applied during the optimization.  The figure of
merit (FOM) used for optimization is $S/\sqrt{S+B}$. The signal, $S$,
is taken from inclusive $\BDln X$ and $\LbLcln X$~Monte Carlo (see
Sec.~\ref{sec:mc}).  The background, $B$, is taken from the sidebands
of the charm signal.  In order for the FOM to accurately reflect the
significance of the signals in data, $S$ is scaled to the expected
data signal with a set of nominal cuts obtained by first optimizing
each cut individually without applying any other cut.  The cuts are
then optimized a second time applying all optimal cuts from the prior
optimization except the cut being optimized.  After two or three
successive iterations, a stable optimal cut point is reached for all cuts.

\begin{figure}
\centerline{
\includegraphics[width=0.7\hsize]{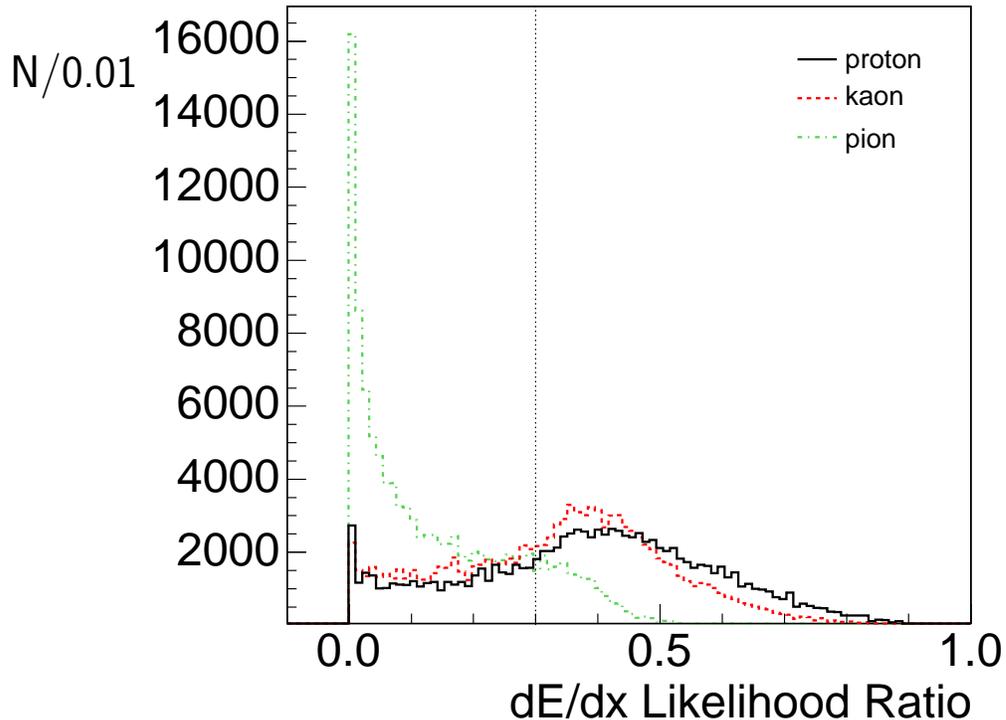}
\put(-380,260){\Large\sf N/0.01}
}
\caption{\label{fig:dedx} 
\dedx~${\cal LR}$ distribution for protons from
$\Lambda^0\ra p\pi^-$ and kaons and pions from $\Dst\ra\Dz\,(\ra
K^-\pi^+)\,\pi^+$ with the proton hypothesis applied.  Tracks with
${\cal LR}$ to the right of the dashed vertical line are identified as
protons.}
\end{figure}

A particle identification cut using \dedx\ is found useful for
reducing the combinatoric background in the \Lc\ signal.  The
combinatoric background can be significantly reduced by correctly
identifying the proton from the \LcpKpi~decay utilizing the specific
energy loss of the proton track measured in the COT. A
\dedx~likelihood ratio, ${\cal LR}$, requirement is applied to the
proton. The likelihood ratio is defined by the relation ${\cal
LR}(p)\equiv {\cal L}(p)/ [{\cal L}(p)+{\cal L}(K)+{\cal L}(\pi)+{\cal
L}(e)+{\cal L}(\mu)]$, where ${\cal
L}(i)\propto\exp\{-Z_i^2/(2\sigma^2_{Z_i})\}$ and
$Z_i\equiv\ln[\dedx^{\rm meas.}_i/\dedx^{\rm pred.}_i]$.
Figure~\ref{fig:dedx} shows the resulting ${\cal LR}$ distributions
for protons from the $\Lambda^0\ra p\pi^-$ decay and kaons and pions
from the $\Dst\ra\Dz\,(\ra K^-\pi^+)\,\pi^+$ decay with the proton
hypothesis applied.  Muons are indistinguishable from pions, while
electrons are well-separated from all of the other distributions,
since their mass is so much lower than the mass of the other
particles.  A cut on ${\cal LR}(p)>0.3$, as determined from the
control samples, is applied to reduce background while keeping the
proton efficiency high.  This cut primarily removes pions, since the
\dedx\ separation between protons and kaons is not as good.

To cancel as many differences in signal reconstruction as possible,
the selection criteria are kept as similar as is feasible across charm
channels.  The optimized cuts designed to limit both the combinatoric
and some non-combinatoric backgrounds are unified to minimize the
differences in selections between channels.  However, some cuts, in
which different optimal values are expected due to differences in the
decay kinematics, are not forced to be similar.  For example,
the proper decay time of the \Dp~meson and \Lc~baryon differ by a
factor of about five.  The selection criteria applied to the
lepton-charm decay signatures are listed in Table~\ref{tab:cuts}.
Additional selection requirements to reduce non-combinatoric
backgrounds are discussed next.

\begin{table*}
\caption{\label{tab:cuts}
Signal selection requirements.} 
\begin{ruledtabular}
\begin{tabular}{lccccc}
Selection cuts              & 
$\lDz$       & $\lDst$      & $\lDp$       & $\lDs$       & $\lLc$ \\
    \hline
$ct(D)$~[cm] $\in$          & 
[-0.01,0.10] & [-0.01,0.10] & [-0.01,0.20] & [-0.01,0.10] & [-0.01,0.05] \\
$ct^*(\lD)$~[cm] $>$    &
0.02         & 0.02         & 0.02         & 0.02         & 0.02         \\
$\sigma_{ct^*}(\lD)$~[cm] $<$ &
0.04         & 0.04         & 0.04         & 0.04         & 0.04         \\
$m(\lD)$~[\gevcc] $\in$ &
[2.4,5.1]    & [2.4,5.1]    & [2.4,5.1]    & [2.4,5.1]    & [3.4,5.5]    \\
$p_T(D)$ [\gevc] $>$        &
5.0          & 5.0          & N/A          & N/A          & N/A          \\
$p_T(p)$ [\gevc] $>$        & 
N/A          & N/A          & N/A          & N/A          & 2.0          \\
$p_T(K)$ [\gevc] $>$        & 
0.6          & 0.6          & 0.6          & 0.6          & 0.6          \\
$\chi^2_{2D}(D) <$          &
10           & 10           & 10           & 10           & 5            \\
vertex prob.$(\lD) >$   &
$10^{-7}$    & $10^{-7}$    & $10^{-7}$    & $10^{-7}$    & $10^{-4}$    \\
$\lxy/\sigma_{Lxy}(D) >$    &
4.5          & 4.5          & 11           & 5            & 4.5          \\
$\Delta m(\Dst,\Dz)$~[\gevcc] $\in$ &
N/A          & [0.1440,0.1475] & N/A       & N/A          & N/A          \\
$p_T(\pi_*)$ [\gevc] $>$    & 
N/A          & 0.4          & N/A          & N/A          & N/A          \\
$|m(\phi)-1.019|$~[\gevcc] $<$ & 
N/A          & N/A          & N/A          & 0.0095       & N/A          \\
\dedx~${\cal LR}(p)\ >$     &
N/A          & N/A          & N/A          & N/A          & 0.3          \\
\end{tabular}
\end{ruledtabular}
\end{table*}

\subsubsection{\label{sec:reflec} Reflection Backgrounds} 

\begin{figure*}
\centerline{
\includegraphics[width=0.5\hsize]{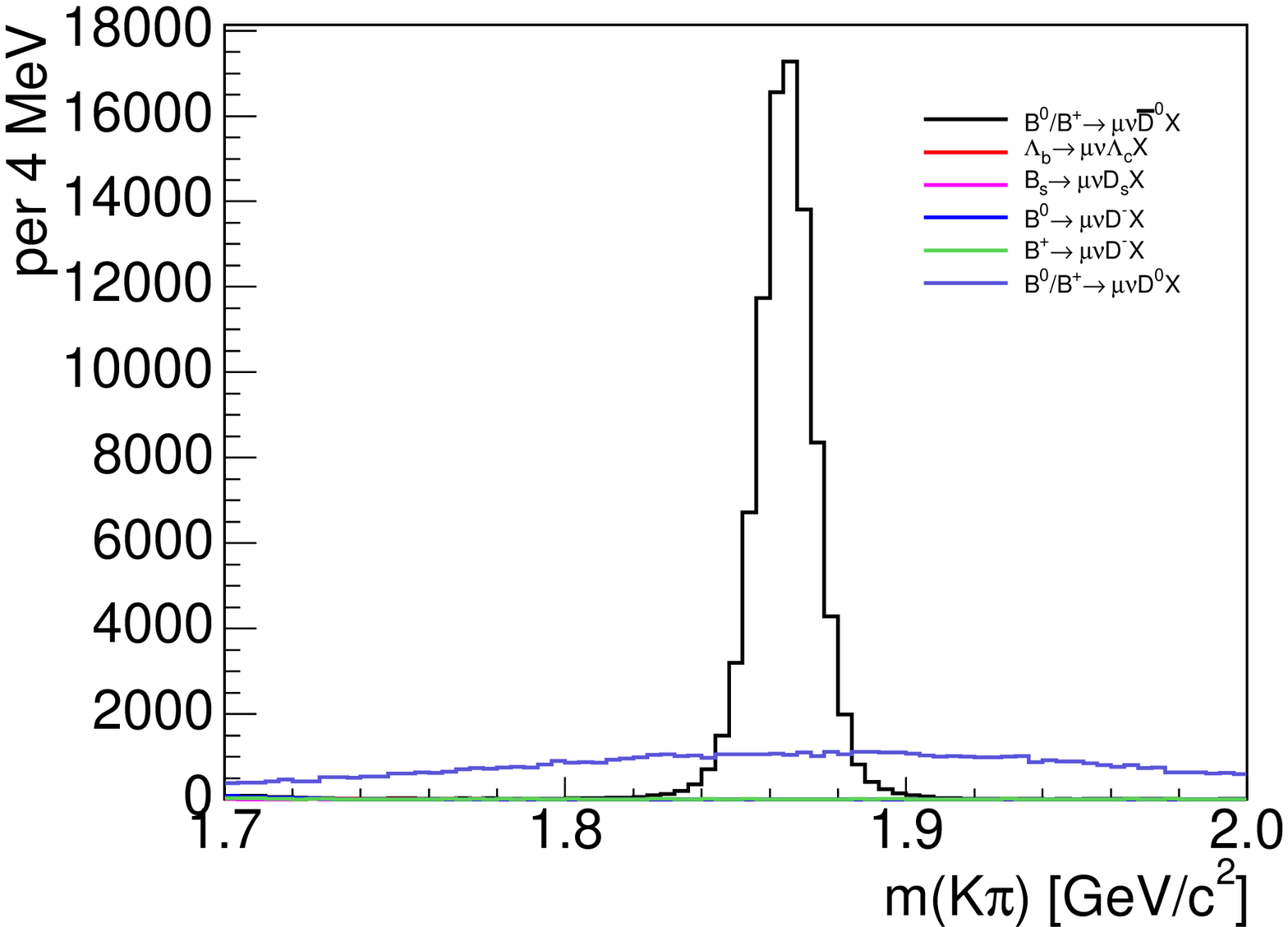}
\includegraphics[width=0.5\hsize]{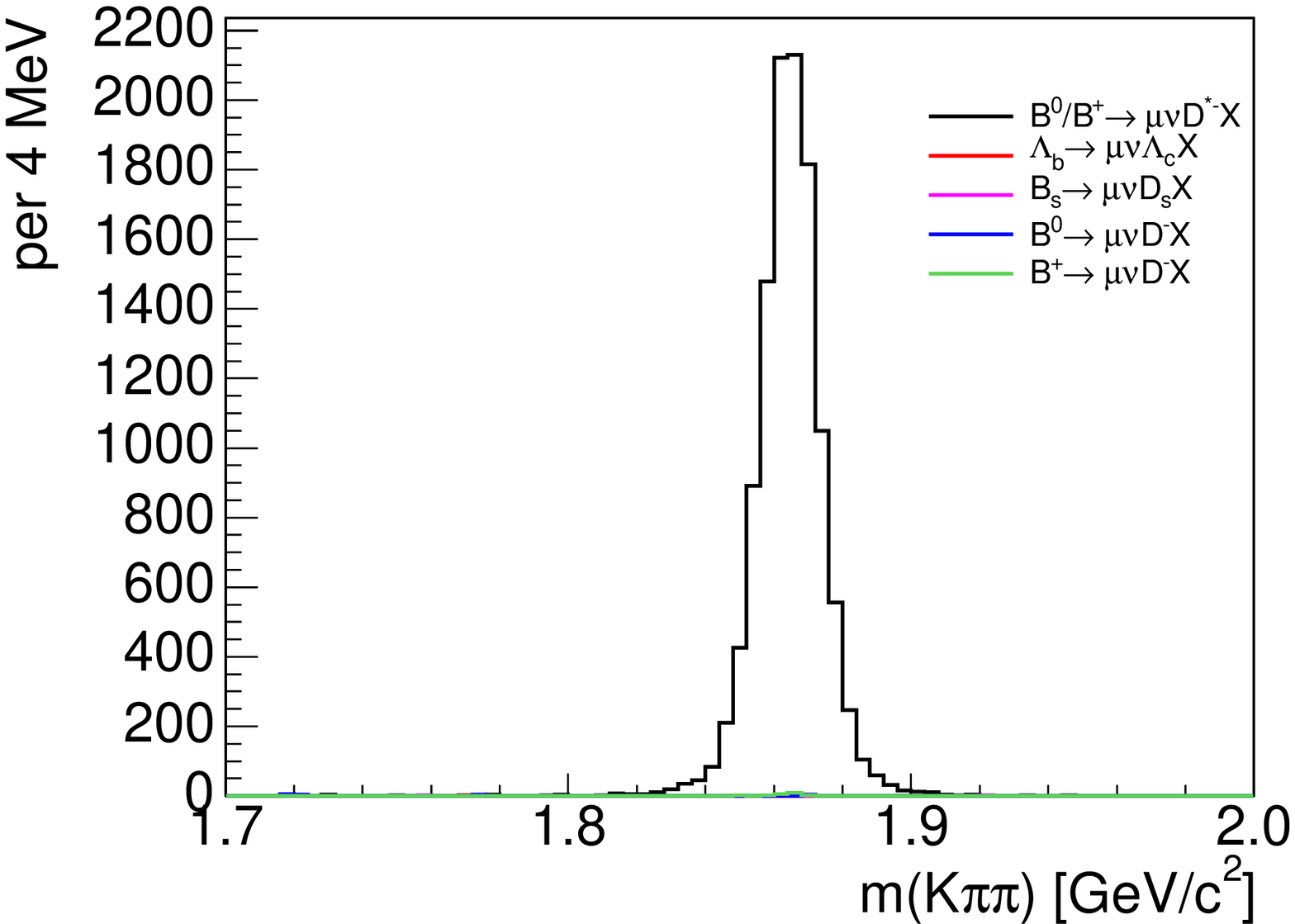}
\put(-415,135){\large\bf (a)}
\put(-177,135){\large\bf (b)}
}
\centerline{
\includegraphics[width=0.5\hsize]{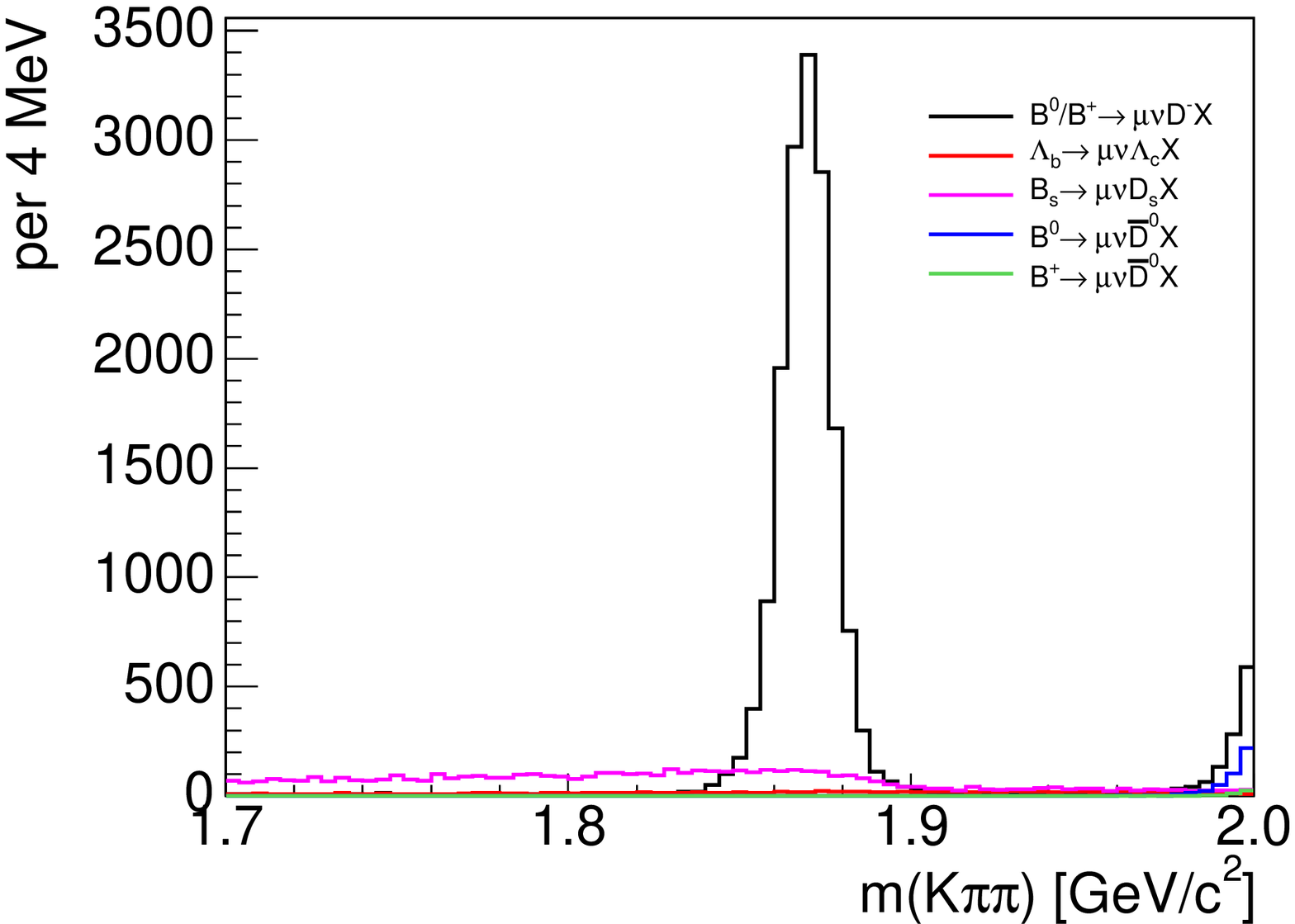}
\includegraphics[width=0.5\hsize]{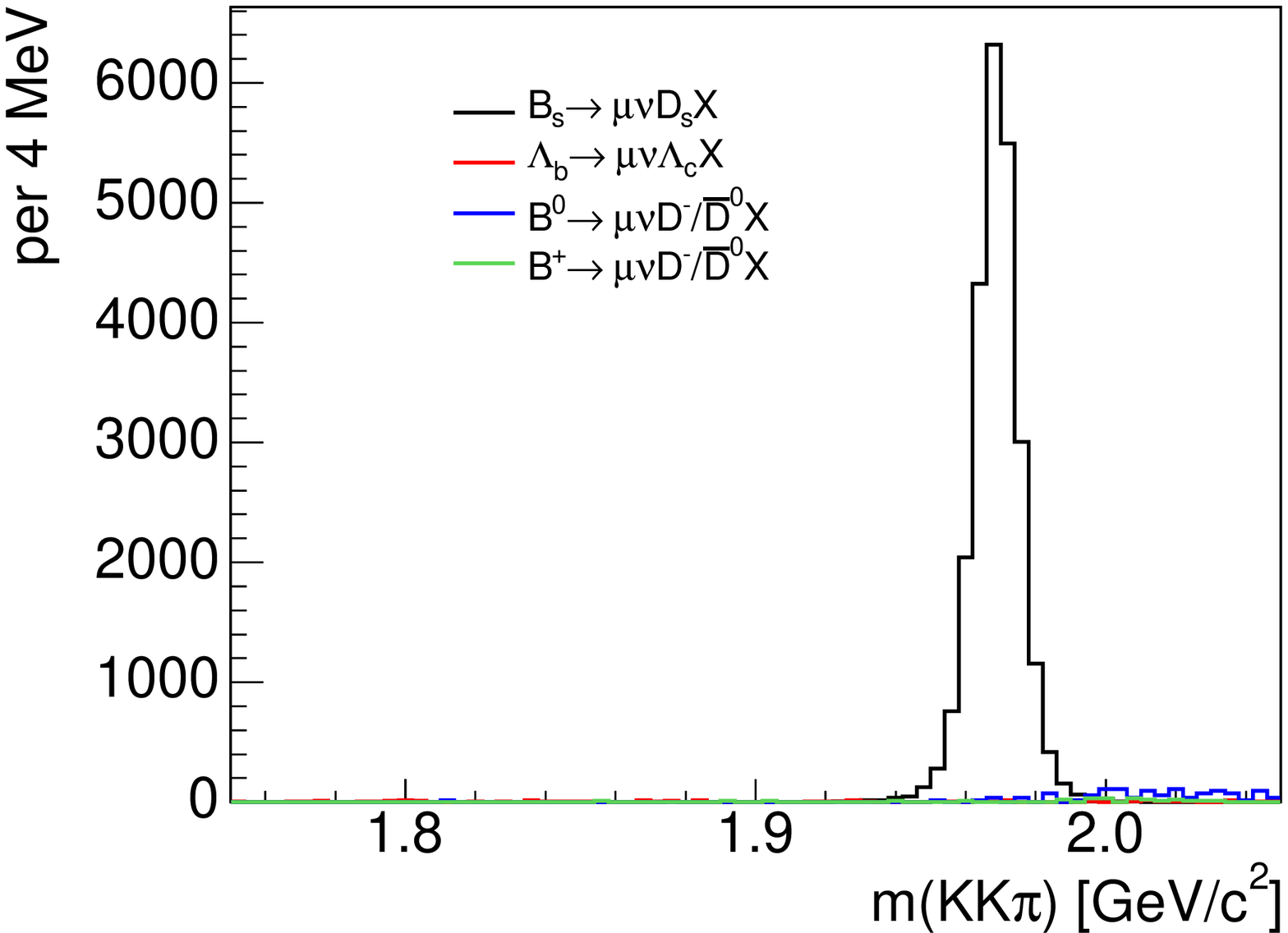}
\put(-415,135){\large\bf (c)}
\put(-177,135){\large\bf (d)}
}
\centerline{
\includegraphics[width=0.5\hsize]{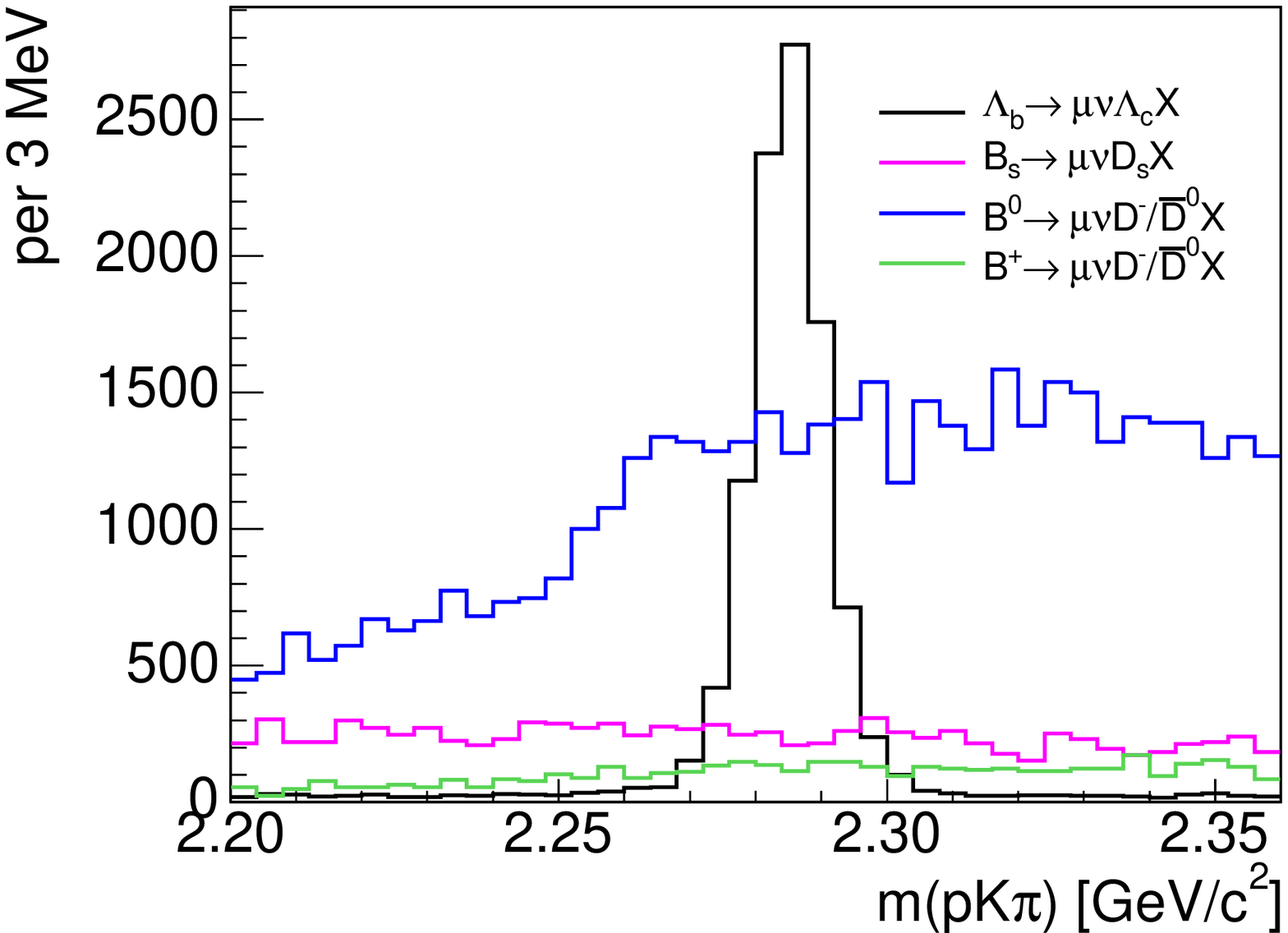}
\put(-175,135){\large\bf (e)}
}
\caption{\label{fig:reflections}
Monte Carlo simulation reflection shapes for (a)~$\Dz$, (b)~$\Dst$, 
(c)~$\Dp$, (d)~$\Ds$, and (e)~$\Lc$. The shapes are normalized to their
expected contributions, assuming $\fu:\fd:\fs:\fb=0.4:0.4:0.1:0.1$, used
for illustrative purposes only.}
\end{figure*}

The selection criteria discussed above (see Sec.~\ref{sec:sig_opt})
optimize the signal sensitivity with respect to the combinatoric
background. However, there are other non-combinatoric backgrounds that
must be considered. This is partially achieved with the $ct^*(\lD)$ and
$p_T(D)$ cuts discussed previously.  Another significant background
arises from reflections, which occur when the particle identifications
in charm decay are mis-assigned.  For example, if the $K^+$ from a
\DsKKpi~decay is assigned the pion mass, the $\pi^+K^-\pi^+$ combination
can contribute to the \Dp~signal. Figure~\ref{fig:reflections} shows the
shapes determined from MC for reflections from (a)~\Dz, (b)~\Dst, 
(c)~\Dp, (d)~\Ds, and (e)~\Lc~decays when these decay channels are
reconstructed as a different charm mode. The shapes are normalized to
their expected contributions, {\it e.g.} assuming
$\fu:\fd:\fs:\fb=0.4:0.4:0.1:0.1$, where these numerical values are for
illustrative purposes only.  The \DsKKpi~decay is the most significant
reflection background below the \Dp~signal, shown in
Fig.~\ref{fig:reflections}(c).  This reflection is particularly 
problematic because the \Ds~reflection begins just underneath the real
\Dp~signal.  Potential $p$-$\pi$ mis-identification is a significant
consideration in the \Lc~signal, shown in Fig.~\ref{fig:reflections}(e).
The \DpKpipi~decay significantly contributes to the background beneath
the \Lc~signal, although its contribution is flat underneath the signal.

\begin{figure}
\centerline{
\includegraphics[width=0.7\hsize]{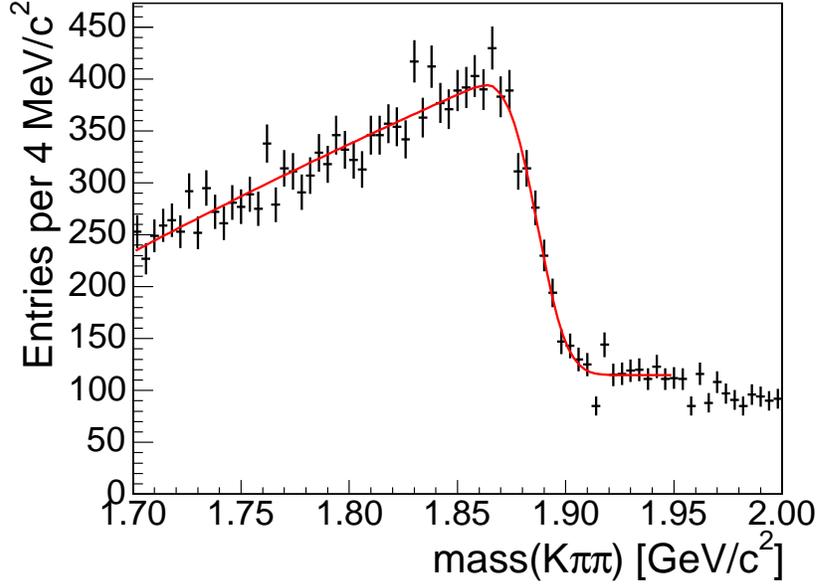}
}
\caption{\label{fig:ds_reflec}
Combined $\Ds e^-$ and $\Ds\mu^-$~reflection into the $K^+\pi^-\pi^-$
invariant mass.  The reflection is determined from an inclusive MC
sample of $\BsDsln X$, where all \Ds~meson decay modes are included.}
\end{figure}

\begin{figure*}
\centerline{
\includegraphics[width=0.5\hsize]{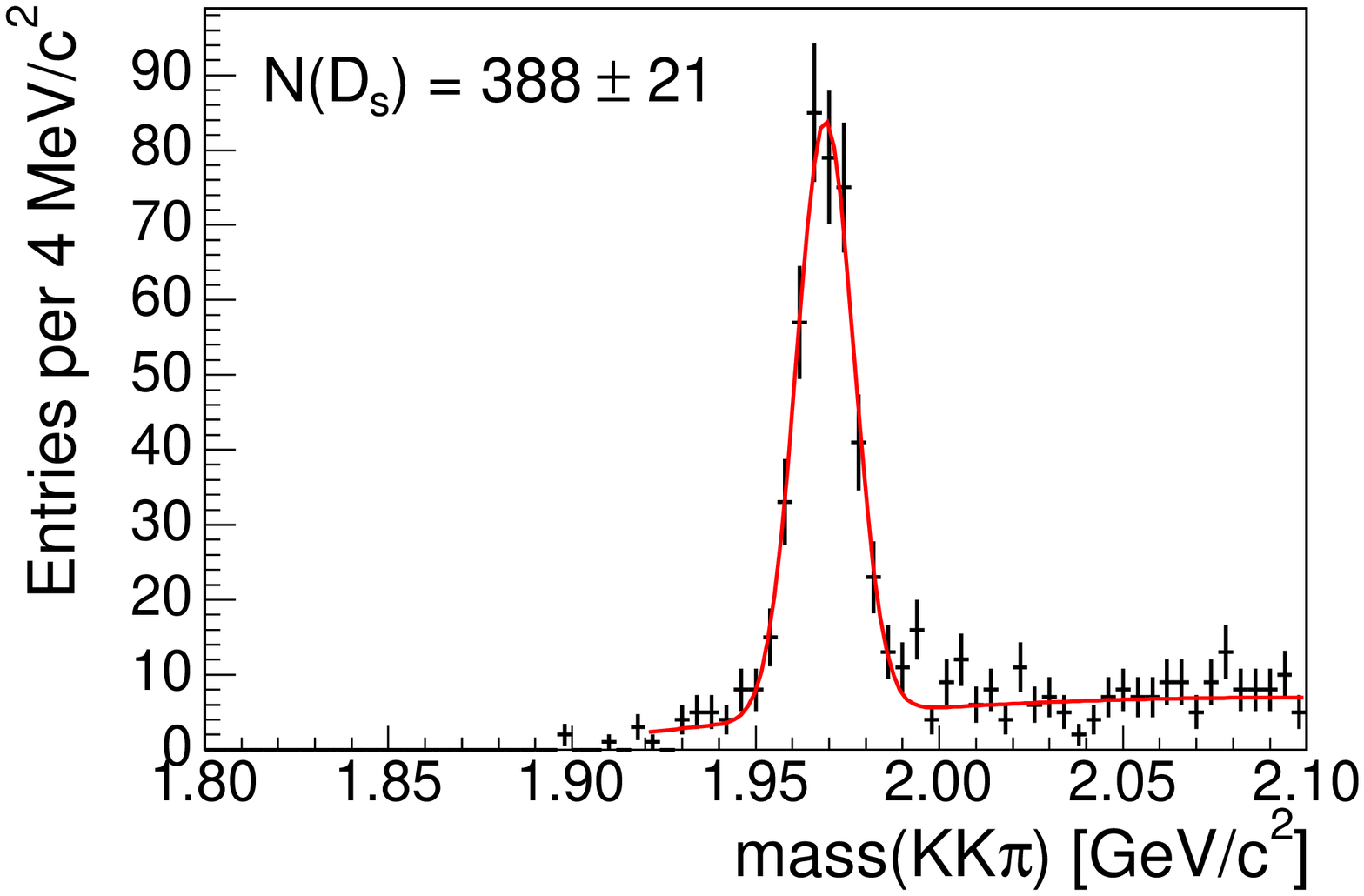}
\includegraphics[width=0.5\hsize]{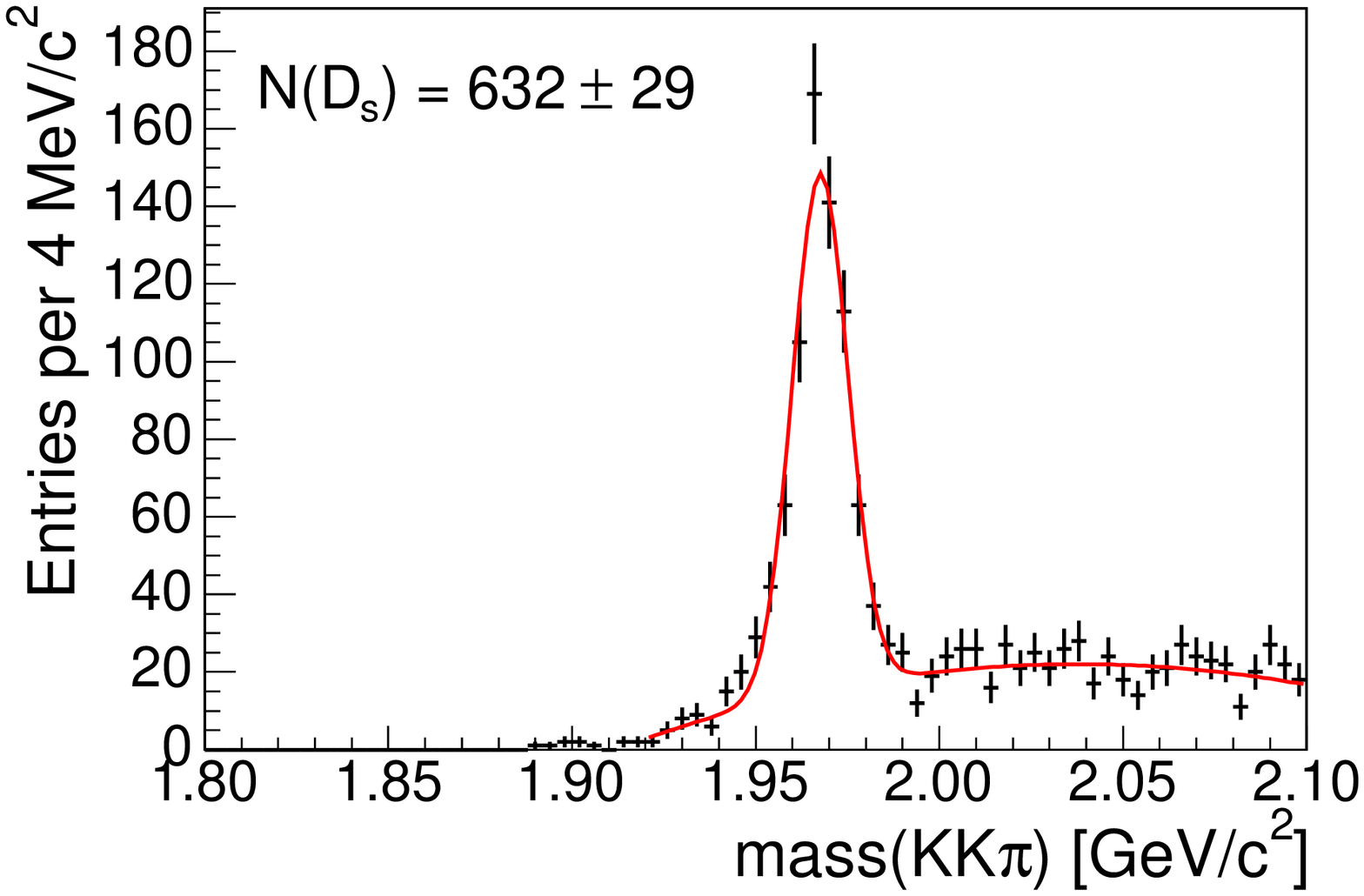}
\put(-280,135){\large\bf (a)}
\put(-45,135){\large\bf (b)}
}
\caption{\label{fig:ds_ref_data}
$N(\Dsphipi)$ reconstructed in $m(K\pi\pi)\in [1.78,1.95]$ in the
(a) $e$+SVT and (b) $\mu$+SVT data.}
\end{figure*}

The shape of the \DsKKpi~reflection background beneath the
\DpKpipi~signal is determined from a Monte Carlo simulation (see
Section~\ref{sec:mc}) study, in which semileptonic $\BsDsln X$~decays
are generated. In these MC events \DpKpipi~candidates are then
reconstructed. The resulting $K^-\pi^+\pi^+$ invariant mass
distribution is shown in Fig.~\ref{fig:ds_reflec}.  The normalization
of the \DsKKpi~reflection shape in the fit to the \Dp~signal is
determined by reconstructing a $\Ds\ra\phi\,(\ra
K^-K^+)\,\pi^+$~signal from the wide signal window, $1.78~\gevcc\leq
m(\DpKpipi)\leq 1.95~\gevcc$, shown in Fig.~\ref{fig:ds_ref_data}.  A
mass cut of $|m(K^+K^-) - 1.019~\gevcc| < 0.0095~\gevcc$, designed to
reduce background to the \Ds~signal, is applied to the $\phi\ra
K^+K^-$~decay.  Monte Carlo simulation is then used to measure the
efficiency of the \Dsphipi~decay relative to the inclusive set of
\DsKKpi~decays that contribute to the reflection.  The converse
\DpKpipi~reflection in the \DsKKpi~signal is negligible due to the
$\phi$ mass cut applied to the $K^+K^-$ invariant mass.

In a manner completely analogous to the way the \Ds~signal yield, $N_{\rm
data}(\Dsphipi)$, is determined in data, the \Ds~candidates decaying to the
$\phi\pi^+$ and $K^-K^+\pi^+$ states, $N_{MC}(\Dsphipi)$ and
$N_{MC}(\DsKKpi)$, respectively, are determined from the Monte Carlo
simulation.  The number of \Ds~mesons expected to contribute to the
\Dp~signal can then be calculated by evaluating
\begin{eqnarray}
N_{\rm data}(\DsKKpi) = \frac{N_{\rm data}(\Dsphipi)}{R_{\phi\pi}},
\end{eqnarray}
where 
\begin{eqnarray}
R_{\phi\pi}\equiv\frac{N_{MC}(\Dsphipi)}{N_{MC}(\DsKKpi)}=0.246\pm 0.016.
\end{eqnarray}
The numbers of \Ds~candidates that contribute to the \Dp~lepton-charm
samples in the wide mass window around the \Dp~signal are
$N_{e}(\DsKKpi)= 1580\pm130$ and $N_{\mu}(\DsKKpi)=2570\pm210$.  The
normalization of the \Ds~reflection in the \Dp~signal is later
constrained to the predicted number of \Ds~reflection events in the
fit to the \Dp~signal (see Sec.~\ref{sec:sig_yield}).

Since the \Dp\ and \Ds~reflections in the \Lc~signal are relatively
flat under the signal region, sideband subtraction is expected to
remove the effect of the \Dp\ and \Ds~reflections on the \Lc~signal
distributions within statistical uncertainty.  Correspondingly, the
event count obtained by fitting the \Lc~signal is not expected to be
significantly influenced by the presence of these
backgrounds. Additionally, the \dedx\ cut applied to the proton
(discussed in the previous section) reduces contamination from pions,
which contribute to the \DpKpipi\ and $\Dst\ra [K^-\pi^+]\pi^+$
reflections.

\begin{figure*}
\centerline{
\includegraphics[width=0.5\hsize]{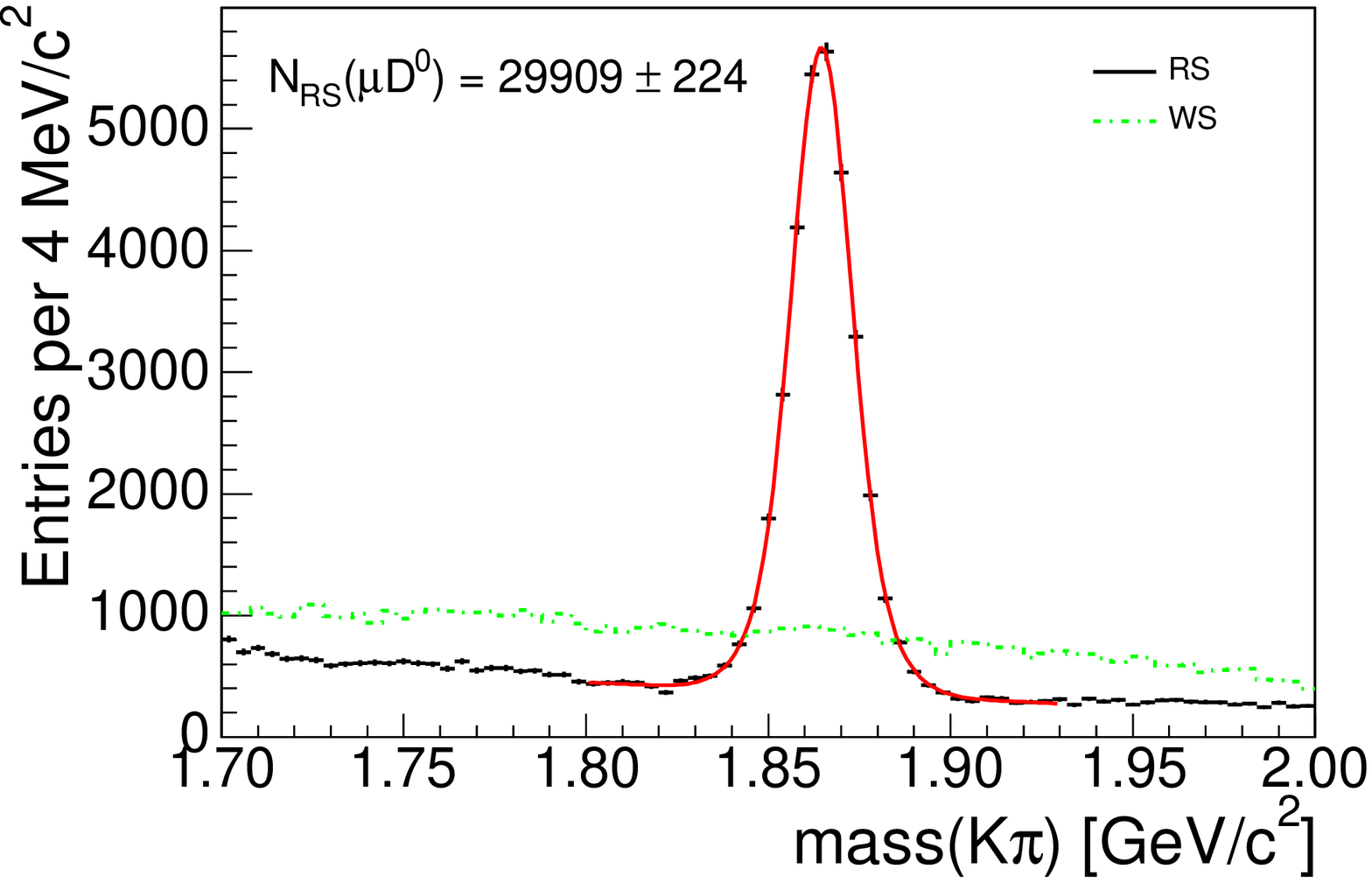}
\includegraphics[width=0.5\hsize]{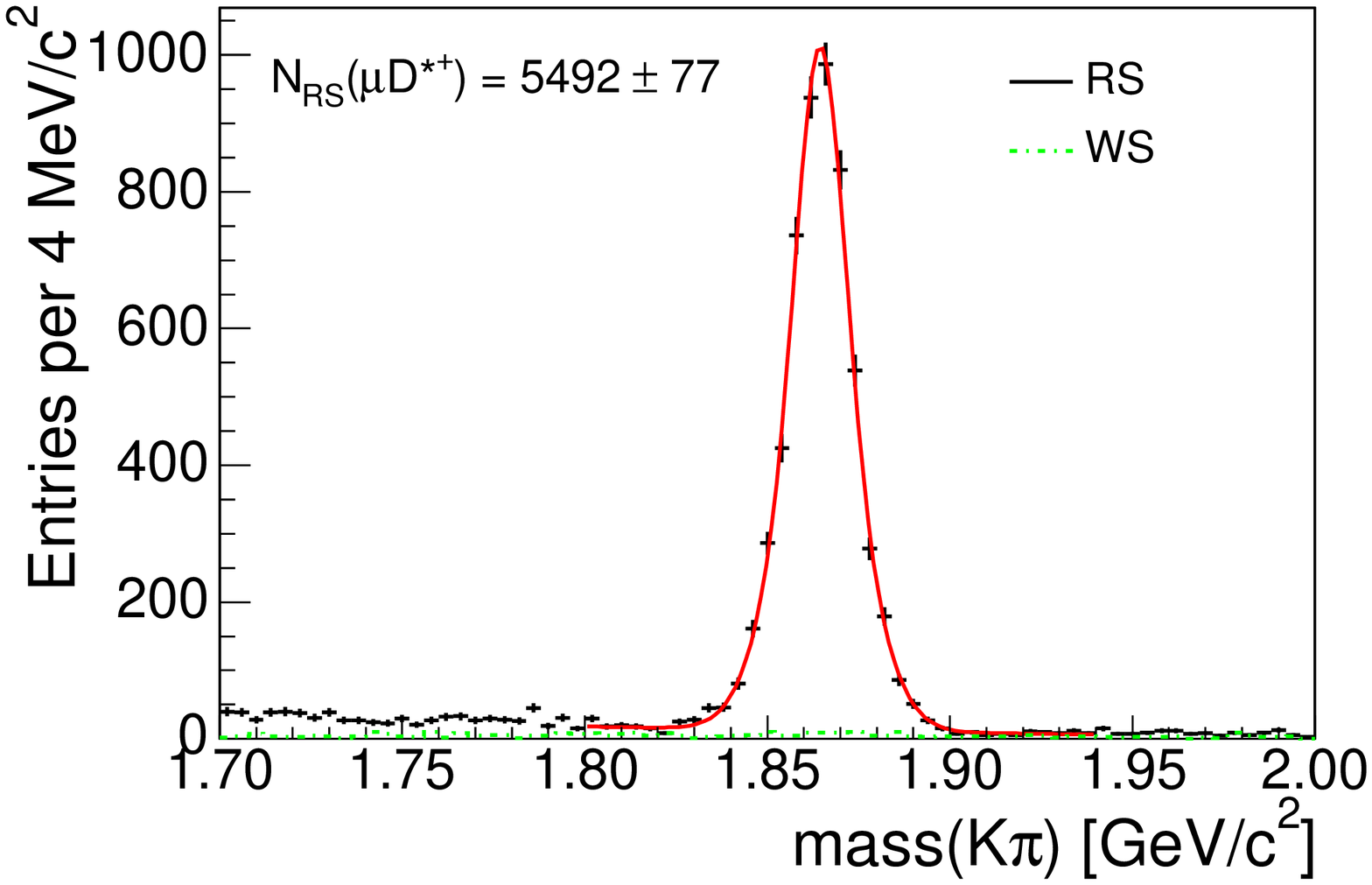}
\put(-415,118){\large\bf (a)}
\put(-177,118){\large\bf (b)}
}
\centerline{
\includegraphics[width=0.5\hsize]{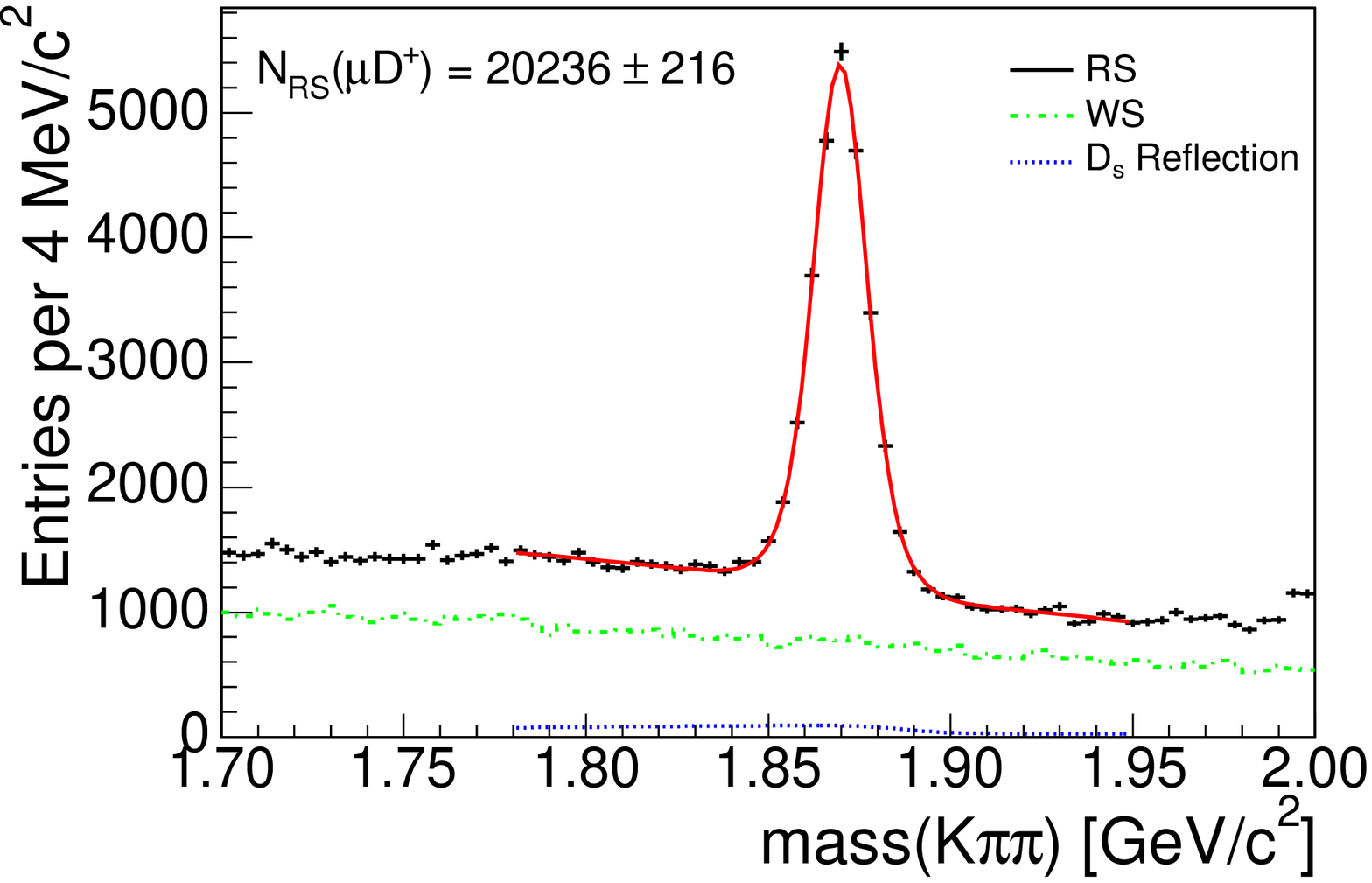}
\includegraphics[width=0.5\hsize]{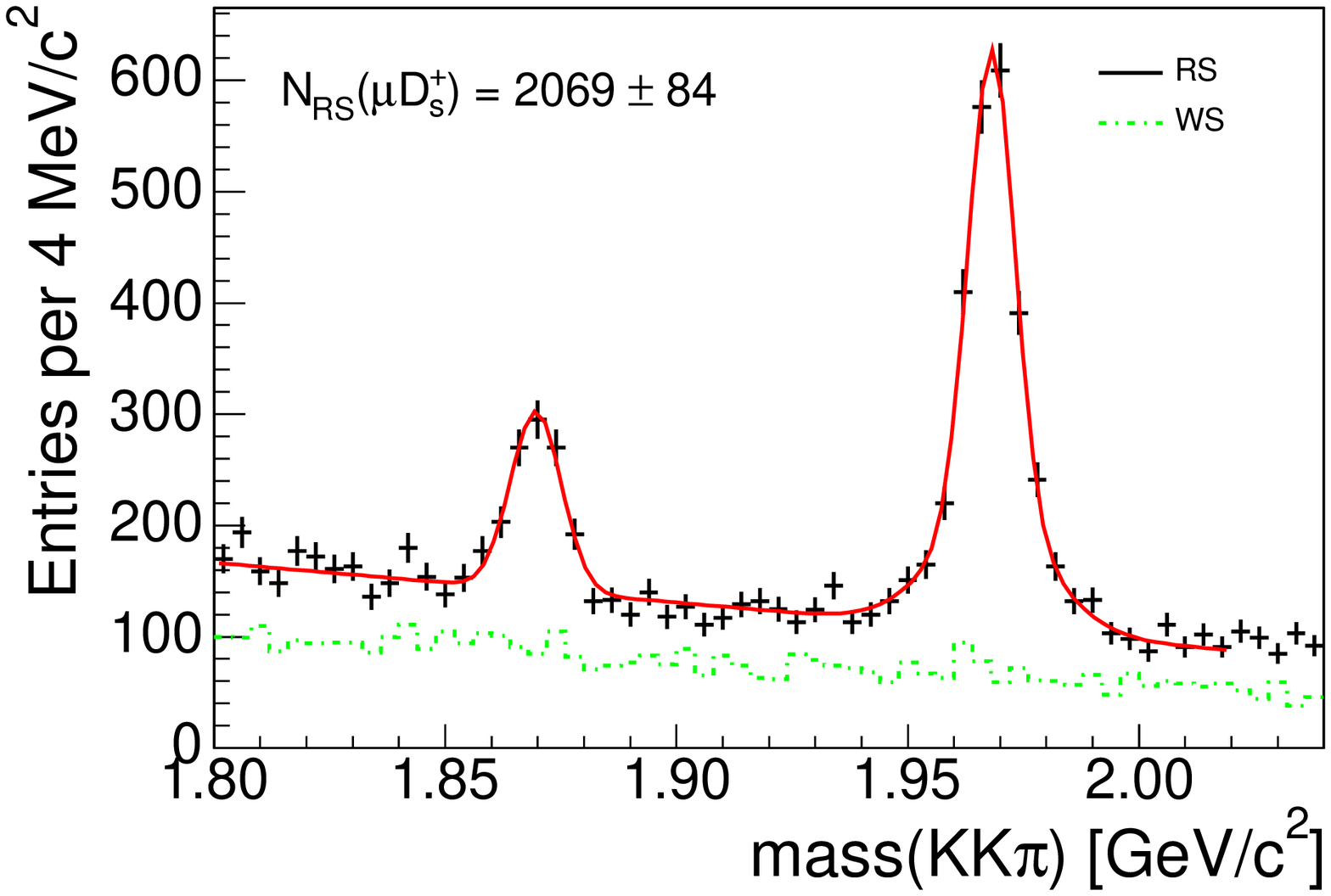}
\put(-415,118){\large\bf (c)}
\put(-177,118){\large\bf (d)}
}
\centerline{
\includegraphics[width=0.5\hsize]{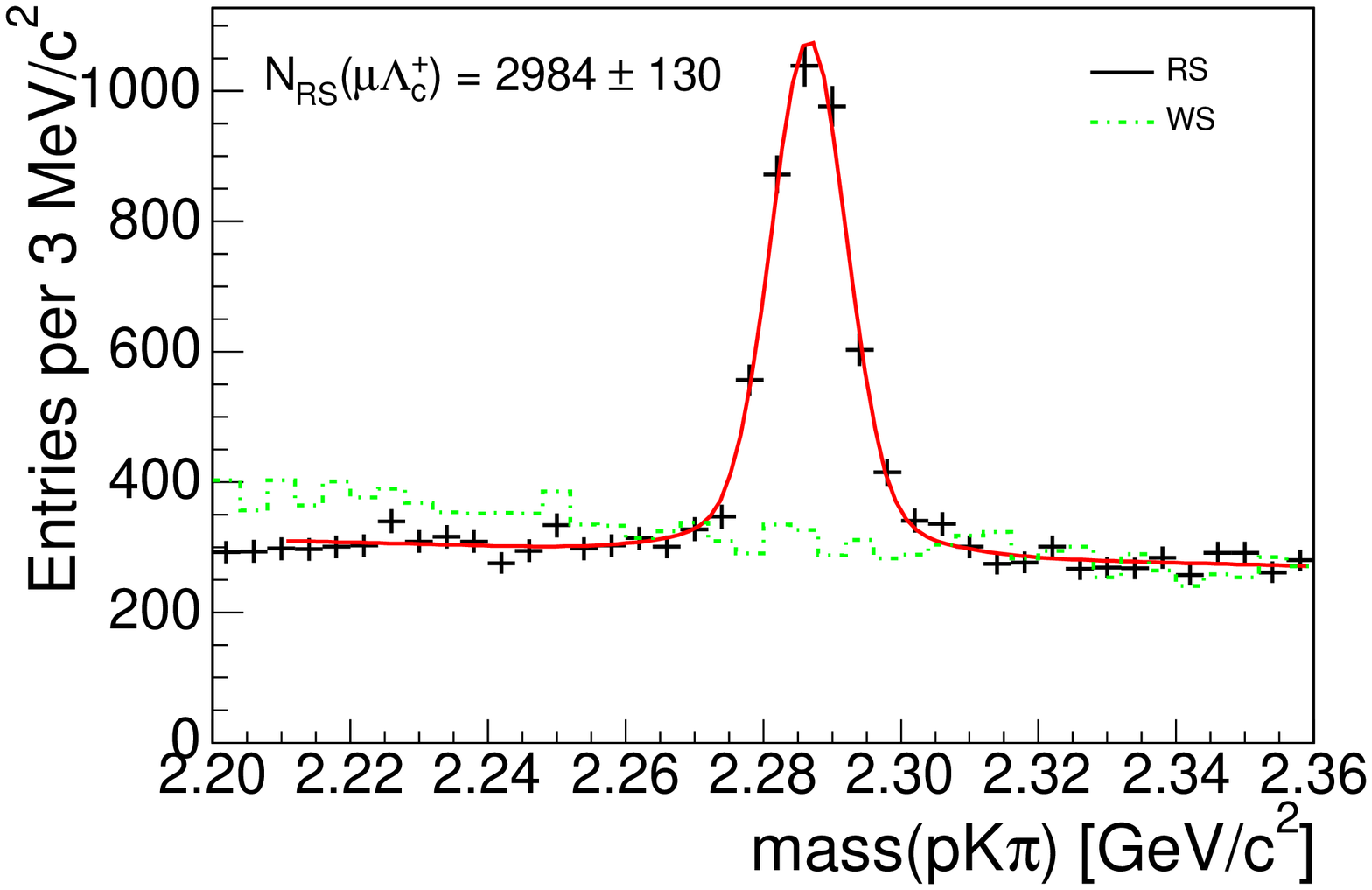}
\includegraphics[width=0.5\hsize]{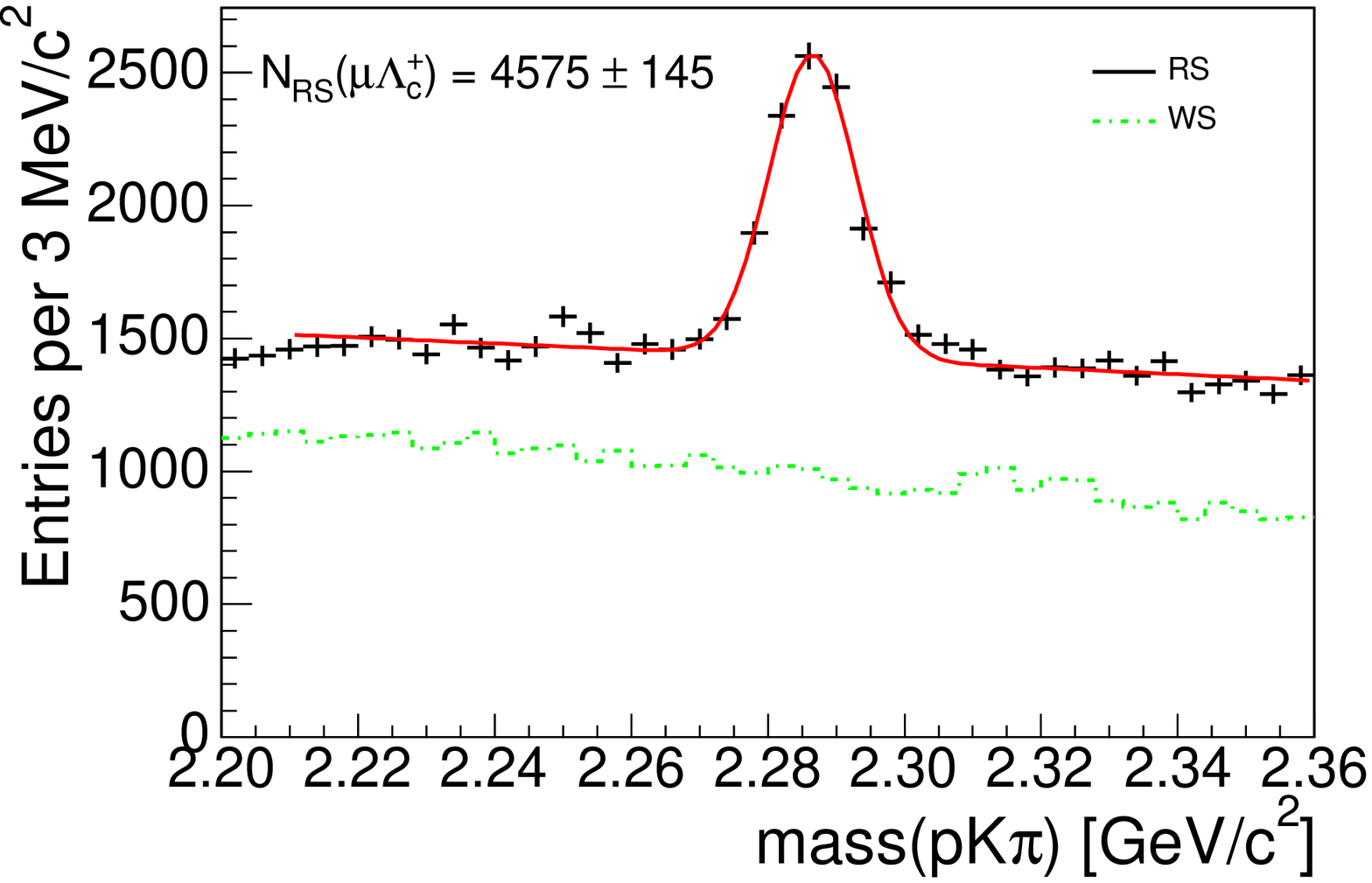}
\put(-415,120){\large\bf (e)}
\put(-177,120){\large\bf (f)}
}
\caption{\label{fig:mu_signals}
$\mu$+SVT right sign (RS) (points with error bars) and wrong sign (WS)
(histogram) invariant mass distribution of (a) $\Dz$, (b) $\Dst$,
(c) \Dp, (d) \Ds,
(e) \Lc~with all cuts applied and (f) without the \dedx\
cut applied. The fit parameterizations described in the
text are overlaid.}
\end{figure*}

\begin{figure*}
\centerline{
\includegraphics[width=0.5\hsize]{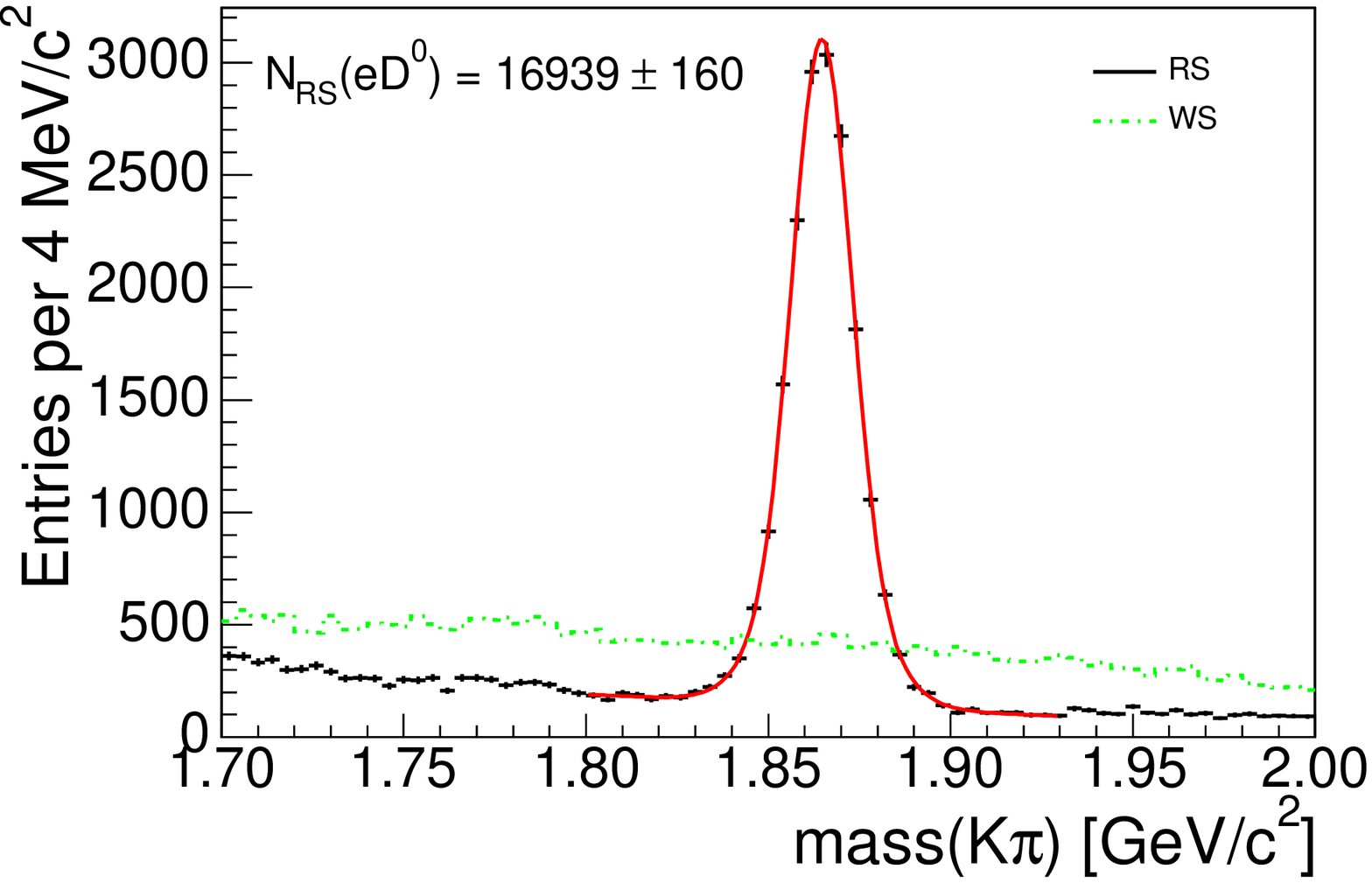}
\includegraphics[width=0.5\hsize]{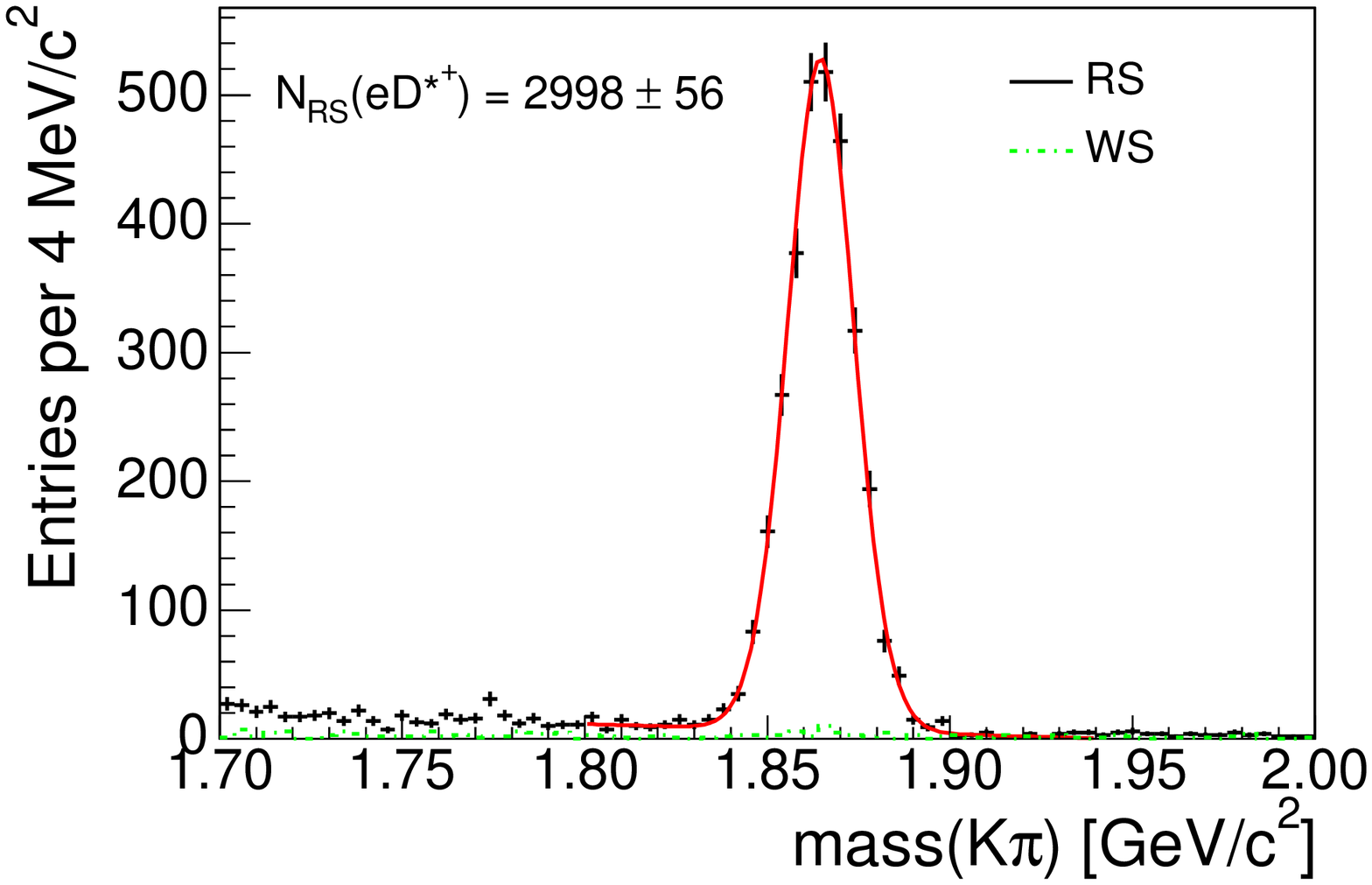}
\put(-415,118){\large\bf (a)}
\put(-177,118){\large\bf (b)}
}
\centerline{
\includegraphics[width=0.5\hsize]{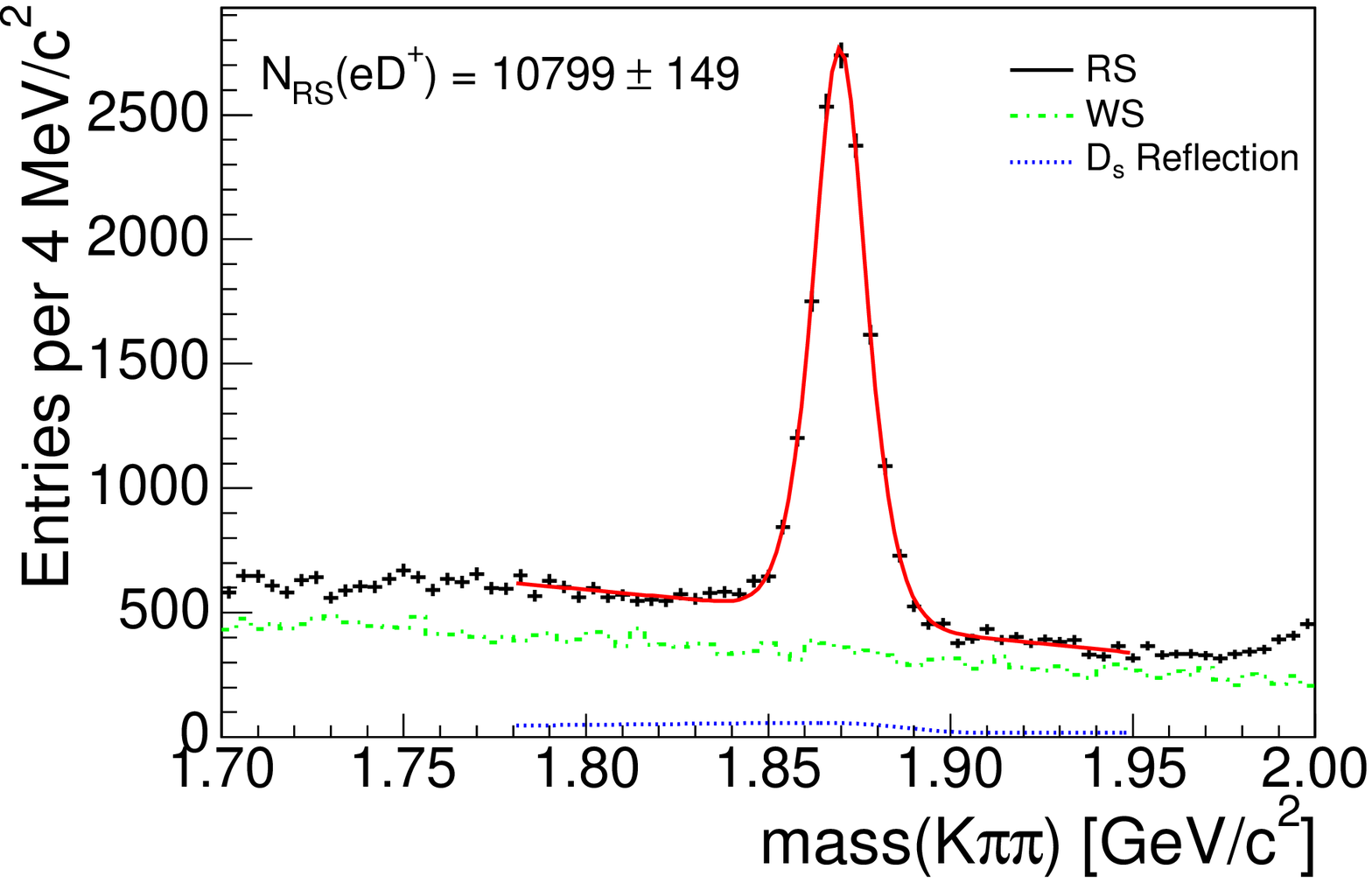}
\includegraphics[width=0.5\hsize]{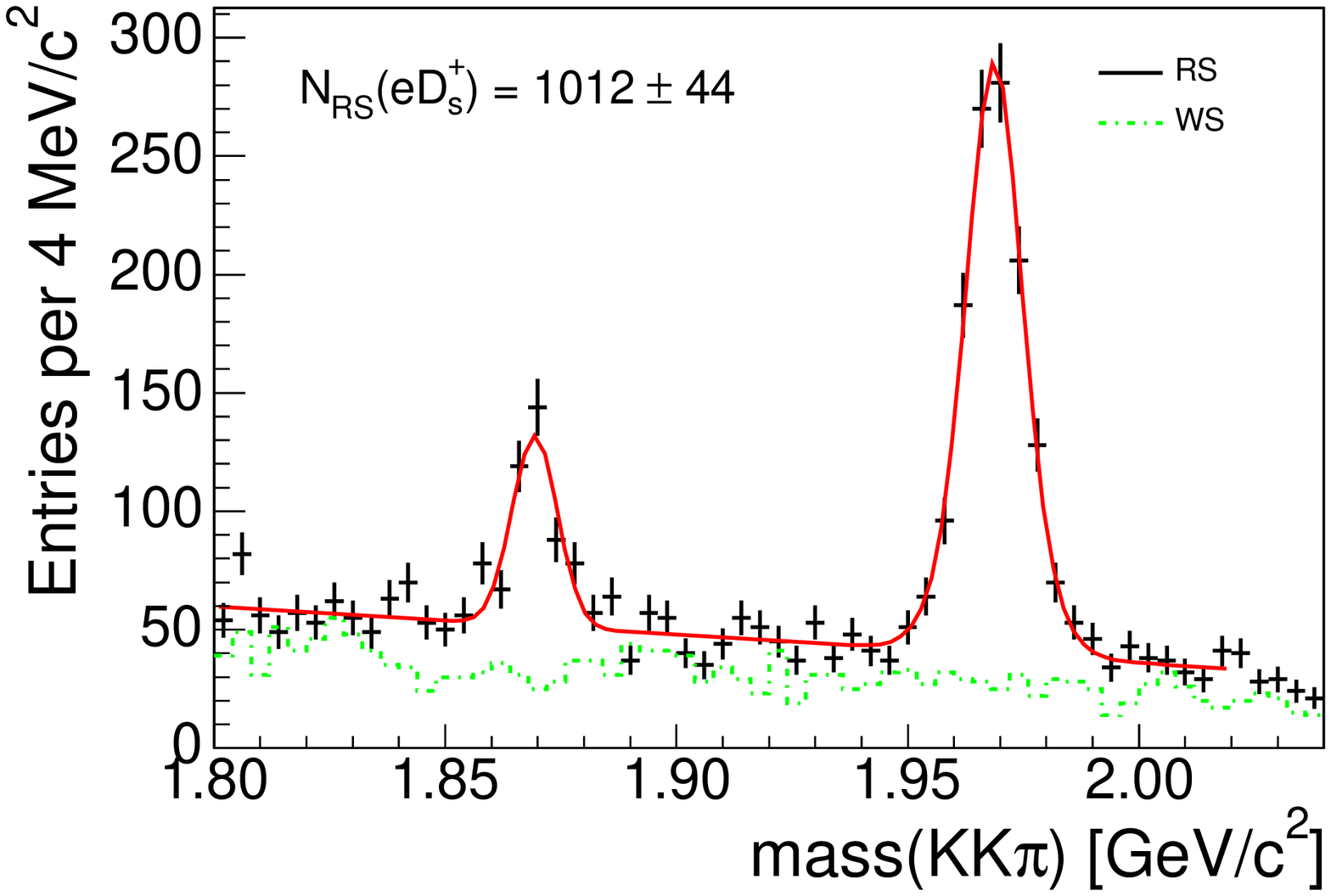}
\put(-415,118){\large\bf (c)}
\put(-177,118){\large\bf (d)}
}
\centerline{
\includegraphics[width=0.5\hsize]{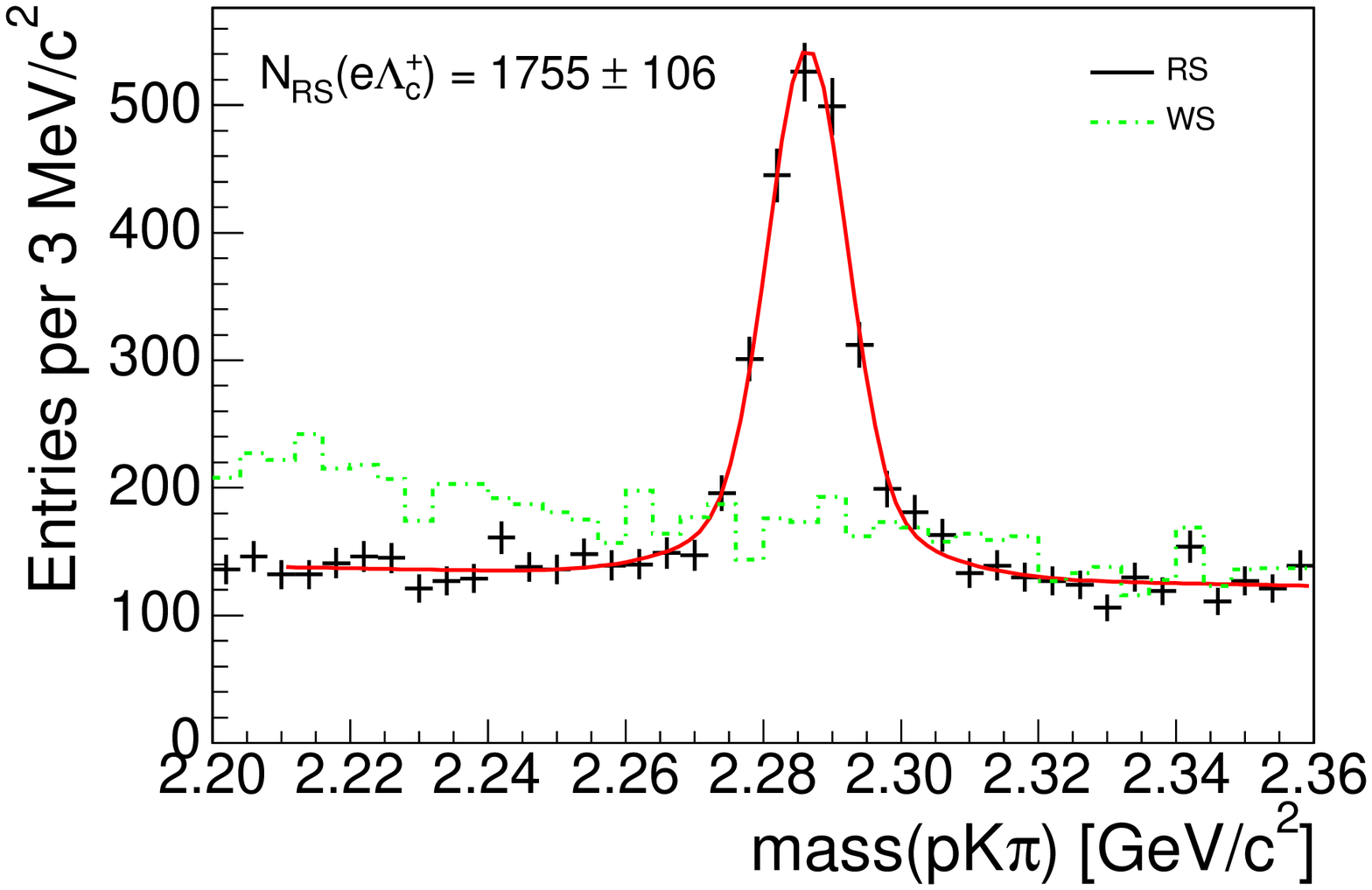}
\includegraphics[width=0.5\hsize]{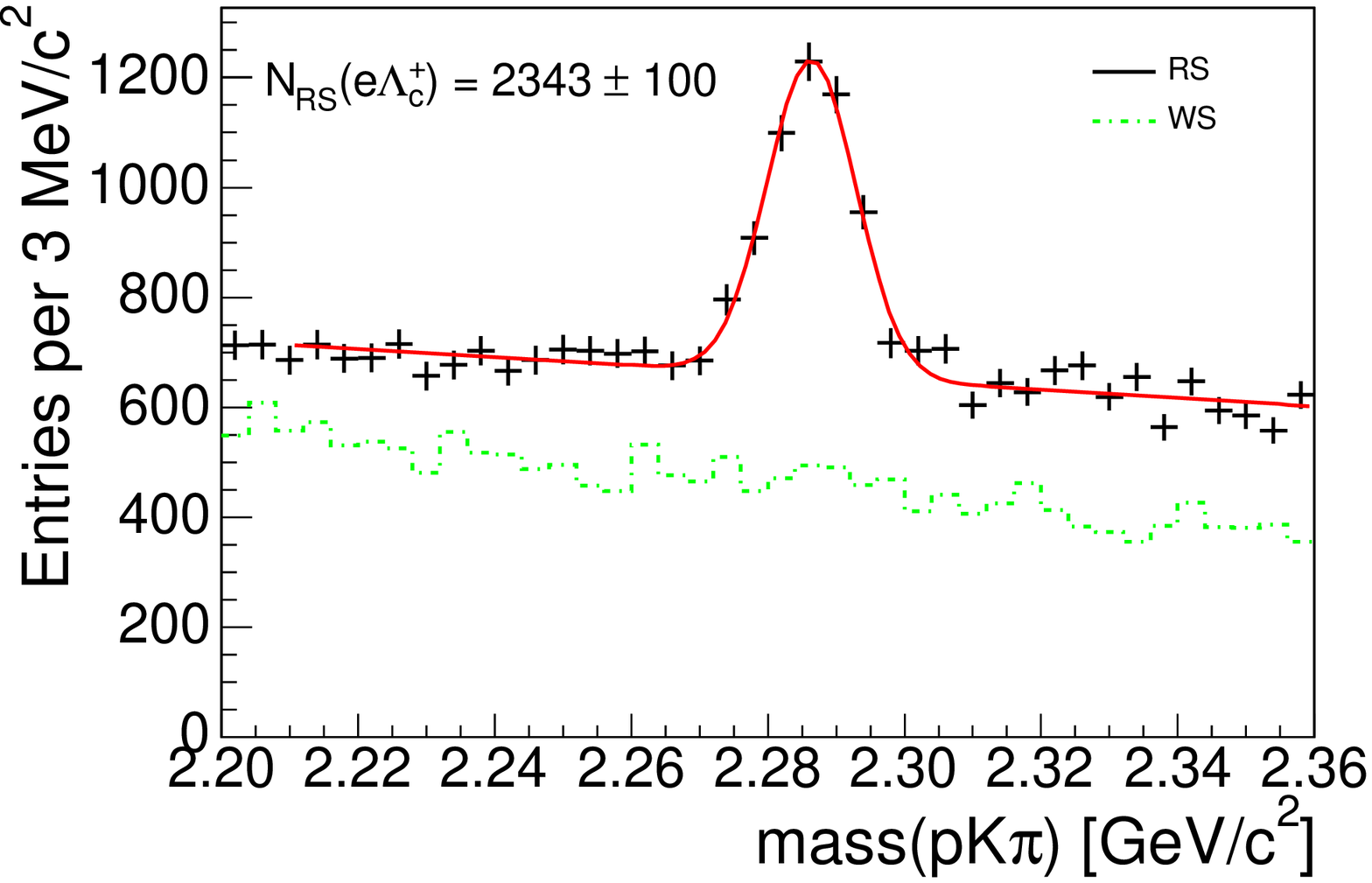}
\put(-415,120){\large\bf (e)}
\put(-177,120){\large\bf (f)}
}
\caption{\label{fig:el_signals}
$e$+SVT right sign (RS) (points with error bars) and wrong sign (WS)
(histogram) invariant mass distribution of (a) $\Dz$, (b) $\Dst$, (c)
\Dp, (d) \Ds, (e) \Lc~with all cuts applied and (f) without the
\dedx\ cut applied. The fit parameterizations described
in the text are overlaid.}
\end{figure*}

\subsection{\label{sec:sig_yield} Signal Yields} 

The $m(K^-\pi^+\pi^+)$, $m(K^-pi^+)$, $m(K^+K^-pi^+)$, and
$m(pK^-pi^+)$ mass spectra are fit to determine the number of
lepton-charm events for the $\lDp$, $\lDz$, $\lDst$, $\lDs$, and
$\lLc$ samples.  The invariant mass distributions of the charm signals
are shown in Fig.~\ref{fig:mu_signals} for the $\mu$+SVT data and in
Fig.~\ref{fig:el_signals} for the $e$+SVT data with all lepton, charm,
and lepton-charm selection criteria applied.  The distributions are
fit with a double Gaussian and linear background shape.  The
reflection of \Ds~decays into the \Dp~final state is included in the
fit to the \Dp~signal.  The normalization of the \Ds~reflection is
constrained to the predicted number of \Ds~reflection events as
described above.  In order to keep the broad Gaussian and reflection
shapes reasonably independent, the double Gaussian means and widths
for the \Dp~are determined before the reflection shape is added to the
fit.  When the combined fit is performed, the parameters of the double
Gaussian are constrained within their uncertainties.  The fits to the
\Dz, \Dst, \Dp, \Ds, and \Lc~charm signals for right sign lepton-charm
pairs are shown in Fig.~\ref{fig:mu_signals} for the $\mu$+SVT data and
in Fig.~\ref{fig:el_signals} for the $e$+SVT data. The invariant mass
distributions for wrong sign combinations of lepton-charm pairs, {\it
e.g.}  $\Dp\ell^+$, are also included in Figs.~\ref{fig:mu_signals} and
\ref{fig:el_signals}, indicating no significant contributions of
possible backgrounds, such as false leptons, to be present in the right
sign signals (see also Sec.~\ref{sec:ws}).  The fitted lepton-charm
yields are listed in Table~\ref{tab:yields}.  The
\Ds~reflection is not included in the \Dp~yield, since the fit shape
to the \Dp~includes a separate shape for the \Ds~reflection, as
discussed in Section~\ref{sec:reflec}. The \dedx\ cut flattens the
background and reduces its overall level by a factor of five, while it
reduces the signal by $\sim$\,35\% in the $\mu$+SVT data and
$\sim$\,28\% in the $e$+SVT data as can be seen in 
Fig.~\ref{fig:mu_signals}(e)-(f) and Fig.~\ref{fig:el_signals}(e)-(f).

\begin{table}
\caption{\label{tab:yields}
Fitted signal yields for lepton-charm final states in 360~pb$^{-1}$.}
\begin{ruledtabular}
\begin{tabular}{l|ccc|ccc}
        & \multicolumn{3}{c|}{$e$+SVT}  & \multicolumn{3}{c}{$\mu$+SVT} \\
Decay   & Yield  & FOM & Fit Prob. [\%] & Yield  & FOM & Fit Prob. [\%] \\     
\hline
$\lDz$  & $16,939\pm160$ & 122  & 64.4 & $29,909\pm224$ & 159  & 12.5 \\
$\lDst$ & $ 2,998\pm56 $ & 54.1 & 1.27 & $ 5,492\pm77 $ & 73.3 & 1.14 \\
$\lDp$  & $10,779\pm149$ & 90.2 & 9.43 & $20,236\pm216$ & 114  & 50.7 \\
$\lDs$  & $ 1,012\pm44 $ & 27.3 & 7.84 & $ 2,069\pm84 $ & 36.6 & 30.2 \\
$\lLc$  & $ 1,755\pm106$ & 32.8 & 33.9 & $ 2,984\pm130$ & 40.9 & 40.9 \\ 
\end{tabular}
\end{ruledtabular}
\end{table}

\section{\label{sec:sample_comp}
Sample Composition Determination Procedure}

This measurement uses flavor SU(3) symmetry to describe the branching
fractions of semileptonic \Bb~meson decays; therefore, the partial
widths of the semileptonic decays of \Bb~mesons are chosen to be
equal, namely
\begin{eqnarray}
\Gamma(\Bd\ra \ell^-\bar{\nu}_{\ell} X)
=\Gamma(\Bu\ra \ell^-\bar{\nu}_{\ell} X)
=\Gamma(\Bs\ra \ell^-\bar{\nu}_{\ell} X),
\end{eqnarray}
where
\begin{eqnarray}
\Gamma(\Bb\ra \ell^-\bar{\nu}_{\ell} X) =
\frac{1}{\tau(\Bb)}{\cal B}(\Bb\ra \ell^-\bar{\nu}_{\ell} X).
\end{eqnarray}

This assumption is referred to as the spectator model, which also
implies that the partial widths of the semileptonic bottom hadron
decays into the pseudoscalar, vector, or higher excited $D$~states are
expected to be equal,
\begin{eqnarray}
\Gamma(\BdDpln)
=\Gamma(\BuDzln)
=\Gamma(\BsDsln)
=\Gamma(\BDln),
\end{eqnarray}
with similar relations holding for $D^*$ and $D^{**}$ decays.  The
additional constraint that
\begin{eqnarray}
\Gamma(\BDln)
+\Gamma^*(\Bb\ra \ell^-\bar{\nu}_{\ell}D^*)
+\Gamma^{**}(\Bb\ra \ell^-\bar{\nu}_{\ell}D^{**}) 
= \Gamma(\Bb\ra\ell^-\bar{\nu}_{\ell} X),
\end{eqnarray}
is also applied to the partial widths.  This constraint includes
non-resonant decays and $b\ra u$ transitions in addition to actual
$D^{**}$ decays in the $D^{**}$ partial width.  Since excited
\Lb~semileptonic decays are not necessarily well-described by the
spectator model, fixed branching fractions are used to
describe those decays~\cite{Ref:Yu} (see Table~\ref{tab:Lb}).

A simplified example illustrating the procedure used to extract the
sample composition follows.  Assuming that the only source of
\lDp~combinations is from the direct decay of a neutral \Bb~meson,
such as \BdDpln, the number of reconstructed \lDp~events can be
expressed as
\begin{eqnarray}
N(\lDp) &=& N(\Bd)
\times{\cal B}(\BdDpln)
\times{\cal B}(\DpKpipi)\nonumber  \\
&&\times\ \varepsilon(\BdDpln,\, \DpKpipi)\nonumber\\ 
&=& \left[N(\Bb)\cdot \fd\right]
\times \left[\tau(\Bd)\cdot\Gamma(\BdDpln)\right]
\times{\cal B}(\DpKpipi) \nonumber\\
&& \times\ \varepsilon(\BdDpln,\, \DpKpipi).
\label{eqn:sc_Dp_only}
\end{eqnarray}
The number of reconstructed \lDp~combinations, $N(\lDp)$, can be
related to the number of \Bd~mesons, $N(\Bd)$, produced in the
fragmentation process by the branching fraction of the
\BdDpln~decay, the branching fraction of the \DpKpipi~charm decay, and the
detection and reconstruction efficiencies for the entire decay chain.
$N(\Bd)$ and $N(\Bb)$ represent the number of \Bd~and generic bottom
hadrons produced, respectively.

\begin{figure}
\centerline{
\includegraphics[width=0.6\hsize]{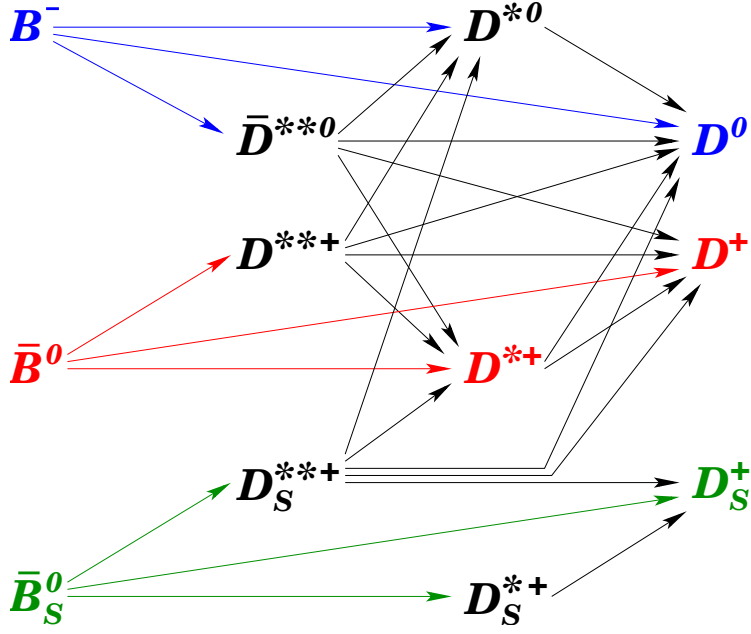}
}
\caption{\label{fig:Sample_combo_sketch}
Illustration of cross talk between $B$-meson species.}
\end{figure}
However, as sketched in Fig.~\ref{fig:Sample_combo_sketch}, cross-talk
between the various bottom hadron species via the excited charm states
necessitates a sample composition parameterization to relate the
lepton-charm signals to the parent bottom~hadrons.  Introducing the
relative fragmentation fractions, the sample composition for the \Bd~can be
written (for illustrative purposes) as:
\begin{eqnarray}
N(\lDp)&=&N(\Bd)\times{\cal B}(\DpKpipi)\nonumber\\
&&\times \left[
\sum_{\Bd\ra \Dp X}{\cal B}(\BdDpln X)\,
\varepsilon(\BdDpln X)\right.\nonumber\\
&&+\frac{\fu}{\fd}
\sum_{\Bu\ra \Dp X}{\cal B}(\Bu\ra\ell^-\bar{\nu}_{\ell}\Dp X)\,
\varepsilon(\Bu\ra\ell^-\bar{\nu}_{\ell}\Dp X)\nonumber\\
&&\left. +\frac{\fs}{\fu+\fd}\left(1+\frac{\fu}{\fd}\right)
\sum_{\Bs\ra \Dp X}{\cal B}(\Bs\ra\ell^-\bar{\nu}_{\ell}\Dp X)\,
\varepsilon(\Bs\ra\ell^-\bar{\nu}_{\ell}\Dp X)\right].
\label{eqn:sc_Dp}
\end{eqnarray}

The sample composition parameterization requires knowledge of the
branching fractions of the charm hadrons, which are determined from
the world average values as compiled by the Particle Data Group
(PDG)~\cite{Ref:PDG_2004}, when available, and from theoretical
predictions and symmetry principles~\cite{Ref:EvtGen}, when not
available in the PDG.  The parameterization also requires knowledge of
the efficiency of reconstructing the decay, which is primarily
determined from Monte Carlo simulations.  In addition to the primary
decays that contribute to the semileptonic signal, indirect
semileptonic decays ({\it e.g.}  $\Bb\ra D\bar{D}X,\ \bar D\ra\ell^-
X$) contributing to the \lD~final state are also included in the
parameterization of the sample composition.  All of the decays
considered in the bottom hadron sample composition procedure are
listed in Table~\ref{tab:sc} (see also
Fig.~\ref{fig:Sample_combo_sketch}).

\linespread{0.8}
\begin{table*}
\caption{\label{tab:sc}
Bottom hadron semileptonic sample composition used in the
measurement. NR refers to non-resonant decays.}
\begin{ruledtabular}
\begin{tabular}{llll}
\Bd                       &\Bu                          &\Bs                                &\Lb                      \\ 
\hline									
$\ell^-\bar{\nu}\Dp$	  &$\ell^-\bar{\nu}\Dz$		&$\ell^-\bar{\nu}\Ds$	            &$\ell^-\bar{\nu}\Lc$       \\	
$\ell^-\bar{\nu}\Dst$	  &$\ell^-\bar{\nu} D^{*0}$	&$\ell^-\bar{\nu} D_s^{*+}$         &$\ell^-\bar{\nu}\Lambda_c(2593)^+$ \\
$\hsps\ra\Dz\pi^+$	  &$\hsps\ra\Dz\piZg$  	        &$\hsps\ra\Ds\gamma$                &$\hsps\ra\Sigma_c(2455)^{++}\pi^-$ \\
$\hsps\hsps\Dp\piZg$      &     			&		                    &$\hsps\hsps\hra\Lc\pi^+$   \\
			  &                             &                                   &$\hsps\ra\Sigma_c(2455)^0\pi^+$  	\\
$\ell^-\bar{\nu} D_1^+$	  &$\ell^-\bar{\nu} D_1^0$	&$\ell^-\bar{\nu} D_{s1}^{+}(2460)$ &$\hsps\hsps\hra\Lc\pi^-$   \\
$\hsps\ra D^{*0}\pi^+$    &$\hsps\ra D^{*0}\pi^0$	&$\hsps\ra D_{s0}^{*+}\pi^0$        &$\hsps\ra\Sigma_c(2455)^+\pi^0$  	\\
$\hsps\hsps\hra\Dz\piZg$  &$\hsps\hsps\hra\Dz\piZg$     &$\hsps\hsps\hra\Ds\pi^0$           &$\hsps\hsps\hra\Lc\pi^0$   \\
$\hsps\ra\Dst\pi^0$       &$\hsps\ra\Dst\pi^-$          &$\hsps\ra\Ds\gamma$                &$\hsps\ra\Lc\pi^+\pi^-$    \\
$\hsps\hsps\hra\Dz\pi^+$  &$\hsps\hsps\hra\Dz\pi^+$     &                                   &$\hsps\ra\Lc\pi^0\pi^0$    \\
$\hsps\hsps\hsps\Dp\piZg$ &$\hsps\hsps\hsps\Dp\piZg$    &                                   &$\hsps\ra\Lc\gamma$        \\
			  &                             &	 		            &                           \\
$\ell^-\bar{\nu} D_0^{*+}$&$\ell^-\bar{\nu} D_0^{*0}$	&$\ell^-\bar{\nu} D_{s0}^{*+}(2317)$&$\ell^-\bar{\nu}\Lambda_c(2625)^+$ \\
$\hsps\ra\Dz\pi^+$	  &$\hsps\ra\Dz\pi^0$	        &$\hsps\ra\Ds\pi^0$                 &$\hsps\ra\Lc\pi^+\pi^-$    \\
$\hsps\hsps\Dp\pi^0$      &$\hsps\hsps\Dp\pi^-$	        &                                   &$\hsps\ra\Lc\pi^0\pi^0$    \\
			  &                             &                                   &$\hsps\ra\Lc\gamma$        \\
$\ell^-\bar{\nu} D_1^{'+}$&$\ell^-\bar{\nu} D_1^{'0}$	&$\ell^-\bar{\nu} D_{s1}^{'+}(2535)$&                           \\
$\hsps\ra D^{*0}\pi^+$    &$\hsps\ra D^{*0}\pi^0$	&$\hsps\ra\Dst K^0$                 &$\ell^-\bar{\nu}\Sigma_c(2455)^{++}\pi^-$\\  
$\hsps\hsps\hra\Dz\piZg$  &$\hsps\hsps\hra\Dz\piZg$     &$\hsps\hsps\hra\Dz\pi^+$           &$\hsps\ra\Lc\pi^+$         \\	  
$\hsps\ra\Dst\pi^0$       &$\hsps\ra\Dst\pi^-$	        &$\hsps\hsps\hsps\Dp\piZg$          &                           \\	  
$\hsps\hsps\hra\Dz\pi^+$  &$\hsps\hsps\hra\Dz\pi^+$     &$\hsps\ra D^{*0}K^+$               &$\ell^-\bar{\nu}\Sigma_c(2455)^0\pi^+$\\ 	  
$\hsps\hsps\hsps\Dp\piZg$ &$\hsps\hsps\hsps\Dp\piZg$    &$\hsps\hsps\hra\Dz\piZg$           &$\hsps\ra\Lc\pi^-$         \\  	  
			  &                             &                                   &                           \\	  
$\ell^-\bar{\nu} D_2^{*+}$&$\ell^-\bar{\nu} D_2^{*0}$	&$\ell^-\bar{\nu} D_{s2}^{'+}(2573)$&$\ell^-\bar{\nu}\Sigma_c(2455)^+\pi^0$\\ 	  
$\hsps\ra D^{*0}\pi^+$    &$\hsps\ra D^{*0}\pi^0$	&$\hsps\ra\Dst K^0$	            &$\hsps\ra\Lc\pi^0$         \\  	  
$\hsps\hsps\hra\Dz\piZg$  &$\hsps\hsps\hra\Dz\piZg$     &$\hsps\hsps\hra\Dz\pi^+$           &                           \\  	  
$\hsps\ra\Dst\pi^0$       &$\hsps\ra\Dst\pi^-$	        &$\hsps\hsps\hsps\Dp\piZg$          &$\ell^-\bar{\nu}\Lc f_0$   \\  	  
$\hsps\hsps\hra\Dz\pi^+$  &$\hsps\hsps\hra\Dz\pi^+$     &$\hsps\ra D^{*0}K^+$               &                           \\  	  
$\hsps\hsps\hsps\Dp\piZg$ &$\hsps\hsps\hsps\Dp\piZg$    &$\hsps\hsps\hra\Dz\piZg$           &$\ell^-\bar{\nu}\Lc\pi^+\pi^-$(NR) \\  
$\hsps\ra\Dz\pi^0$	  &$\hsps\ra\Dz\pi^0$	        &$\hsps\ra\Dp K^0$                  &                           \\  	  
$\hsps\ra\Dp\pi^-$	  &$\hsps\ra\Dp\pi^-$	        &$\hsps\ra\Dz K^+$                  &$\ell^-\bar{\nu}\Lc\pi^0\pi^0$(NR) \\   
                          &                             &                                   &                           \\	  
$\ell^-\bar{\nu}\Dst\pi^0$(NR)&$\ell^-\bar{\nu}\Dst\pi^-$(NR)&$\ell^-\bar{\nu} D_s^{*+}\pi^0$(NR)&           		\\	  
$\hsps\ra\Dz\pi^+$	  &$\hsps\ra \Dz\pi^+$	        &$\hsps\ra\Ds\gamma$                &                           \\	  
$\hsps\hsps\Dp\piZg$      &$\hsps\hsps\Dp\piZg$         &                                   &                           \\
			  &     			&$\ell^-\bar{\nu}\Ds\pi^0$(NR)      &                           \\
$\ell^-\bar{\nu} D^{*0}\pi^+$(NR) &$\ell^-\bar{\nu} D^{*0}\pi^0$(NR) &                      &                           \\
$\hsps\ra\Dz\piZg$	  &$\hsps\ra\Dz\piZg$	        &                                   &                           \\
			  &     			&                                   &                           \\
$\ell^-\bar{\nu}\Dp\pi^0$(NR) &$\ell^-\bar{\nu}\Dp\pi^-$(NR)	&                           &                           \\
$\ell^-\bar{\nu}\Dz\pi^+$(NR) &$\ell^-\bar{\nu}\Dz\pi^0$(NR)	&                           &                           \\
                          &                             &                                   &                           \\
$D^{(*)}\bar{D}^{(*)}K$   &$D^{(*)}\bar{D}^{(*)}K$      &$D^{(*)}\bar{D}^{(*)}K$            &                           \\
$D^{(*)+}{D}^{(*)-}$      &                             &                                   &                           \\
$D_s^{(*)}{D}^{(*)}X$     &$D_s^{(*)}D^{(*)}X$          &$ D_s^{(*)}{D}^{(*)}X$             &$\tau^-\nu\Lc$             \\ 
                          &                             &$D_s^{(*)}D_s^{(*)}X$              &$\tau^-\nu\Lambda_c(2593)^+$\\
$\tau^-\nu D^{+(*),(**)}$ &$\tau^-\nu D^{0(*),(**)}$    &$ \tau^-\nu D_s^{+(*),(**)}$       &$\tau^-\nu\Lambda_c(2625)^+$\\
\end{tabular}
\end{ruledtabular}
\end{table*}
\linespread{1.2}

The bottom meson branching fractions are included via the partial
widths, listed in Table~\ref{tab:gamma}, and adjusted by the lifetime
of the respective bottom hadron, given in Table~\ref{tab:lifetimes}.
The branching fractions used for the \Lb~semileptonic decays are
estimated from measurements of the branching fractions made in other
CDF measurements ~\cite{Ref:Yu}, as shown in Table~\ref{tab:Lb}.  The
ground state charm branching fractions used in this measurement are
listed in Table~\ref{tab:charm}.

\begin{table}
\caption{\label{tab:gamma}
Partial widths of the \Bb~mesons used (from Ref.~\cite{Ref:PDG_2004}).} 
\begin{ruledtabular}
\begin{tabular}{lc}
\Bb~Decay                            & Partial Width [ps$^{-1}$] \\
\hline				       
$\Gamma(\BDln)$			                    & $0.0134\pm0.0009$ \\
$\Gamma^*(\Bb\ra\ell^-\bar{\nu}_{\ell}D^*)$         & $0.0372\pm0.0017$ \\
$\Gamma^{**}(\Bb\ra\ell^-\bar{\nu}_{\ell}D^{**})$   & $0.0141\pm0.0010$  \\
\end{tabular}
\end{ruledtabular}
\end{table}

\begin{table}
\caption{\label{tab:lifetimes}
Bottom hadron lifetimes used in the measurement (from Ref.~\cite{Ref:PDG_2004}).}
\begin{ruledtabular}
\begin{tabular}{lc}
\Bb~Lifetimes             & [ps]  \\
\hline
$\tau (\Bd)$              &  $1.536\pm0.014$       \\
$\tau (\Bu)$              &  $1.671\pm0.018$       \\
$\tau (\Bu)/\tau (\Bd)$   &  $1.086\pm0.017$       \\
$\tau (\Bs)$              &  $1.461\pm0.057$       \\
$\tau (\Lb)$              &  $1.229\pm0.080$       \\
\end{tabular}
\end{ruledtabular}
\end{table}

\begin{table}
\caption{\label{tab:Lb}
\Lb~branching fractions ${\cal B}$ used in the measurement (from
Ref.~\cite{Ref:Yu}). The lack of a quoted uncertainty indicates an
assumption made for ${\cal B}$. NR refers to non-resonant decays.}
\begin{ruledtabular}
\begin{tabular}{ll}
\Lb~Decay	  	                          &${\cal B}$                     \\
\hline				                  
$\ell^-\bar{\nu}_{\ell}\Lc X$                     &(9.2$\pm$2.1)\% 		   \\
				                  
$\ell^-\bar{\nu}_{\ell}\Lc$	                  &(6.54$\pm$0.22)\% 		   \\
$\ell^-\bar{\nu}_{\ell}\Lambda_c(2593)^+$	  &(3.07$\pm$1.02)$\times 10^{-3}$ \\
$\ell^-\bar{\nu}_{\ell}\Lambda_c(2625)^+$	  &(5.14$\pm$0.99)$\times 10^{-3}$ \\
$\ell^-\bar{\nu}_{\ell}\Sigma_c(2455)^{++}\pi^-$  &(2.7$\pm$1.0)$\times 10^{-3}$   \\
$\ell^-\bar{\nu}_{\ell}\Sigma_c(2455)^+\pi^0$	  &(2.7$\pm$1.0)$\times 10^{-3}$   \\
$\ell^-\bar{\nu}_{\ell}\Sigma_c(2455)^0\pi^+$	  &(2.7$\pm$1.0)$\times 10^{-3}$   \\
$\ell^-\bar{\nu}_{\ell}\Lambda_c f_0$		  &2.6$\times 10^{-3}$	   \\
$\ell^-\bar{\nu}_{\ell}\Lambda_c\pi^+\pi^-$ (NR)  &5.2$\times 10^{-3}$	   \\
$\ell^-\bar{\nu}_{\ell}\Lambda_c\pi^0\pi^0$ (NR)  &2.6$\times 10^{-3}$	   \\
\end{tabular}
\end{ruledtabular}
\end{table}

\begin{table}
\caption{\label{tab:charm}
Ground state charm branching fractions ${\cal B}$ used in the
measurement (from Ref.~\cite{Ref:PDG_2004}).}
\begin{ruledtabular}
\begin{tabular}{lc}
Charm Decay                       & ${\cal B}\ [\%]$\\
\hline
$\Dp\ra K^-\pi^+\pi^+$            & $8.8\pm0.6$\\

$\Dz\ra K^-\pi^+$                 & $3.80\pm0.09$\\
$\Dst\ra \Dz\pi^+$                & $67.7\pm0.5$\\
$\Ds\ra \phi\pi^+$                & $3.6\pm0.9$\\
\quad $\phi\ra K^+K^-$   & $49.1\pm0.6$\\

$\Lc\ra pK^-\pi^+$                & $5.0\pm1.3$\\
\end{tabular}
\end{ruledtabular}
\end{table}

\section{\label{sec:eff}
Efficiencies}

Since the bottom hadron fragmentation fractions are measured relative
to each other, most efficiencies in the measurements are expected to
cancel.  Many of the remaining relative efficiencies are determined
from Monte Carlo simulated data in which the trigger and all detector
calibrations are configured just as they are determined for a given
run in real data.  Comparisons between the data and the inclusive
simulation samples, discussed in the subsequent section, validate the
use of simulation to estimate the relative efficiencies between
\Bb~semileptonic decays and the lepton-charm signals.  A few absolute
efficiencies, such as the different XFT trigger efficiencies for $K$,
$\pi$, and $p$ are not properly described in the Monte Carlo
simulation.  These efficiencies must be determined from data and are
discussed in Section~\ref{sec:eff_data}.

\subsection{\label{sec:mc} Monte Carlo Simulation} 

Monte Carlo simulation is used at various points throughout the
measurement.  Although the simulation utilized in the sample
composition process is generated both for exclusive bottom hadron
decays and inclusive \Bb~semileptonic decays, all of the Monte Carlo
simulation samples used in the measurement have the same parameters
for generation.  The simulated events are passed through the 
{\sc geant}-based~\cite{Ref:Geant1,Ref:Geant2} CDF\,II detector
simulation~\cite{Ref:cdfsim}.  {\sc geant\,3} simulates the
passage of the long-lived particles through the material of the
detector and includes multiple scattering effects.  All simulated
samples are generated with a ``realistic'', rather than parametric,
simulation.  A tuned magnetic field and {\textsc geant} material
description~\cite{Ref:BField} are applied in order to correct for
regions of the detector where the material is under-represented in the
simulation.

\begin{figure*}
\centerline{
\includegraphics[width=0.5\hsize]{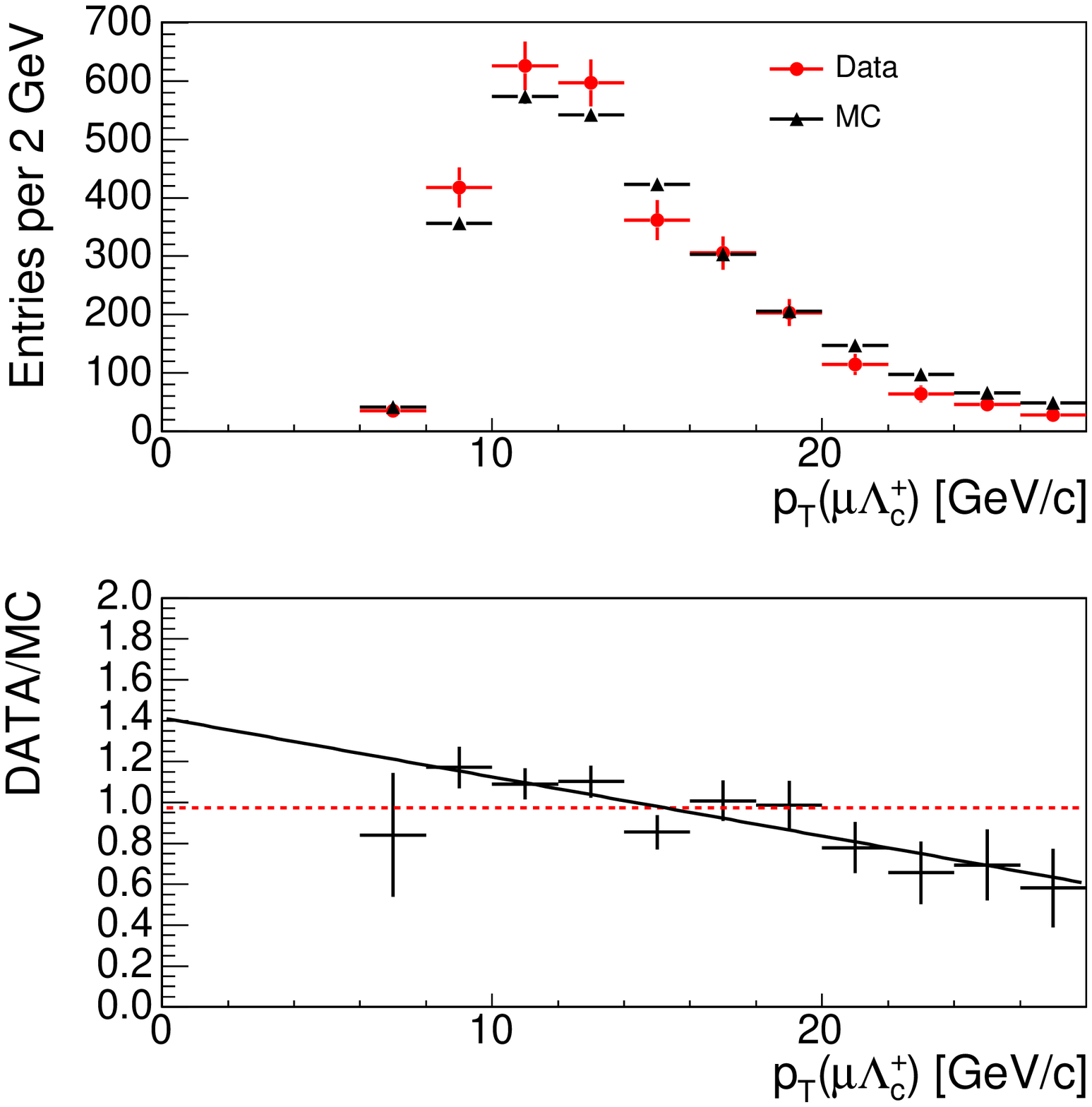}
\includegraphics[width=0.5\hsize]{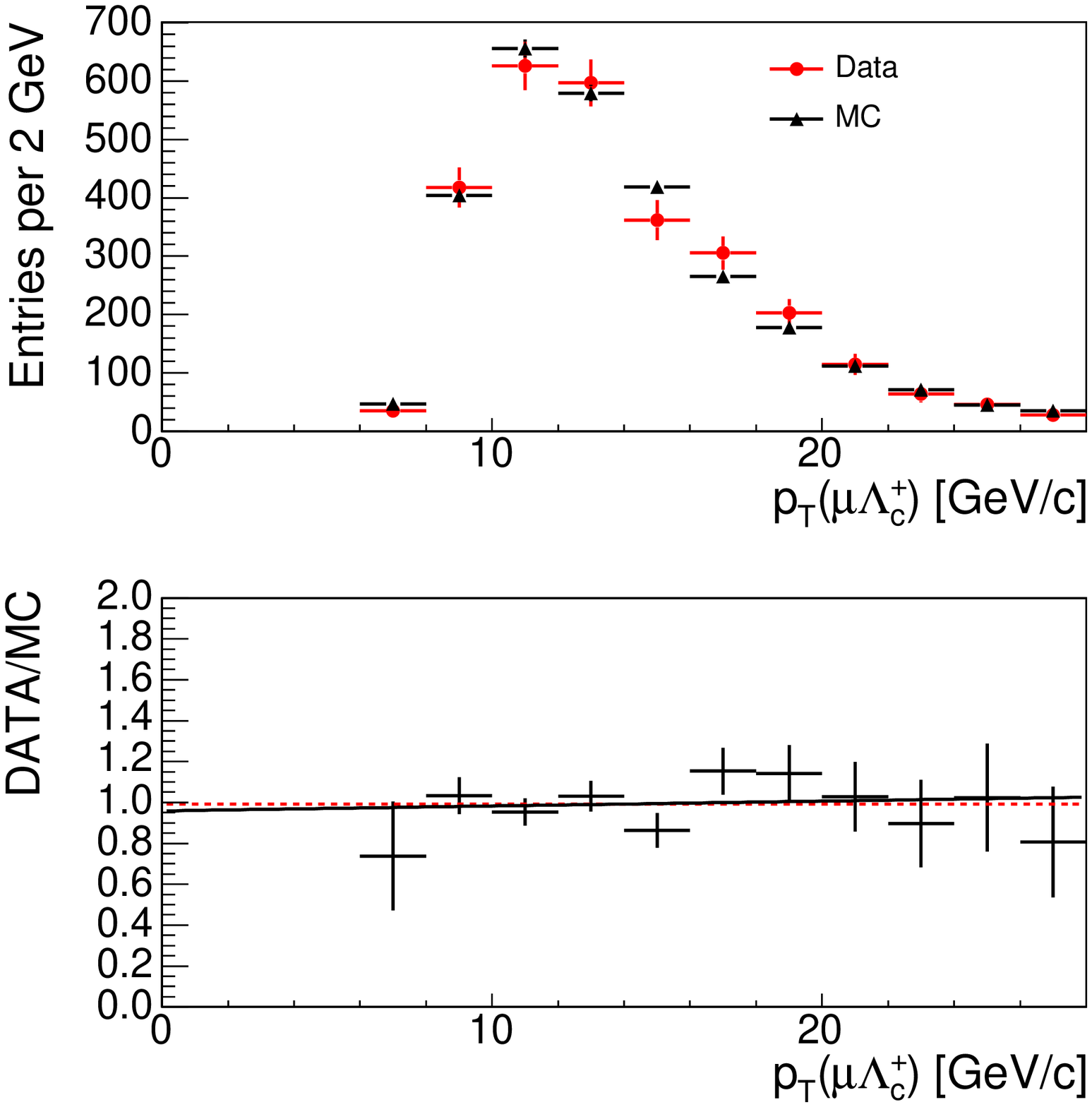}
\put(-420,195){\large\bf (a)}
\put(-180,195){\large\bf (b)}
}
\caption{\label{fig:lc_pt_spec}
Comparisons of $\mu^-\Lc$ transverse momentum spectrum between data
and Monte Carlo simulation generated according to the $p_T$~spectrum
inferred from (a) the inclusive $J/\psi$~cross-section measurement and
(b) the tuned semileptonic $\Lb$ spectrum. The corresponding bottom
plots show the ratio of data over MC with fits of a constant (dotted)
and straight line (solid) overlaid.}
\end{figure*}

A single bottom hadron is generated according to an input transverse
momentum and rapidity spectrum, which have been determined from data.
The $p_T$~spectrum, obtained in the inclusive $J/\psi$~cross-section
measurement~\cite{Ref:CDFdet}, is used as the input \Bb~meson
spectrum.  However, it appears that the momentum distributions of the
\Lb~decay products, in particular the \lLc~momentum spectrum, are not
well described using the same spectrum as is used for the \Bb~mesons,
as can be seen in Figure~\ref{fig:lc_pt_spec}(a).
The Monte Carlo simulation generated $\mu^-\Lc$ transverse momentum
spectrum is observed to be harder than the measured semileptonic
spectrum.  This indicates a potential difference in the momentum
dependence of $b$~baryon and \Bb~meson fragmentation processes.
Consequently, instead of using the spectrum used for the mesons, a
$p_T$~spectrum derived from the semileptonic \lLc\ data is used in the
measurement.  This tuned spectrum~\cite{Ref:Karenthesis} shows good
agreement between data and Monte Carlo simulation, as shown in
Fig.~\ref{fig:lc_pt_spec}(b).  The tuned \lLc\
spectrum is obtained by re-weighting the bottom hadron $p_T$~spectrum
measured from the inclusive $J/\psi$~cross-section
measurement~\cite{Ref:CDFdet}. The re-weighting function is determined
from the disagreement between the
\lLc\ data and the generated spectrum, which is fit to a first order
polynomial, $w=b+m\cdot p_T$, where $b = 1.43\pm 0.08$ and $m =
-0.026\pm 0.007$ are the values of the fit averaged between the
$e$+SVT and $\mu$+SVT data.  The tuned spectrum is then varied by
$\pm$2\,$\sigma$ of the uncertainty on the slope ($\sigma_m$) to bound
the uncertainty on this spectrum (see Sec.~\ref{sec:Lbspectrum}).  All
of the Monte Carlo simulation events are generated with an input $p_T$
threshold of $p_T(\Bb)>5~\gevc$ and $|\eta(\Bb)|<1.1$.

After the bottom hadron is generated, it is then decayed using the
{\sc EvtGen} decay package~\cite{Ref:EvtGen}, which decays the
particles according to a user-specified decay chain and theoretical
decay models.  The \Bb~meson form factors used in the Monte Carlo
simulation for this analysis are taken from various models based on
heavy quark effective theory (HQET)~\cite{Ref:ISGW2,Ref:Goity}.  The
ISGW2~\cite{Ref:ISGW2} model implemented in {\sc EvtGen} governs the
\Bb~meson semileptonic decays to the ground state and doubly excited
charm mesons, while the HQET decay model, implemented in {\sc EvtGen},
is used for the \Bb~meson semileptonic decays to excited charm states.
Non-resonant $D^{**}$ meson decays are described by the model
developed by Goity and Roberts~\cite{Ref:Goity}.  The \Lb~baryon
semileptonic decay model is newly implemented~\cite{Ref:Karenthesis}
into the {\sc EvtGen} package for this measurement.  The baryon form
factors for the primary semileptonic $\LbLcln$$^{(*,**)}$ decays are
taken from constituent quark model calculations made by Pervin {\it et
al.}~\cite{Ref:Pervin}.  These results agree with the large N$_c$
predictions by Leibovich and Stewart~\cite{Ref:Leibovich} to order
${\cal O}(1/m_{Q})$.  Non-resonant \Lb~decays, which are expected to
contribute comparatively little to the total \Lb~semileptonic width,
are described by a phase space decay model.

\subsubsection{\label{sec:data_mc} Data - Simulation Comparison} 

\begin{figure*}
\centerline{
\includegraphics[width=0.5\hsize]{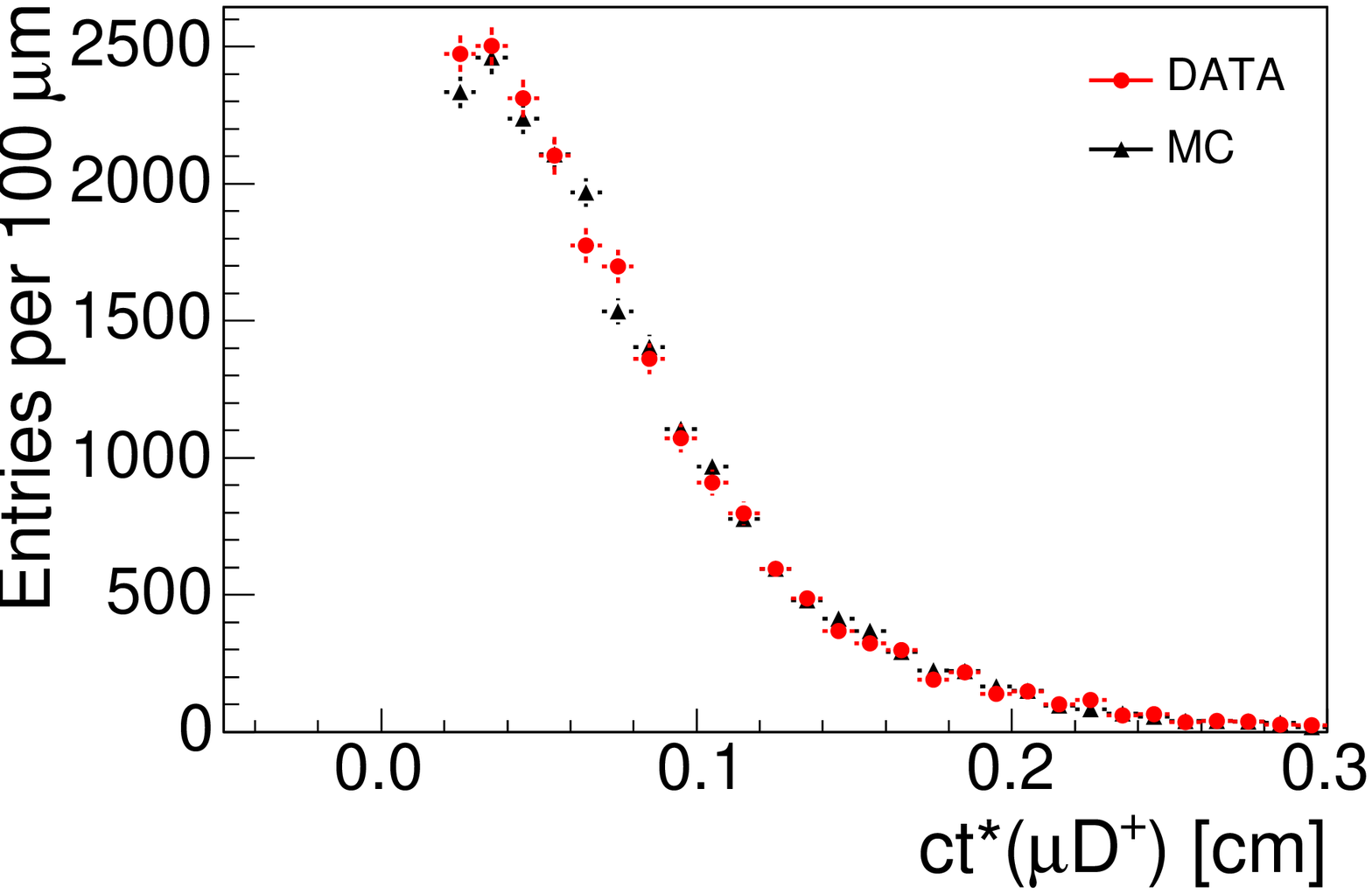}
\includegraphics[width=0.5\hsize]{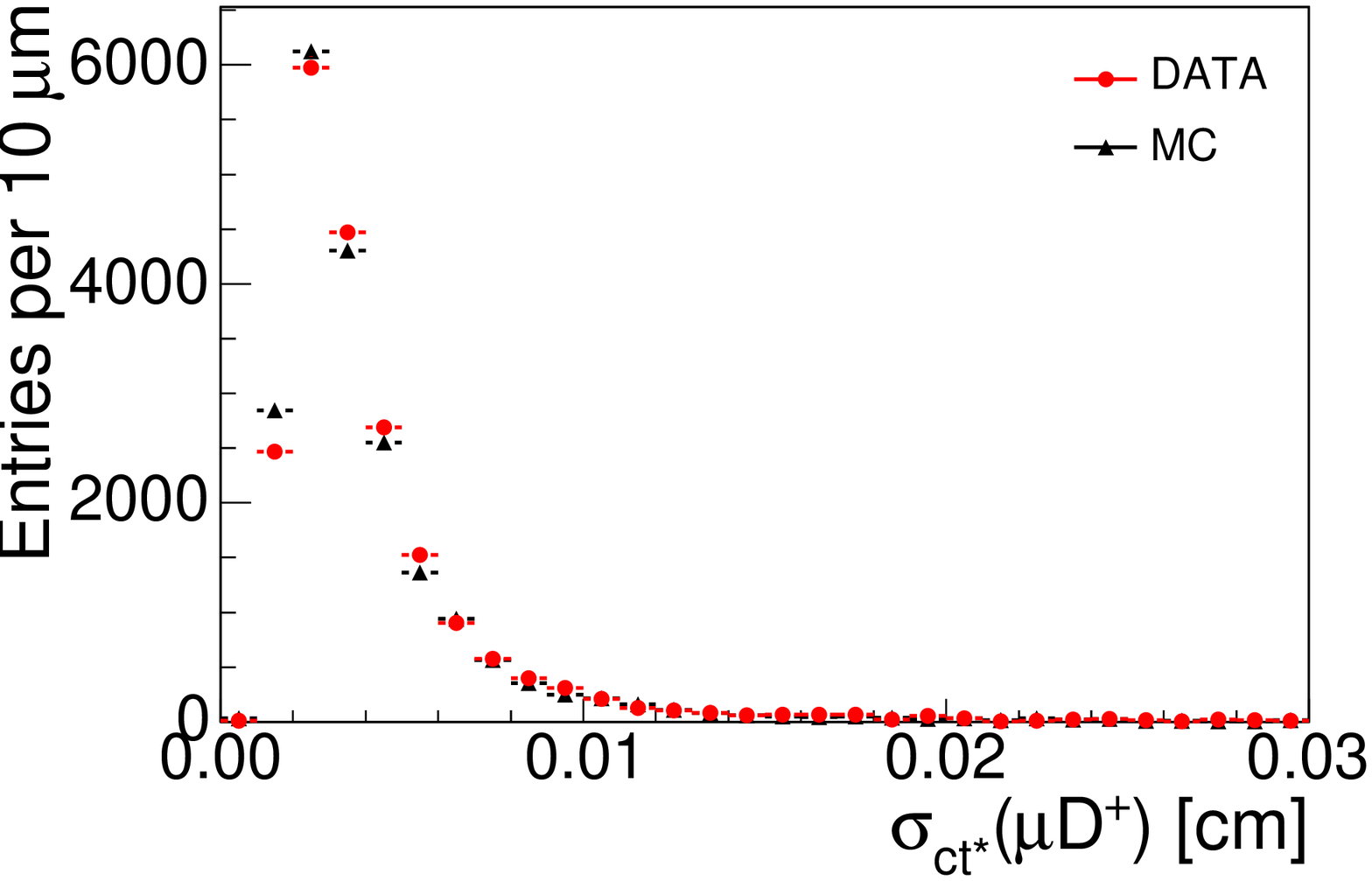}
\put(-290,105){\large\bf (a)}
\put(-50,105){\large\bf (b)}
}
\centerline{
\includegraphics[width=0.5\hsize]{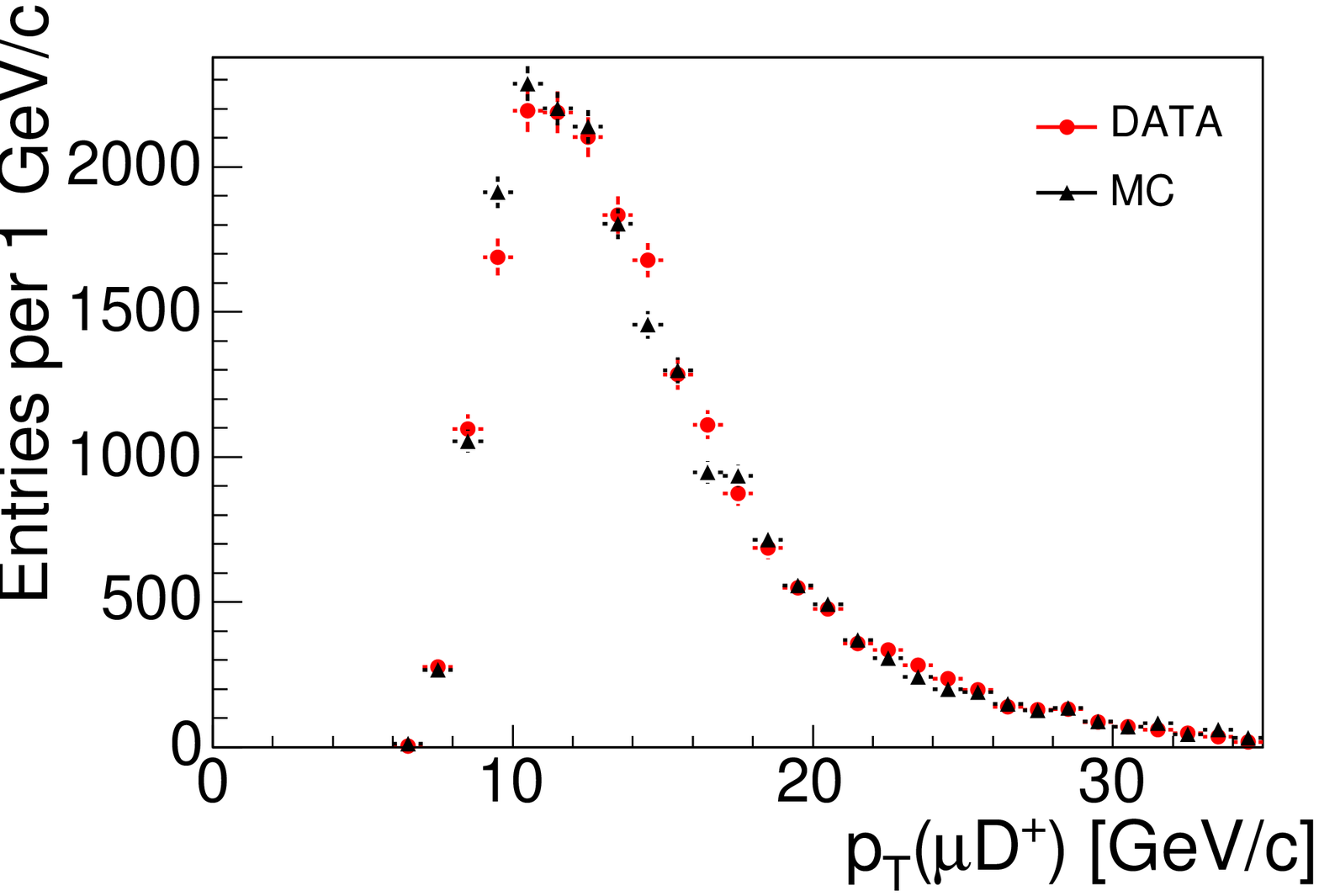}
\includegraphics[width=0.5\hsize]{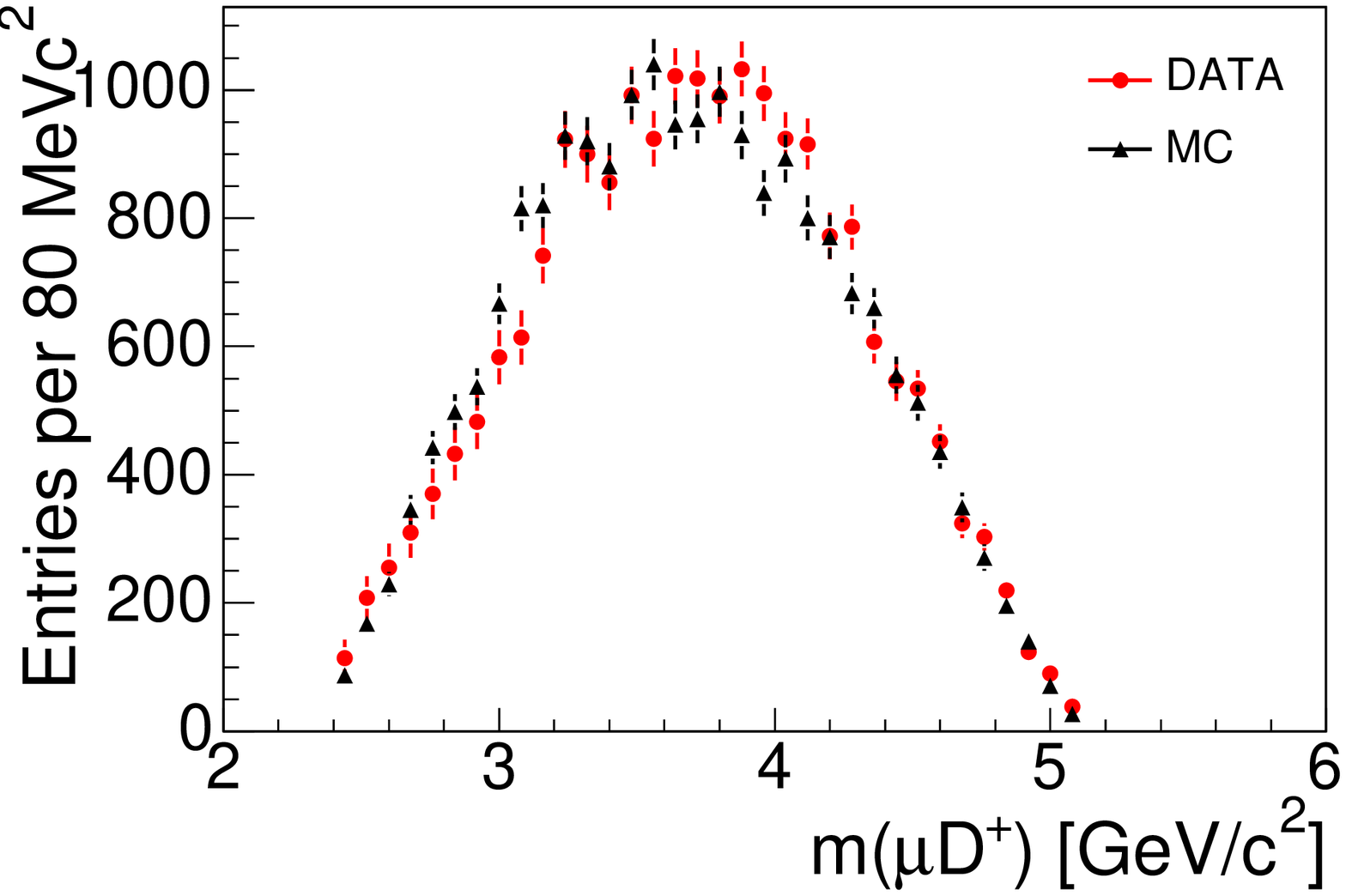}
\put(-290,105){\large\bf (c)}
\put(-50,105){\large\bf (d)}
}
\caption{\label{fig:dp_mu_b}
Data-MC comparisons for $\mu^-\Dp$ of (a) $ct^*(\mu^-\Dp)$, 
(b)~$\sigma_{ct^*}(\mu^-\Dp)$, (c) $p_T(\mu^-\Dp)$, and (d)
$m(\mu^-\Dp)$.}
\end{figure*}

Four inclusive Monte Carlo simulation samples, $\Bd\ra
\ell^-\bar{\nu}_{\ell}D^{0,+} X$, $\Bu\ra
\ell^-\bar{\nu}_{\ell}D^{0,+} X$, $\BsDsln X$, and $\LbLcln X$ are
generated to validate the use of simulation to determine the kinematic
efficiencies of the bottom hadron semileptonic decays used in the
measurement.  The agreement between the data and the Monte Carlo
should not be very sensitive to variations in the $D^{**}$ branching
fractions between the default {\sc EvtGen} table and the one to be
later determined in the fit for the fragmentation fractions.  The
agreement between data and Monte Carlo simulation is checked for
quantities used in the signal selection (listed in
Table~\ref{tab:cuts}).  In general, the agreement between data and
simulation is good in both the $\mu$+SVT and $e$+SVT data. A typical
example of comparisons between data and Monte Carlo in the $\mu$+SVT
sample is shown in Fig.~\ref{fig:dp_mu_b} for (a)~$ct^*(\mu^-\Dp)$,
(b)~$\sigma_{ct^*}(\mu^-\Dp)$, (c)~$p_T(\mu^-\Dp)$, and
(d)~$m(\mu^-\Dp)$.  A complete set of comparisons between data and MC
can be found in Ref.~\cite{Ref:Karenthesis}.  The area of the
simulation distribution is normalized to the corresponding area of the
data distribution for this comparison.  The quality of the comparisons
are quantified by fitting the ratio of the data to the simulation by
both a first order polynomial and a constant.  The former indicates
potential biases between the two distributions ({\it i.e.} whether the
simulation distribution is too hard or soft relative to the data),
while the latter gives a measure of overall agreement between the
distributions. No significant disagreement, determined from the fit to
a constant line, is observed between data and simulation in the
quantities used for the signal selection.

\subsection{\label{sec:eff_data} Relative Efficiency Determination from Data} 

Many efficiencies in the measurement of the relative fragmentation
fractions are expected to cancel, and many of the remaining relative
efficiencies are determined from the Monte Carlo simulation.  The few
efficiencies that are not well described by the simulation are
determined from the data and discussed next.

\subsubsection{\label{sec:xft} XFT Efficiency} 

Differences in the XFT efficiencies of kaons, pions, and protons are
expected due to the stringent hit requirement placed on COT tracks by
the XFT trigger.  Since the \dedx~of kaons and protons is lower than
the \dedx\ of pions for $p_t>2~\gevc$, the COT hit requirement leads
to a lower efficiency for kaons and protons relative to the pion XFT
efficiency.  These efficiencies are difficult to describe in the Monte
Carlo simulation due to varying COT operating conditions during the
data-taking period of this measurement.  Therefore, they are derived
from the data.  The SVT efficiencies, which contribute to the triggers
used in this measurement, depend directly on the XFT efficiencies.
The species dependence of the SVT efficiencies originate entirely from
the XFT, since the energy loss between $K$, $\pi$, and $p$ are
negligible in the silicon detector relative to the drift chamber.  The
differences in the SVT efficiencies between reconstructed lepton-charm
channels are therefore described by the dependence of XFT efficiencies
on particle species.

\begin{figure*}
\centerline{
\includegraphics[width=0.5\hsize]{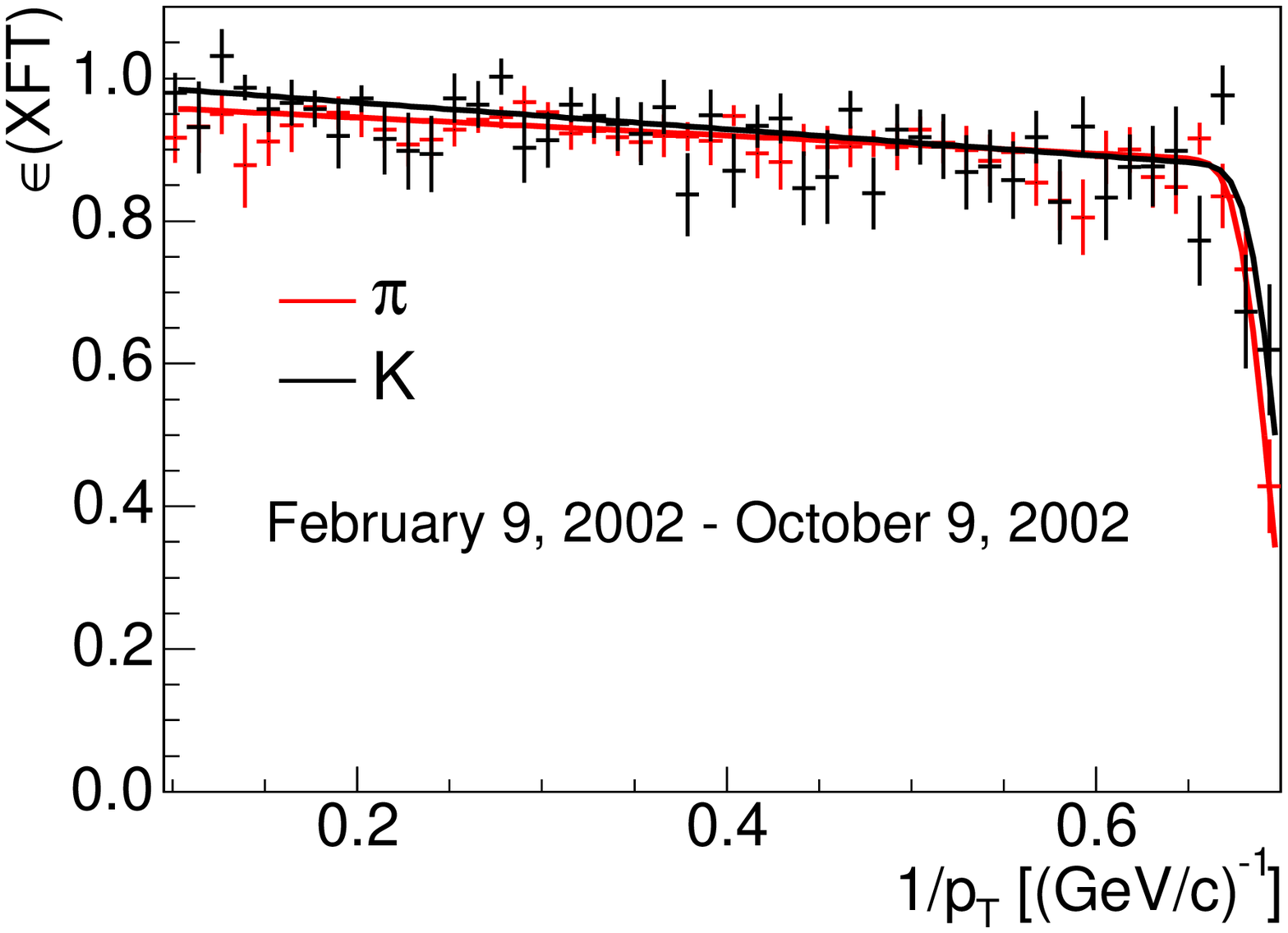}
\includegraphics[width=0.5\hsize]{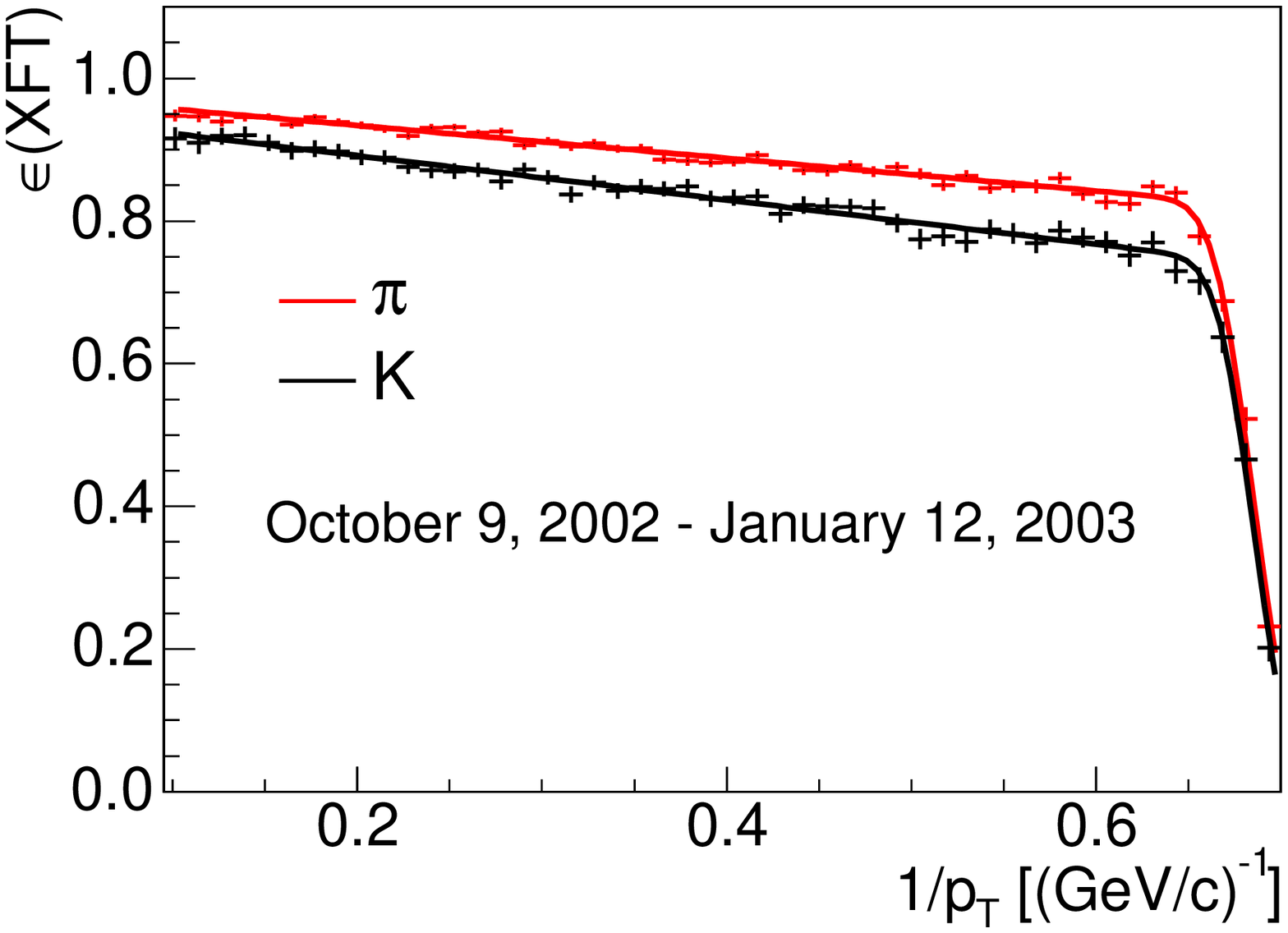}
\put(-420,45){\large\bf (a)}
\put(-165,45){\large\bf (b)}
}
\centerline{
\includegraphics[width=0.5\hsize]{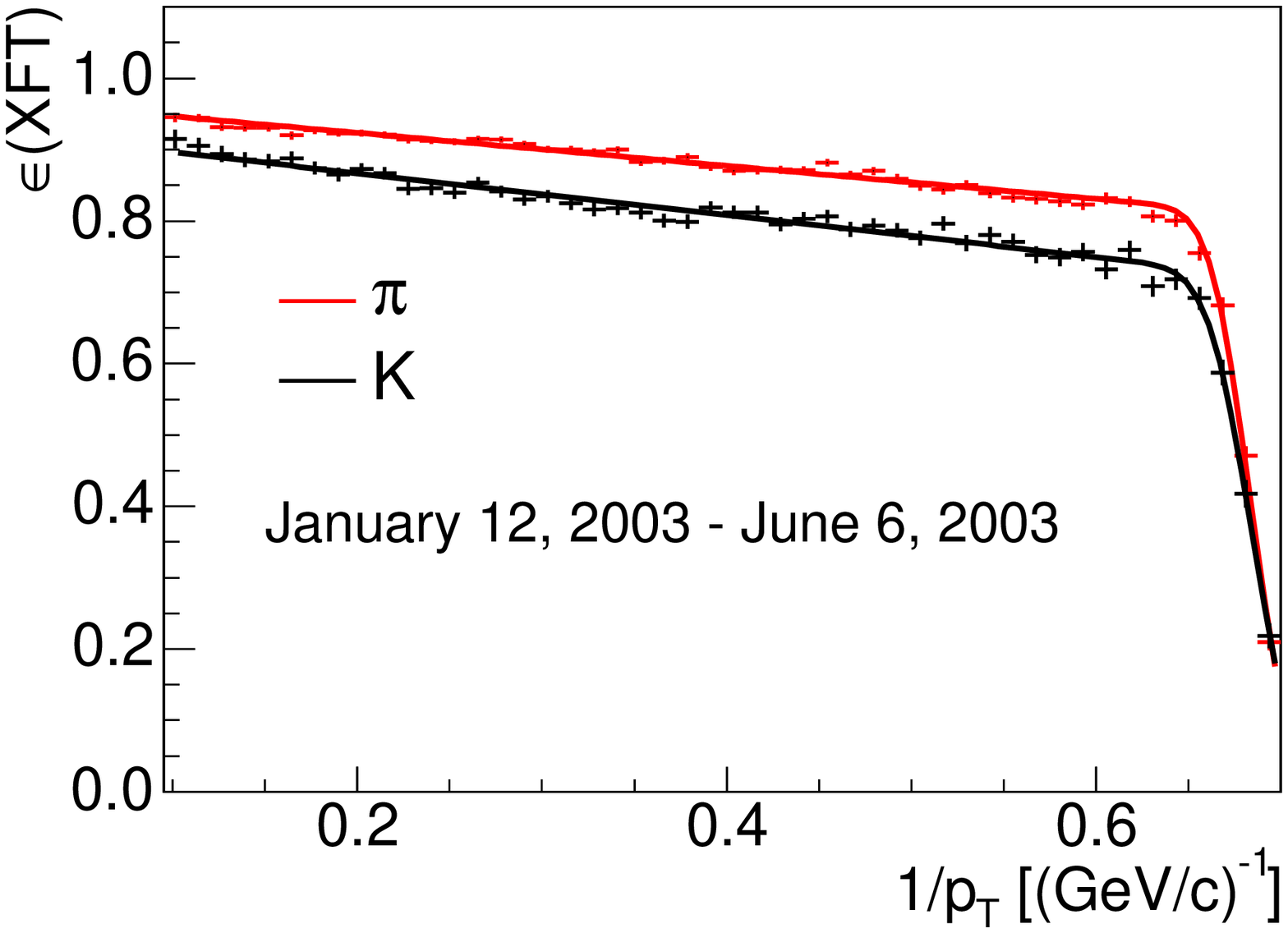}
\includegraphics[width=0.5\hsize]{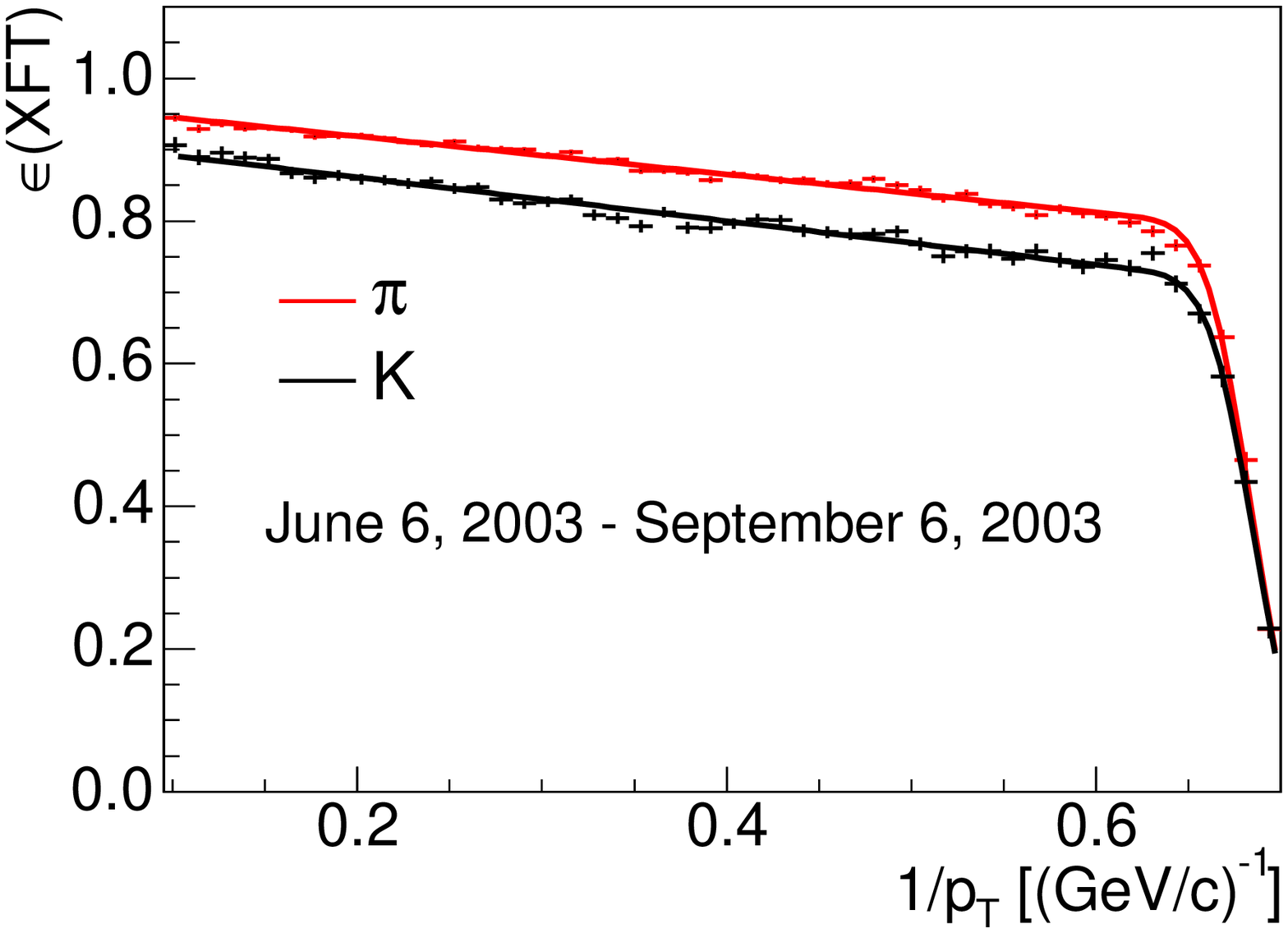}
\put(-420,45){\large\bf (c)}
\put(-175,45){\large\bf (d)}
}
\centerline{
\includegraphics[width=0.5\hsize]{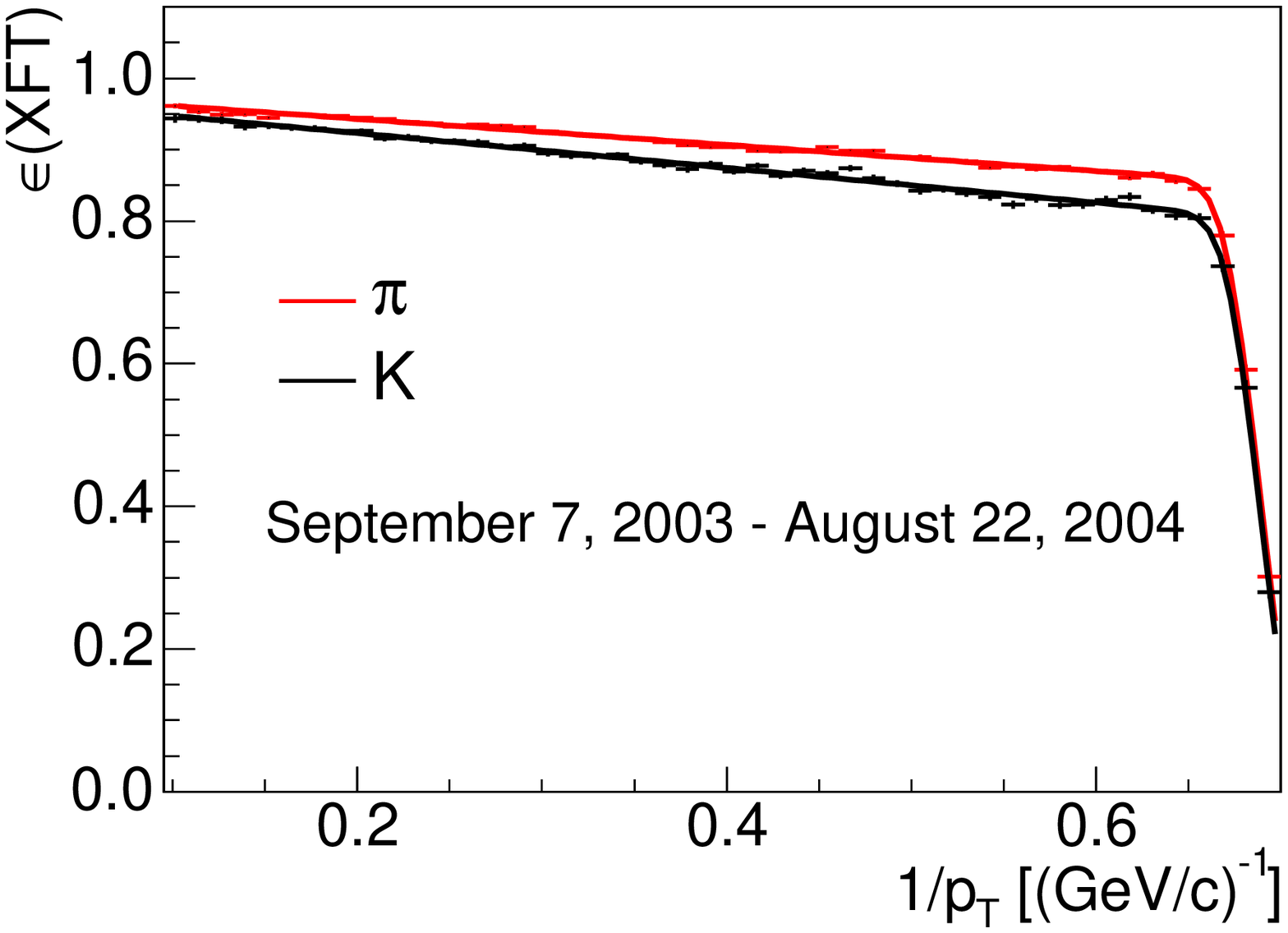}
\put(-180,45){\large\bf (e)}
}
\caption{\label{fig:xft_data}
XFT efficiency $\varepsilon_{\rm XFT}$ as a function of $p_T^{-1}$ for
pions (upper curve) and kaons (lower curve) for various run ranges
indicated in distributions (a) through (e).}
\end{figure*}

\begin{figure*}
\centerline{
\includegraphics[width=0.5\hsize]{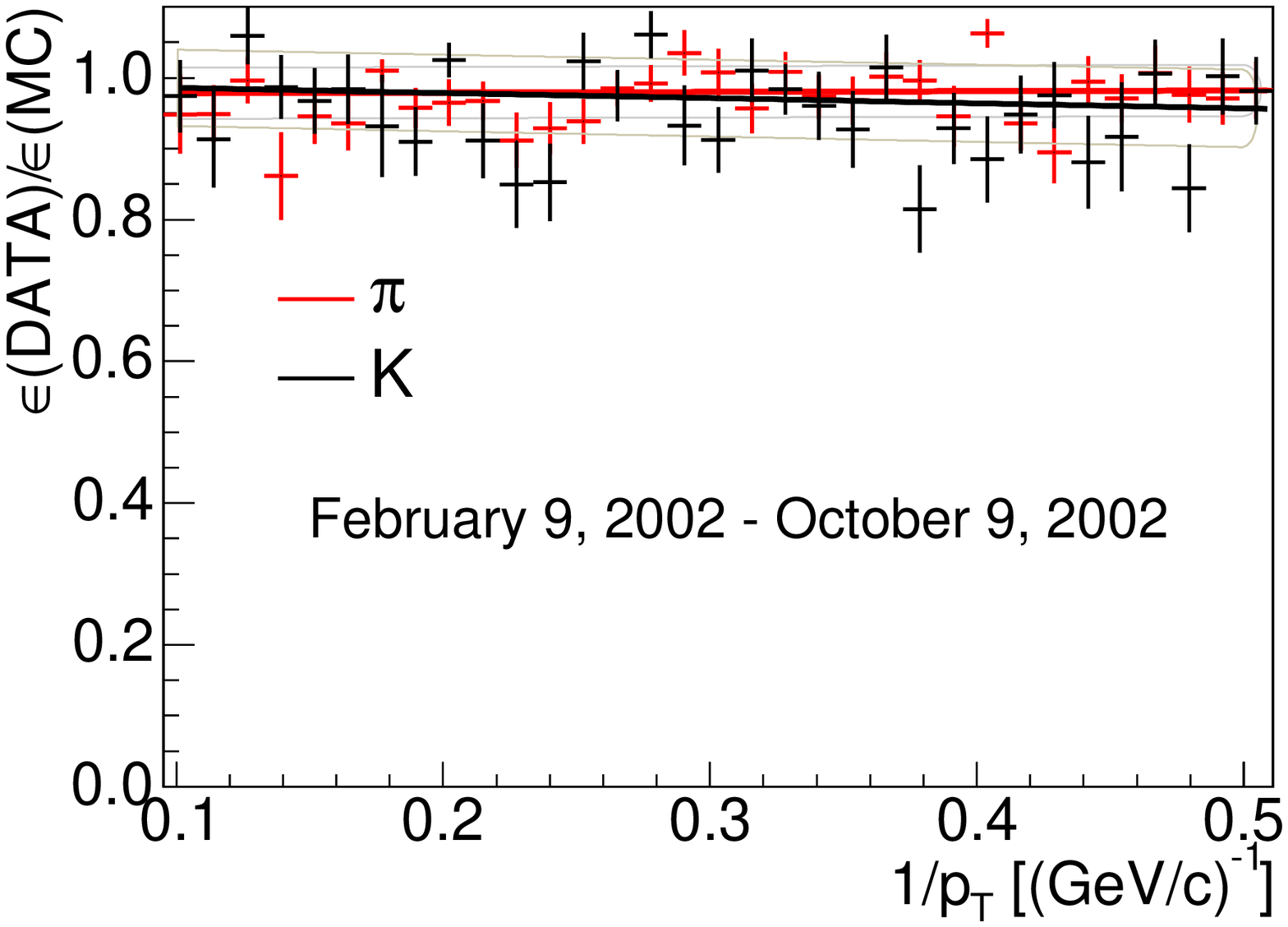}
\includegraphics[width=0.5\hsize]{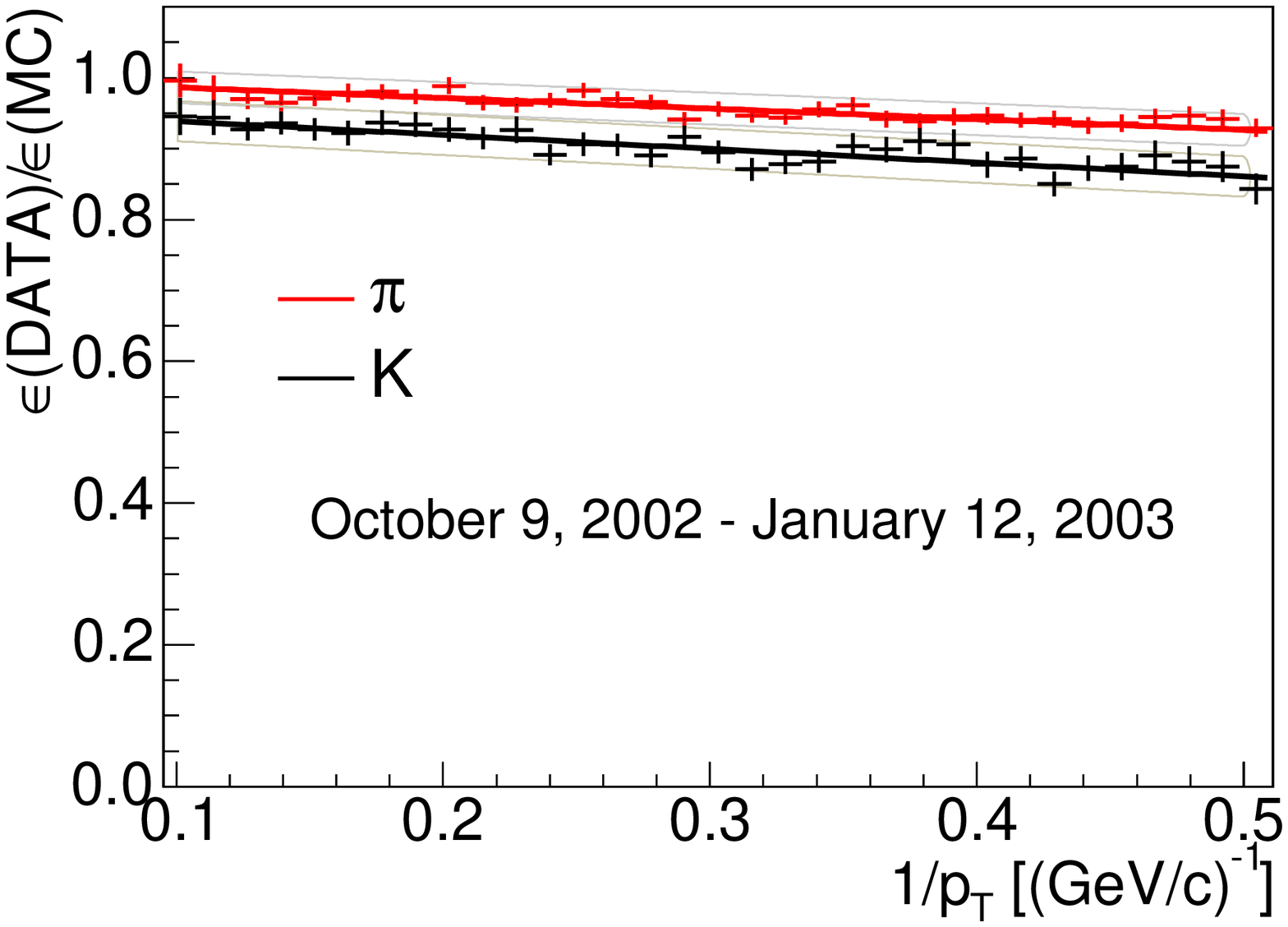}
\put(-420,45){\large\bf (a)}
\put(-165,45){\large\bf (b)}
}
\centerline{
\includegraphics[width=0.5\hsize]{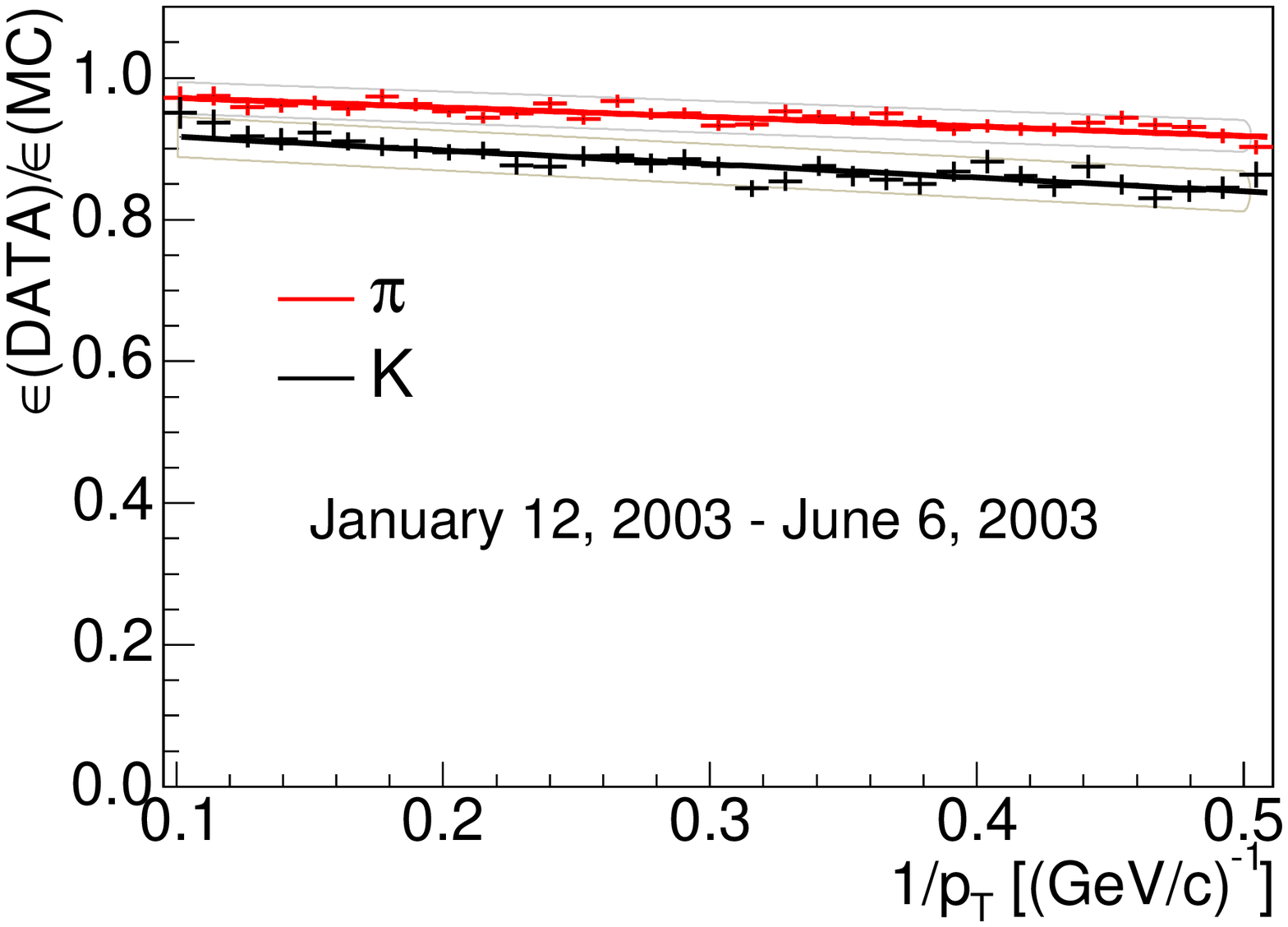}
\includegraphics[width=0.5\hsize]{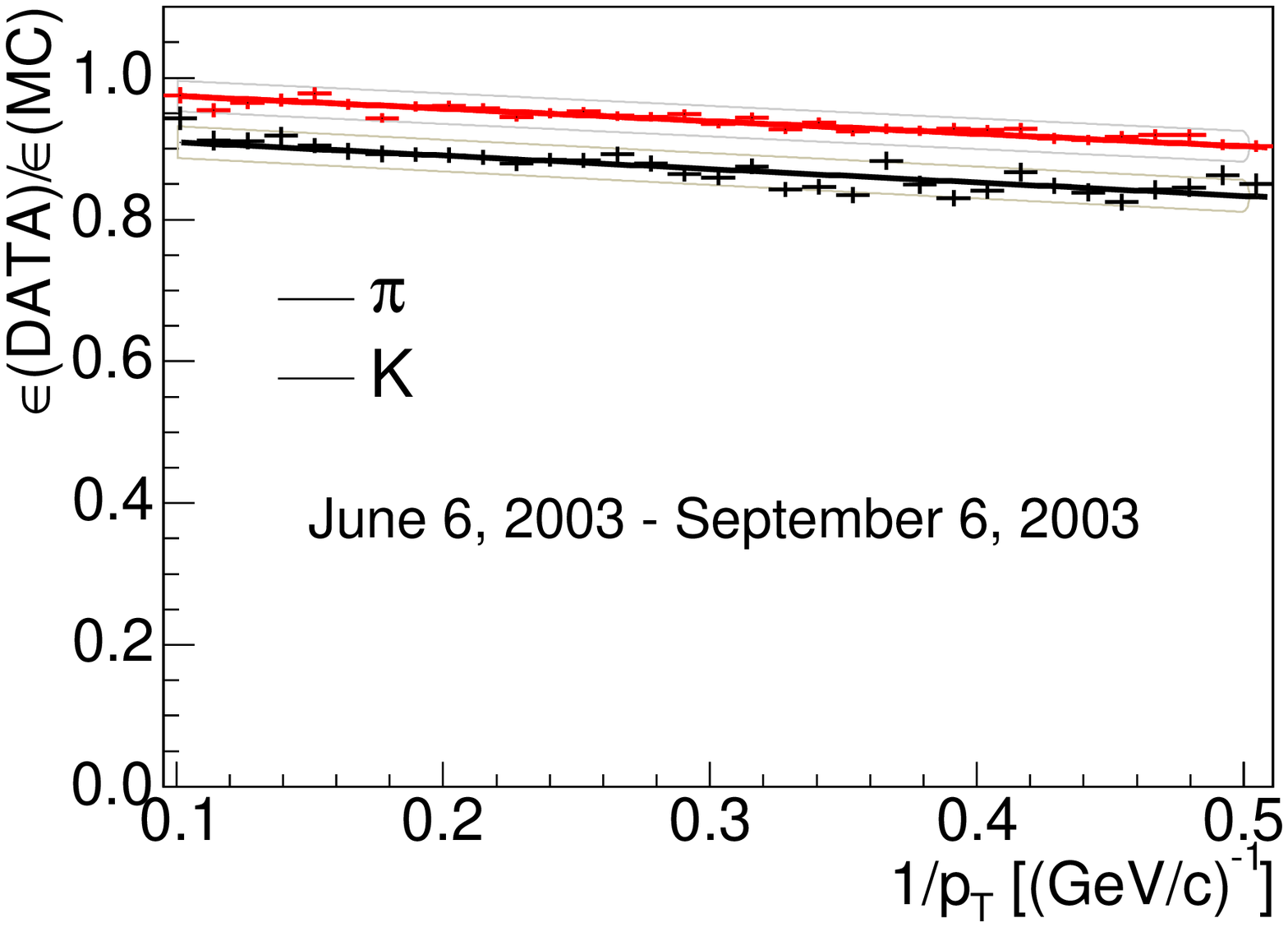}
\put(-420,45){\large\bf (c)}
\put(-175,45){\large\bf (d)}
}
\centerline{
\includegraphics[width=0.5\hsize]{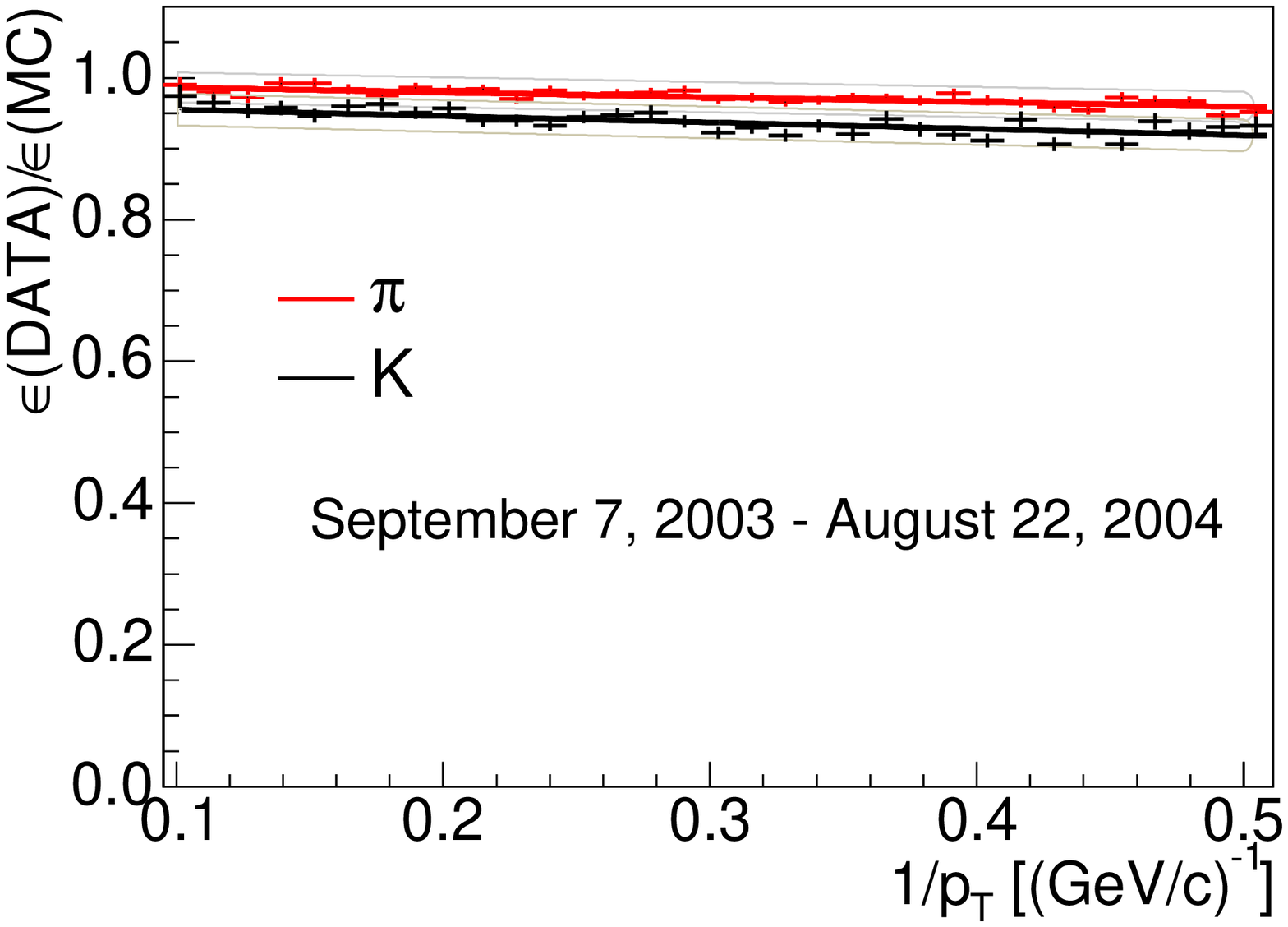}
\put(-180,45){\large\bf (e)}
}
\caption{\label{fig:xft_data_mc}
Ratio $\varepsilon_{\rm XFT}({\rm DATA})/\varepsilon_{\rm XFT}({\rm MC})$ 
as a function of $p_T^{-1}$ for pions (upper line) and kaons (lower
line) for various run ranges indicated in distributions (a) through (e).}
\end{figure*}

The XFT efficiencies for $K$ and $\pi$ are measured by reconstructing the
\DpKpipi\ decay mode in the two-track trigger (TTT) data sample, where two
tracks are required to match to the SVT trigger and no lepton
requirement is made. Two of the final state $K$ or $\pi$ tracks are
matched to the SVT tracks.  The track that is not matched to an SVT
track is treated as the unbiased track, which is then examined to
determine whether it could have fired the XFT trigger.  Tracks that
could have passed the XFT trigger are included in the numerator of the
efficiency, while all unbiased tracks are included in the denominator.
A similar procedure is carried out for the proton XFT efficiency,
using $\Lambda^0\ra p\pi^-$ events reconstructed in data collected
with the TTT, where two other tracks in the event are required to have
fired the SVT trigger.  These efficiencies, binned in time to span the
data set used, are shown in Fig.~\ref{fig:xft_data} for $K$ and $\pi$.
The ratio of the $K$ and $\pi$ efficiencies determined in the data
relative to those determined in the corresponding Monte Carlo
simulation are shown in Fig.~\ref{fig:xft_data_mc}. These corrections
are parametrized by linear functions of the form $a_0+a_1/p_T$, and
the obtained fit parameters are listed in Table~\ref{tab:XFT}.
Details of the fits for the proton XFT efficiency can be found in
Ref.~\cite{Ref:Karenthesis}.

\begin{table}[tbp]
\caption{\label{tab:XFT}
Parameterizations for the XFT Monte Carlo simulation corrections.}
\begin{ruledtabular}
\begin{tabular}{lcccc}
                &\multicolumn{2}{c}{$K$}              &\multicolumn{2}{c}{$\pi$}    	\\
Data period     &$a_0$             &$a_1$             &$a_0$             &$a_1$   	\\
\hline
Feb'02-Oct'02 &0.9931$\pm$0.05   &-0.0725$\pm$0.02  &0.9772$\pm$0.03  &0.00968$\pm$0.01 \\
Oct'02-Jan'03 &0.9584$\pm$0.02   &-0.1952$\pm$0.007 &1.0016$\pm$0.01  &-0.1501$\pm$0.005\\ 
Jan'03-Jun'03 &0.9359$\pm$0.02   &-0.1919$\pm$0.007 &0.9851$\pm$0.01  &-0.1341$\pm$0.004\\
Jun'03-Sep'03 &0.9282$\pm$0.01   &-0.1897$\pm$0.005 &0.9921$\pm$0.008 &-0.1776$\pm$0.004\\
Sep'03-Aug'04 &0.9643$\pm$0.01   &-0.0907$\pm$0.004 &0.9931$\pm$0.007 &-0.0678$\pm$0.003\\
\hline
\hline
                &\multicolumn{4}{c}{proton}  \\
		&$a_0$            &\multicolumn{1}{c}{$a_1$}            &$a_2$             &$a_3$   \\
\hline
Feb'02-Jun'03   &1.063$\pm$0.090  &\multicolumn{1}{c}{-1.326$\pm$0.963} &3.198$\pm$3.218   &-2.203$\pm$3.391  \\
\end{tabular}
\end{ruledtabular}
\end{table}

\subsubsection{\label{sec:dedx} \dedx~Efficiency} 

The \dedx~efficiency of the ${\cal LR}$ cut applied to the proton,
discussed previously in Section~\ref{sec:sig_opt}, is also evaluated
from data and the Monte Carlo simulation is adjusted accordingly. The
$\Lambda^0\ra p\pi^-$ control sample is used to evaluate the
efficiency of the ${\cal LR}(p) > 0.3$ cut applied to the proton from
the \Lc~decay.  The \dedx~efficiency is obtained by dividing the
number of protons that pass the \dedx~${\cal LR}(p)$ cut by all
protons in bins of proton transverse momentum. The shape of the
efficiency is parameterized by two functional forms: a third order
polynomial plus a constant and using only a constant.  Both
parameterizations fit the data well; the former is used as the default
parameterization in the measurement, while the latter is used in the
evaluation of systematic uncertainties (see Sec.~\ref{sec:sys}).

\subsubsection{\label{sec:single_track} Single Track Efficiency} 

The difference in efficiency between the two track charm topology in
the \DzKpi\ decay and a three track topology such as the \DpKpipi~decay
arises from the efficiency of reconstructing an additional single
track. Since the Monte Carlo simulation contains only the decay
products of the generated bottom hadron and no additional tracks from
the fragmentation process or underlying event, the efficiency of
reconstructing a track in a simulation event is different from that in
data. Thus the single track efficiency is determined from the data
relative to the simulation.  In order to evaluate this efficiency, the
lepton plus four track state $\mu^-\Dz$ with $\Dz\ra K^-\pi^+\pi^-\pi^+$, is
reconstructed and normalized to the $\mu^-\Dz$~decay with $\Dz\ra K^-\pi^+$
in both the data and the simulation.  This represents the square of
the efficiency, $\varepsilon_{trk}^2$, to find a single track in the
data relative to the simulation, assuming that the reconstruction of
the third and fourth tracks in the $\Dz\ra K^-\pi^+\pi^-\pi^+$~decay
are uncorrelated.  This assumption will be treated as a source of
systematic uncertainty, as discussed in Sec.~\ref{sec:sys}.  This
procedure yields the single track efficiency in data relative to the
same efficiency in MC, $\varepsilon_{trk} = 87.8\pm 0.8\,{\rm
(stat)}\, ^{+1.9}_{-0.9}\,({\cal B})\%$, where the systematic
uncertainty is dominated by the knowledge of the relative branching
fraction ${\cal B}(\Dz\ra K^-\pi^+\pi^-\pi^+)/{\cal BR}(\DzKpi) = 2.10\pm
0.03\pm0.06$~\cite{Ref:CLEO-c}, which is used to adjust the generated
Monte Carlo simulation samples for both \Dz~decay modes.

\subsubsection{\label{sec:tot_eff} Total Relative Efficiency} 

The relative efficiencies $\varepsilon_{\rm rel}$ that are included in
the sample composition are the product of the acceptance
$\varepsilon_{\rm accep}$, lepton plus SVT trigger efficiency
$\varepsilon_{\rm trigg}$, analysis efficiency $\varepsilon_{\rm an}$,
and adjusted XFT efficiency $\varepsilon_{p/K/\pi\ {\rm trigg}}$,
\begin{eqnarray}
\varepsilon_{\rm rel}(\lDp) &=&
\varepsilon_{\rm accep}{\rm (MC)}\times\varepsilon_{\rm trigg}{\rm (MC)}
\times\varepsilon_{\rm an}{\rm (MC)} \nonumber \\
&&\times\, \varepsilon_{p/K/\pi\ {\rm trigg}}{\rm (data/MC)}.
\end{eqnarray}
The relative efficiencies are similar for the other lepton-charm
signals.  An extra factor of $\varepsilon^{-1}_{\rm track}$(data/MC)
is needed for $\varepsilon_{\rm rel}(\lDz)$ and $\varepsilon_{\rm
rel}(\lDst)$ to adjust the two track charm topology relative to the
three track charm states, while the \lLc\ relative efficiency requires
an additional efficiency correction $\varepsilon_{\dedx}$(data) for
the \dedx~cut imposed on the proton candidate.

Monte Carlo yields in each channel are determined by fitting the
simulation signal 
with a double Gaussian, analogous to the fits to the data.  A single
Gaussian is used to obtain the yield in the indirect lepton-charm
decays, where the lepton originates from $\Bb\ra\tau^- DX,\
\tau^-\ra\ell^- X$, or another charm ({\it e.g.}  $\Bb\ra D\bar{D}X,\
\bar D\ra\ell^- X$), because the yields are generally quite low in
these channels and are poorly described by a double Gaussian.  The
total efficiencies derived for \lDp, \lDz, \lDst, and \lDs\ channels
are listed in Tables~\ref{tab:Dp_eff}-\ref{tab:Ds_eff} for each
\Bb~mode considered in the sample composition. The efficiencies are
calculated relative to a reference channel where the yield out of
$10^7$ generated events is listed for each corresponding reference
channel.  All of the decays shown in the tables are generated
separately and for each decay $10^7$ events are generated.  The
relative efficiencies for the primary semileptonic decays do not
include the branching fractions.  The efficiencies for the ``physics
backgrounds'' include the semileptonic branching fractions of the $D$
or $\tau$, as well as the $\Bb\ra DDX$ sample composition.  Since a
fixed sample composition is used for the $\LbLcln X$~decays, the
efficiencies quoted in Table~\ref{tab:Lc_eff} have the excited charm
baryon branching fractions from Table~\ref{tab:Lb} applied.

\begin{table}
\caption{\label{tab:Dp_eff}
Efficiencies from $\lDp$ Monte Carlo simulation. For display purpose,
the efficiencies are given relative to the ground state charm mode with
the yield of the semileptonic decay into the ground state charm mode
included in parentheses.}
\begin{ruledtabular}
\begin{tabular}{lcc}
Decays		                                                        &$e+$SVT		&$\mu+$SVT  	        \\
\hline											    						
$\Bd\ra\ell^-\bar{\nu}\Dp$						&$1.089\cdot10^{-3}$ (10,890$\pm$110)	&$1.307\cdot10^{-3}$ (13,070$\pm$120) 	\\
														      			
$\hspace*{1.2cm}\ell^-\bar{\nu} \Dst(\Dp\piZg)$				&1.061$\pm$0.015  	&1.100$\pm$0.013  	\\
											       			     	   	 	
$\hspace*{1.2cm}\ell^-\bar{\nu} D_1^+(\pi^0\Dst)(\Dp\piZg)$		&0.672$\pm$0.011 	&0.750$\pm$0.010  	\\
										       			     	   	 	
$\hspace*{1.2cm}\ell^-\bar{\nu} D_0^{*+}(\Dp\pi^0)$			&0.625$\pm$0.010	&0.712$\pm$0.010	\\
											       			     	   	 	
$\hspace*{1.2cm}\ell^-\bar{\nu} D_1^{'+}(\pi^0\Dst)(\Dp\piZg)$		&0.680$\pm$0.010   	&0.748$\pm$0.010 	\\
											       			     	   	 	
$\hspace*{1.2cm}\ell^-\bar{\nu} D_2^{*+}(\pi^0\Dst)(\Dp\piZg)$		&0.673$\pm$0.011  	&0.753$\pm$0.010  	\\

$\hspace*{1.2cm}\ell^-\bar{\nu} D_2^{*+}(\Dp\pi^-)$			&0.696$\pm$0.011 	&0.783$\pm$0.011 	\\
											       			     	   	 	
$\hspace*{1.2cm}\ell^-\bar{\nu} \Dst\pi^0(\Dp\piZg)$ (NR)		&0.485$\pm$0.008	&0.638$\pm$0.009  	\\
											       			     	   	 	
$\hspace*{1.2cm}\ell^-\bar{\nu} \Dp\pi^0$ (NR)				&0.544$\pm$0.009	&0.764$\pm$0.010  	\\
														                        
$\hspace*{1.2cm}D^{(*)}\bar{D}^{(*)}K(\ell \Dp X)$			&0.0012$\pm$0.0002	&0.0044$\pm$0.0003	\\

$\hspace*{1.2cm}D^{(*)+}D^{(*)-}(\ell \Dp X)$				&0.0092$\pm$0.0004	&0.0160$\pm$0.0005	\\
														  			
$\hspace*{1.2cm}D_s^{(*)}D^{(*)}X(\ell \Dp X)$				&0.0027$\pm$0.0002	&0.0069$\pm$0.0003	\\
														  			
$\hspace*{1.2cm}\tau^-\bar{\nu} D^{(*,**)}(\ell \Dp X)$		        &0.0212$\pm$0.0005	&0.0282$\pm$0.0007	\\
														                        
\hline																	
$\Bu\ra\ell^-\bar{\nu} D_1^0(\pi^-\Dst)(\Dp\piZg)$			&0.658$\pm$0.011 	&0.754$\pm$0.011 	\\
											       			     	   	 	
$\hspace*{1.2cm}\ell^-\bar{\nu} D_0^{*0}(\Dp\pi^-)$			&0.622$\pm$0.010	&0.727$\pm$0.010 	\\
											       			     	   	 	
$\hspace*{1.2cm}\ell^-\bar{\nu} D_1^{'0}(\pi^-\Dst)(\Dp\piZg)$		&0.671$\pm$0.011  	&0.753$\pm$0.011 	\\
											       			     	   	 	
$\hspace*{1.2cm}\ell^-\bar{\nu} D_2^{*0}(\pi^-\Dst)(\Dp\piZg\pi^-)$	&0.646$\pm$0.011 	&0.759$\pm$0.011  	\\

$\hspace*{1.2cm}\ell^-\bar{\nu} D_2^{*0}(\Dp\pi^-)$			&0.666$\pm$0.011 	&0.763$\pm$0.011 	\\
											       			     	   	 	
$\hspace*{1.2cm}\ell^-\bar{\nu} \Dst\pi^-(\Dp\piZg)$ (NR)		&0.491$\pm$0.010	&0.626$\pm$0.009	\\
											       			     	   	 	
$\hspace*{1.2cm}\ell^-\bar{\nu} \Dp\pi^-$ (NR)				&0.534$\pm$0.009	&0.767$\pm$0.010 	\\
														                        
$\hspace*{1.2cm}D^{(*)}\bar{D}^{(*)}K(\ell \Dp X)$			&0.0005$\pm$0.0001	&0.0013$\pm$0.0001	\\

$\hspace*{1.2cm}D_s^{(*)}D^{(*)}X(\ell \Dp X)$				&0.0008$\pm$0.0001	&0.0038$\pm$0.0002	\\

$\hspace*{1.2cm}\tau^-\bar{\nu} D^{(*,**)}(\ell \Dp X)$			&0.0031$\pm$0.0002	&0.0045$\pm$0.0002	\\
														                        
\hline														                        
$\Bs\ra\ell^-\bar{\nu} D_{s1}^{'+}(2535)(K^0\Dst)(\Dp\piZg)$		&0.612$\pm$0.010 &0.695$\pm$0.010	\\
														     	   	 	
$\hspace*{1.2cm}\ell^-\bar{\nu} D_{s2}^{*+}(2573)(K^0\Dst)(\Dp\piZg)$	&0.575$\pm$0.009	&0.665$\pm$0.009	\\

$\hspace*{1.2cm}\ell^-\bar{\nu} D_{s2}^{*+}(2573)(\Dp K^0)$		&0.592$\pm$0.010	&0.691$\pm$0.009	\\
														                        
$\hspace*{1.2cm}D^{(*)}\bar{D}^{(*)}K(\ell \Dp X)$			&0.0024$\pm$0.0002	&0.0073$\pm$0.0003	\\
														   			
$\hspace*{1.2cm}D_s^{(*)}D^{(*)}X(\ell \Dp X)$				&0.0011$\pm$0.0001	&0.0037$\pm$0.0002	\\
\end{tabular}
\end{ruledtabular}
\end{table}

\linespread{1.0}
\begin{table}
\caption{\label{tab:D0_eff}
Efficiencies in $\lDz$ Monte Carlo simulation.}
\begin{ruledtabular}
\begin{tabular}{lcc}
Decays                                                                  &$e+$SVT 		&$\mu+$SVT  		\\
\hline														                        
$\Bd\ra\ell^-\bar{\nu} \Dst(\Dz\pi^+)$					&0.970$\pm$0.012 	&1.028$\pm$0.012  	\\
											       			      	   	 	
$\hspace*{1.2cm}\ell^-\bar{\nu} D_1^+(\pi^0\Dst)(\Dz\pi^+)$		&0.616$\pm$0.008	&0.702$\pm$0.009	\\
$\hspace*{1.2cm}\ell^-\bar{\nu} D_1^+(\pi^+D^{*0})(\Dz\piZg)$		&0.605$\pm$0.009	&0.682$\pm$0.009	\\
											       			      	   	 	
$\hspace*{1.2cm}\ell^-\bar{\nu} D_0^{*+}(\Dz\pi^+)$			&0.577$\pm$0.008	&0.679$\pm$0.009	\\
											       			      	   	 	
$\hspace*{1.2cm}\ell^-\bar{\nu} D_1^{'+}(\pi^0\Dst)(\Dz\pi^+)$		&0.640$\pm$0.009	&0.700$\pm$0.009	\\
$\hspace*{1.2cm}\ell^-\bar{\nu} D_1^{'+}(\pi^+D^{*0})(\Dz\piZg)$	&0.557$\pm$0.008	&0.631$\pm$0.008	\\
											       			      	   	 	
$\hspace*{1.2cm}\ell^-\bar{\nu} D_2^{*+}(\pi^0\Dst)(\Dz\pi^+)$		&0.613$\pm$0.009	&0.707$\pm$0.009	\\
$\hspace*{1.2cm}\ell^-\bar{\nu} D_2^{*+}(\pi^+D^{*0})(\Dz\piZg)$	&0.622$\pm$0.009	&0.696$\pm$0.009	\\
$\hspace*{1.2cm}\ell^-\bar{\nu} D_2^{*+}(\Dz\pi^0)$			&0.640$\pm$0.009 	&0.745$\pm$0.010	\\
											       			      	   	 	
$\hspace*{1.2cm}\ell^-\bar{\nu} \Dst\pi^0(\Dz\pi^+)$ (NR)		&0.461$\pm$0.007	&0.562$\pm$0.008	\\
											       			      	   	 	
$\hspace*{1.2cm}\ell^-\bar{\nu} D^{*0}\pi^+(\Dz\piZg)$ (NR)		&0.451$\pm$0.007	&0.578$\pm$0.008	\\
											       			      	   	 	
$\hspace*{1.2cm}\ell^-\bar{\nu} \Dz\pi^+$ (NR)				&0.518$\pm$0.008	&0.698$\pm$0.009	\\
												                        
$\hspace*{1.2cm}D^{(*)}\bar{D}^{(*)}K(\ell \Dz X)$			&0.0024$\pm$0.0002	&0.0084$\pm$0.0003	\\
$\hspace*{1.2cm}D^{(*)+}D^{(*)-}(\ell \Dz X)$				&0.0033$\pm$0.0002	&0.0074$\pm$0.0003	\\
														                        
$\hspace*{1.2cm}D_s^{(*)}D^{(*)}X(\ell \Dz X)$				&0.0026$\pm$0.0001	&0.007$\pm$0.0002	\\
														                        
$\hspace*{1.2cm}\tau^-\bar{\nu} D^{(*,**)}(\ell \Dz X)$			&0.0225$\pm$0.0005	&0.0311$\pm$0.0007	\\
														                        
\hline														                        
$\Bu\ra\ell^-\bar{\nu} \Dz$						&$1.376\cdot10^{-3}$ (13,760$\pm$120)	&$1.535\cdot10^{-3}$ (15,350$\pm$130) 	\\
													                        
$\hspace*{1.2cm}\ell^-\bar{\nu} D^{*0}(\Dz\piZg)$			&1.024$\pm$0.012  	&1.057$\pm$0.012  	\\
											       			      	        	
$\hspace*{1.2cm}\ell^-\bar{\nu} D_1^0(\pi^0D^{*0})(\Dz\piZg)$		&0.643$\pm$0.009	&0.717$\pm$0.009	\\
$\hspace*{1.2cm}\ell^-\bar{\nu} D_1^0(\pi^-\Dst)(\Dz\pi^+)$		&0.607$\pm$0.009	&0.710$\pm$0.009	\\
											       			      	        	
$\hspace*{1.2cm}\ell^-\bar{\nu} D_0^{*0}(\Dz\pi^0)$			&0.619$\pm$0.009	&0.716$\pm$0.009	\\
											       			      	        	
$\hspace*{1.2cm}\ell^-\bar{\nu} D_1^{'0}(\pi^0D^{*0})(\Dz\piZg)$	&0.648$\pm$0.009	&0.742$\pm$0.009	\\
$\hspace*{1.2cm}\ell^-\bar{\nu} D_1^{'0}(\pi^-\Dst)(\Dz\pi^+)$		&0.609$\pm$0.009	&0.709$\pm$0.009	\\
											       			      	        	
$\hspace*{1.2cm}\ell^-\bar{\nu} D_2^{*0}(\pi^0D^{*0})(\Dz\piZg)$	&0.638$\pm$0.009	&0.726$\pm$0.009	\\
$\hspace*{1.2cm}\ell^-\bar{\nu} D_2^{*0}(\pi^-\Dst)(\Dz\pi^+)$		&0.598$\pm$0.009	&0.710$\pm$0.009	\\
$\hspace*{1.2cm}\ell^-\bar{\nu} D_2^{*0}(\Dz\pi^0)$			&0.662$\pm$0.009 	&0.757$\pm$0.009	\\
											       			      	        	
$\hspace*{1.2cm}\ell^-\bar{\nu} \Dst\pi^-(\Dz\pi^+)$ (NR)		&0.456$\pm$0.007	&0.580$\pm$0.008	\\
											       			      	        	
$\hspace*{1.2cm}\ell^-\bar{\nu} D^{*0}\pi^0(\Dz\piZg)$ (NR)		&0.474$\pm$0.007	&0.597$\pm$0.008	\\
											       			      	        	
$\hspace*{1.2cm}\ell^-\bar{\nu} \Dz\pi^0$ (NR)				&0.565$\pm$0.008	&0.745$\pm$0.009	\\
														                        
$\hspace*{1.2cm}D^{(*)}\bar{D}^{(*)}K(\ell \Dz X)$			&0.0033$\pm$0.0002	&0.0104$\pm$0.0004	\\
														                        
$\hspace*{1.2cm}D_s^{(*)-}D^{(*)0}(\ell \Dz X)$				&0.0044$\pm$0.0003	&0.0109$\pm$0.0003	\\
														                        
$\hspace*{1.2cm}\tau^-\bar{\nu} D^{(*,**)}(\ell \Dz X)$			&0.0403$\pm$0.0007	&0.0523$\pm$0.0009	\\
														                        
\hline														                        
$\Bs\ra\ell^-\bar{\nu} D_{s1}^{'+}(2535)(K^0\Dst)(\Dz\pi^+)$		&0.541$\pm$0.008	&0.638$\pm$0.008	\\
$\hspace*{1.2cm}\ell^-\bar{\nu} D_{s1}^{'+}(2535)(K^+D^{*0})(\Dz\piZg)$ &0.519$\pm$0.008	&0.633$\pm$0.008	\\
														      	   	 	
$\hspace*{1.2cm}\ell^-\bar{\nu} D_{s2}^{*+}(2573)(K^0\Dst)(\Dz\pi^+)$	&0.513$\pm$0.008	&0.614$\pm$0.008	\\
$\hspace*{1.2cm}\ell^-\bar{\nu} D_{s2}^{*+}(2573)(K^+D^{*0})(\Dz\piZg)$ &0.501$\pm$0.008	&0.621$\pm$0.008	\\
$\hspace*{1.2cm}\ell^-\bar{\nu} D_{s2}^{*+}(2573)(\Dz K^+)$		&0.537$\pm$0.008	&0.639$\pm$0.008	\\
														                        
$\hspace*{1.2cm}D^{(*)}\bar{D}^{(*)}K(\ell \Dz X)$			&0.0022$\pm$0.0002	&0.0072$\pm$0.0003	\\
														                        
$\hspace*{1.2cm}D_s^{(*)}D^{(*)}X(\ell \Dz X)$				&0.0014$\pm$0.0001	&0.0056$\pm$0.0002	\\
\end{tabular}
\end{ruledtabular}
\end{table}
\linespread{1.2}

\begin{table}
\caption{\label{tab:Dst_eff}
Efficiencies in $\lDst$ Monte Carlo simulation.}
\begin{ruledtabular}
\begin{tabular}{lcc}
Decays									&$e+$SVT 		&$\mu+$SVT 		\\
\hline														                        
$\Bd\ra\ell^-\bar{\nu} \Dst(\Dz\pi^+)$					&$0.888\cdot10^{-3}$ (8,880$\pm$100) 	&$1.068\cdot10^{-3}$ (10,680$\pm$100)	\\
														                        
$\hspace*{1.2cm}\ell^-\bar{\nu} D_1^+(\pi^0\Dst)(\Dz\pi^+)$		&0.642$\pm$0.011 	&0.696$\pm$0.010 	\\
											       				   	 	
$\hspace*{1.2cm}\ell^-\bar{\nu} D_1^{'+}(\pi^0\Dst)(\Dz\pi^+)$		&0.680$\pm$0.011 	&0.691$\pm$0.011 	\\
											       				   	 	
$\hspace*{1.2cm}\ell^-\bar{\nu} D_2^{*+}(\pi^0\Dst)(\Dz\pi^+)$		&0.642$\pm$0.011 	&0.698$\pm$0.011 	\\
											       				   	 	
$\hspace*{1.2cm}\ell^-\bar{\nu} \Dst\pi^0(\Dz\pi^+)$ (NR)		&0.484$\pm$0.009  	&0.552$\pm$0.009	\\
														                        
$\hspace*{1.2cm}D^{(*)}\bar{D}^{(*)}K(\ell \Dst X)(\Dz\pi^+)$		&0.0010$\pm$0.0003	&0.0028$\pm$0.0002	\\
$\hspace*{1.2cm}D^{(*)+}D^{(*)-}(\ell \Dst X)(\Dz\pi^+)$		&0.0033$\pm$0.0003	&0.0064$\pm$0.0004	\\
														                        
$\hspace*{1.2cm}D_s^{(*)}D^{(*)}X(\ell \Dst X)(\Dz\pi^+)$		&0.00105$\pm$0.00004	&0.0023$\pm$0.0002	\\
														                        
$\hspace*{1.2cm}\tau^-\bar{\nu} D^{(*,**)}(\ell \Dst X)(\Dz\pi^+)$	&0.018$\pm$0.003	&0.0235$\pm$0.0007	\\
														                        
\hline														                        
$\Bu\ra\ell^-\bar{\nu} D_1^0(\pi^-\Dst)(\Dz\pi^+)$			&0.646$\pm$0.011 	&0.696$\pm$0.011 	\\
											       				   	 	
$\hspace*{1.2cm}\ell^-\bar{\nu} D_1^{'0}(\pi^-\Dst)(\Dz\pi^+)$		&0.640$\pm$0.011 	&0.701$\pm$0.011	\\
										       				   	 	
$\hspace*{1.2cm}\ell^-\bar{\nu} D_2^{*0}(\pi^-\Dst)(\Dz\pi^+)$		&0.620$\pm$0.011 	&0.688$\pm$0.011 	\\
											       				   	 	
$\hspace*{1.2cm}\ell^-\bar{\nu} \Dst\pi^-(\Dz\pi^+)$ (NR)		&0.466$\pm$0.009	&0.577$\pm$0.009	\\
														                        
$\hspace*{1.2cm}D^{(*)}\bar{D}^{(*)}K(\ell \Dst X)(\Dz\pi^+)$		&0.0006$\pm$0.0001	&0.0013$\pm$0.0002	\\
														                        
$\hspace*{1.2cm}D_s^{(*)}D^{(*)}X(\ell \Dst X)(\Dz\pi^+)$		&0.00028$\pm$0.00006	&0.00021$\pm$0.00007	\\
														                        
$\hspace*{1.2cm}\tau^-\bar{\nu} D^{(*,**)}(\ell \Dst X)(\Dz\pi^+)$	&0.002$\pm$0.001	&0.0024$\pm$0.0009	\\
														                        
\hline														                        
$\Bs\ra\ell^-\bar{\nu} D_{s1}^{'+}(2535)(K^0\Dst)(\Dz\pi^+)$		&0.553$\pm$0.010	&0.620$\pm$0.010	\\
														                        
$\hspace*{1.2cm}\ell^-\bar{\nu} D_{s2}^{*+}(2573)(K^0\Dst)(\Dz\pi^+)$	&0.525$\pm$0.010	&0.593$\pm$0.009	\\
														                        
$\hspace*{1.2cm}D^{(*)}\bar{D}^{(*)}K(\ell \Dst X)(\Dz\pi^+)$		&0.0007$\pm$0.0003	&0.0022$\pm$0.0003	\\
														                        
$\hspace*{1.2cm}D_s^{(*)}D^{(*)}X(\ell \Dst X)(\Dz\pi^+)$		&0.00024$\pm$0.00008	&0.0012$\pm$0.0002	\\
\end{tabular}
\end{ruledtabular}
\end{table}

\begin{table}[tbp]
\caption{\label{tab:Ds_eff}
Efficiencies in $\lDs$ Monte Carlo simulation.}
\begin{ruledtabular}
\begin{tabular}{lcc}
Decays									&$e+$SVT 		&$\mu+$SVT		\\
\hline														                        
$\Bs\ra\ell^-\bar{\nu} \Ds$						&$0.998\cdot10^{-3}$ (9,980$\pm$100)	&$1.201\cdot10^{-3}$ (12,010$\pm$110)	\\
 														                        
$\hspace*{1.2cm}\ell^-\bar{\nu} D_s^{*+}(\Ds\gamma)$			&1.035$\pm$0.014  	&1.11$\pm$0.014  	\\
											       				   	 	
$\hspace*{1.2cm}\ell^-\bar{\nu} D_{s0}^{*+}(2317)(\Ds\pi^0)$		&0.684$\pm$0.011 	&0.773$\pm$0.011  	\\
											       				   	 	
$\hspace*{1.2cm}\ell^-\bar{\nu} D_{s1}^{+}(2460)(\pi^0 D_{s0}^{*+})(\Ds\pi^0)$&0.709$\pm$0.011  &0.786$\pm$0.011 	\\
$\hspace*{1.2cm}\ell^-\bar{\nu} D_{s1}^{+}(2460)(\Ds\gamma)$		&0.710$\pm$0.011 	&0.781$\pm$0.011 	\\
											       				   	 	
$\hspace*{1.2cm}\ell^-\bar{\nu} D_s^{*+}\pi^0(\Ds\gamma)$ (NR)		&0.436$\pm$0.008	&0.591$\pm$0.009	\\
											       				   	 	
$\hspace*{1.2cm}\ell^-\bar{\nu} \Ds\pi^0$ (NR)				&0.479$\pm$0.008 	&0.722$\pm$0.010	\\
														                        
$\hspace*{1.2cm}D_s^{(*)}D^{(*)}X(\ell \Ds X)$				&0.0023$\pm$0.0003	&0.0086$\pm$0.0007	\\
														                        
$\hspace*{1.2cm}D_s^{(*)+}D_s^{(*)-}(\ell \Ds X)$			&0.0075$\pm$0.0003	&0.0175$\pm$0.0005	\\
														                        
$\hspace*{1.2cm}\tau^-\bar{\nu} D_s^{(*,**)+}(\ell \Ds X)$		&0.034$\pm$0.005	&0.052$\pm$0.001	\\
														                        
\hline														                        
$\Bd\ra D_s^{(*)+}D^{(*)-}(\ell \Ds X)$					&0.0055$\pm$0.0003	&0.0126$\pm$0.0005	\\
\hline														                        
$B^+\ra D_s^{(*)+}D^{(*)-}(\ell \Ds X)$					&0.0055$\pm$0.0003	&0.0109$\pm$0.0005	\\
\end{tabular}
\end{ruledtabular}
\end{table}

\begin{table}[tbp]
\caption{\label{tab:Lc_eff}
Efficiencies in $\lLc$ Monte Carlo simulation.  
}
\begin{ruledtabular}
\begin{tabular}{lcc}
Decays								&$e+$SVT 		&$\mu+$SVT 		\\
\hline														                        
$\Lb\ra\ell^-\bar{\nu} \Lc X$					&$0.629\cdot10^{-3}$ (6,290$\pm$90)	&$0.722\cdot10^{-3}$ (7,220$\pm$100)	\\
$\hspace*{1.0cm}\tau^-\bar{\nu} \Lambda_c^{+} X(\ell \Lc X)$	&0.026$\pm$0.001	&0.033$\pm$0.0007	\\
\end{tabular}
\end{ruledtabular}
\end{table}

\section{\label{sec:fit} 
Fit of Relative Fragmentation Fractions}

A $\chi^2$-fit is used to extract the fragmentation fractions from the
sample composition of the semileptonic bottom hadron decays
reconstructed.  The measured yields in the five lepton-charm signals
are fit to the yields predicted by the sample composition procedure,
and the decay rates ($\Gamma^{(*,**)}$) of the \Bb~meson to the ground
and excited states are constrained within their errors.  The $\chi^2$,
which is minimized, is
\begin{eqnarray}
\chi^2 &=& \sum_{i=1}^5
\left (\frac{N(\ell^-D_i)_{\rm measured}-N(\ell^-D_i)_{\rm predicted}}
{\sigma({N(\ell^-D_i)_{\rm measured}})}\right )^2 \nonumber\\
&&+ \left (\frac{\Gamma-\Gamma_{\rm PDG}}{\sigma_{\Gamma_{\rm PDG}}}\right )^2 
+ \left (\frac{\Gamma^*-\Gamma^*_{\rm PDG}}{\sigma_{\Gamma^*_{\rm PDG}}}\right )^2  
+ \left (\frac{\Gamma^{**}-\Gamma^{**}_{\rm PDG}}{\sigma_{\Gamma^{**}_{\rm PDG}}}\right )^2,  
\label{eqn:chi2}
\end{eqnarray}
where $N_{\rm predicted}$ is determined from the sample composition
process and $N_{\rm measured}$ is obtained from the data (see
Tab.~\ref{tab:yields}).  In order to fit in terms of better measured
quantities, the number of predicted lepton-charm events are expressed in
terms of $N(\Bd)$, which is an overall normalization in the fit and not
indicative of the physical number of \Bd~mesons in the data, and in
terms of the \Bb~meson lifetimes relative to the lifetime of the \Bd.
The predicted number of lepton-charm events used in the fit for the
\Bb~mesons is expressed as:
\begin{eqnarray}
N(\ell^-D_i)_{\rm predicted} &=& \sum_{j=d,u,s} N(\Bd)\frac{f_j}{\fd}
\times \tau(\Bd) \nonumber\\
&&\times\tau(B_j)/\tau(\Bd)\sum_{k}\Gamma_k \times 
\sum_{m}{\cal B}(D_{jkm}\ra D_i)\,{\cal B}(D_i)\,
\varepsilon_{ijkm}\nonumber\\
&=& \sum_{j=d,u,s} N(\Bd) \frac{f_j}{\fu+\fd}\left(1+\frac{\fu}{\fd}\right)
\times\tau(\Bd)\nonumber\\
&&\times\tau(B_j)/\tau(\Bd)\sum_{k}\Gamma_k \times 
\sum_{m}{\cal B}(D_{jkm}\ra D_{i}){\cal B}(D_i)\varepsilon_{ijkm},
\label{eqn:npred_fit_meson}
\end{eqnarray}
where $D_i = {\Dp,\Dz, \Dst, \Ds}$,
$\Gamma_k={\Gamma,\Gamma^*,\Gamma^{**}}$, and the sum over $m$ applies
to $\Gamma^{**}$ if there is more than one $D^{**}$ state or
non-resonant decay that can contribute to the final state $D_i$. In
that case the sum is weighted according to the branching fractions
${\cal BR}(D_{jkm}\ra D_i)$ of the various contributing $D^{**}$
states. The efficiencies $\varepsilon_{ijkm}$ refer to the
corresponding absolute efficiencies of events obtained in a particular
final state normalized to the generated number of MC events with 
$p_T(\Bb)>5~\gevc$. These efficiencies are detailed in
Tables~\ref{tab:Dp_eff} through \ref{tab:Ds_eff}.  In the case of the
\Lb~baryon, where a fixed sample composition is used, the predicted
number of events is given as
\begin{eqnarray}
N(\lLc)_{\rm predicted} &=& N(\Bd) \frac{\fb}{\fu+\fd}
\left (1+\frac{\fu}{\fd}\right )\times{\cal B}(\LbLcln X)\,
{\cal B}(\LcpKpi) \nonumber\\
&&\times\varepsilon(\LbLcln X).
\label{eqn:npred_fit_baryon}
\end{eqnarray}
In this case, the excited charm baryon branching fractions into the 
\Lc~state are included in the efficiency $\varepsilon(\LbLcln X)$ in
Eq.~(\ref{eqn:npred_fit_baryon}), as listed in Table~\ref{tab:Lc_eff}.

There are four free parameters in the fit for the fragmentation
fractions: three relative fragmentation parameters \fufd, \fsoud,
\fboud, and the normalization parameter $N(\Bd)$, plus three
constrained parameters: $\Gamma$, $\Gamma^*$, and $\Gamma^{**}$. The
values of \fs\ and \fb\ are fit relative to $(\fu+\fd)$ to minimize as
many biases in the measurement as possible and to highlight the fact
that the \Bs\ is reconstructed relative to the \Bd\ and
\Bu~signals.  Additionally, performing the fit relative to $\fu+\fd$ limits
any possible inaccuracies in the separation of \Bd\ and \Bu~through
the sample composition procedure into the \lDz\ and \lDp~final
states. As mentioned earlier, the sum of fragmentation fractions \fu,
\fd, \fs, and \fb\ is not constrained to unity in the fit, since
not all $b$~baryons are necessarily accounted for by reconstructing
\lLc~states.

The electron and muon samples are fit separately, since the relative
lepton efficiencies between the electron and muon modes are not
expected to readily cancel.  The fit results are given in
Table~\ref{tab:fit} with statistical errors only indicating good
agreement between the $e$+SVT and $\mu$+SVT data sets.  As mentioned
previously, $f_q$ indicates the fragmentation fraction integrated
above the momentum threshold of sensitivity in the data, $f_q\equiv
f_q(p_T(\Bb)>7~\gevc)$.  Note that about 90\% of the \lLc~combinations
in data have transverse momenta below $\sim20$~\gevc, but none have
$p_T$ less than 7~\gevc.

\begin{table}
\caption{\label{tab:fit}
Fit results with statistical errors only.}
\begin{ruledtabular}
\begin{tabular}{ccc}
Fit Parameter               & $e$+SVT            & $\mu$+SVT         \\
\hline		             					    
\fufd                       & $1.044\pm0.028$    & $1.062\pm0.024$   \\
\fsoud                      & $0.162\pm0.008$    & $0.158\pm0.006$   \\
\fboud                      & $0.292\pm0.020$    & $0.275\pm0.015$   \\
$\Gamma$~[ps$^{-1}$]        & $0.0157\pm0.0007$  & $0.0154\pm0.0007$ \\
$\Gamma^*$~[ps$^{-1}$]      & $0.0327\pm0.0014$  & $0.0331\pm0.0013$ \\
$\Gamma^{**}$~[ps$^{-1}$]   & $0.0145\pm0.0010$  & $0.0146\pm0.0010$ \\
$N(\Bd)\ (10^9)$            & $2.02\pm0.07$      & $2.93\pm0.10$     \\
\end{tabular}
\end{ruledtabular}
\end{table}

\section{\label{sec:sys}
Systematic Uncertainties}

The main uncertainties in the measurement of the relative
fragmentation fractions come from the uncertainties in the branching
fractions of the charm mesons, which contribute both directly and
indirectly to the measurement, and the uncertainty associated with the
baryon $p_T$ spectrum, which affects the simulation-based efficiency.
The uncertainties in the measurement due to the XFT and
\dedx~efficiencies are negligible in comparison with other systematic
uncertainties.  The complete list of systematic uncertainties assigned
to the fragmentation fractions is given in
Table~\ref{tab:systematics}.  A weighted average between the $e$+SVT
and $\mu$+SVT samples is calculated before and after applying a
particular systematic variation in order to determine the systematic
uncertainty for a given quantity.  The determination of the individual
systematic uncertainties is discussed in the following sections.

\subsection{\label{sec:ws} False Lepton Backgrounds}

\begin{figure*}
\centerline{
\includegraphics[width=0.5\hsize]{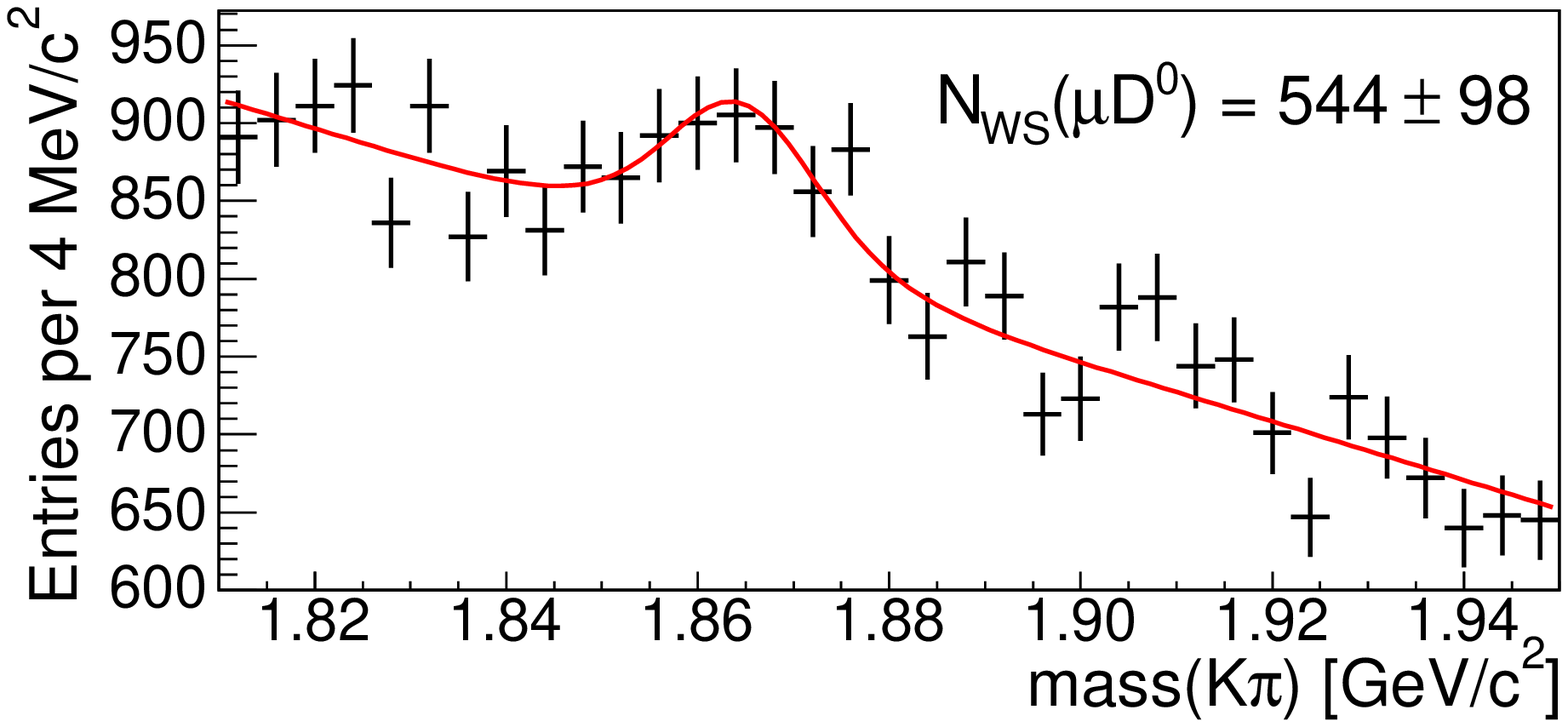}
\includegraphics[width=0.5\hsize]{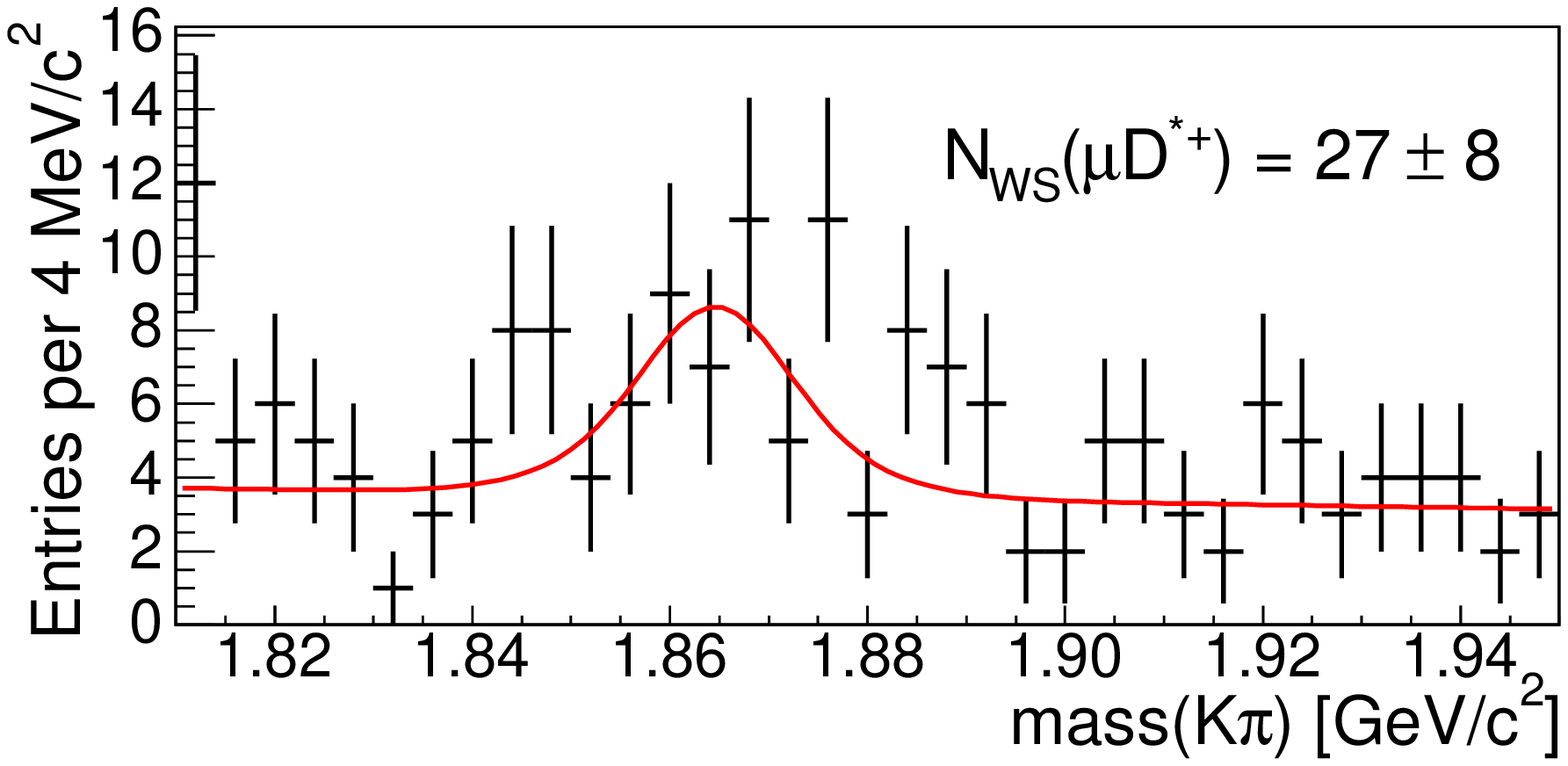}
\put(-425,32){\large\bf (a)}
\put(-185,82){\large\bf (b)}
}
\centerline{
\includegraphics[width=0.5\hsize]{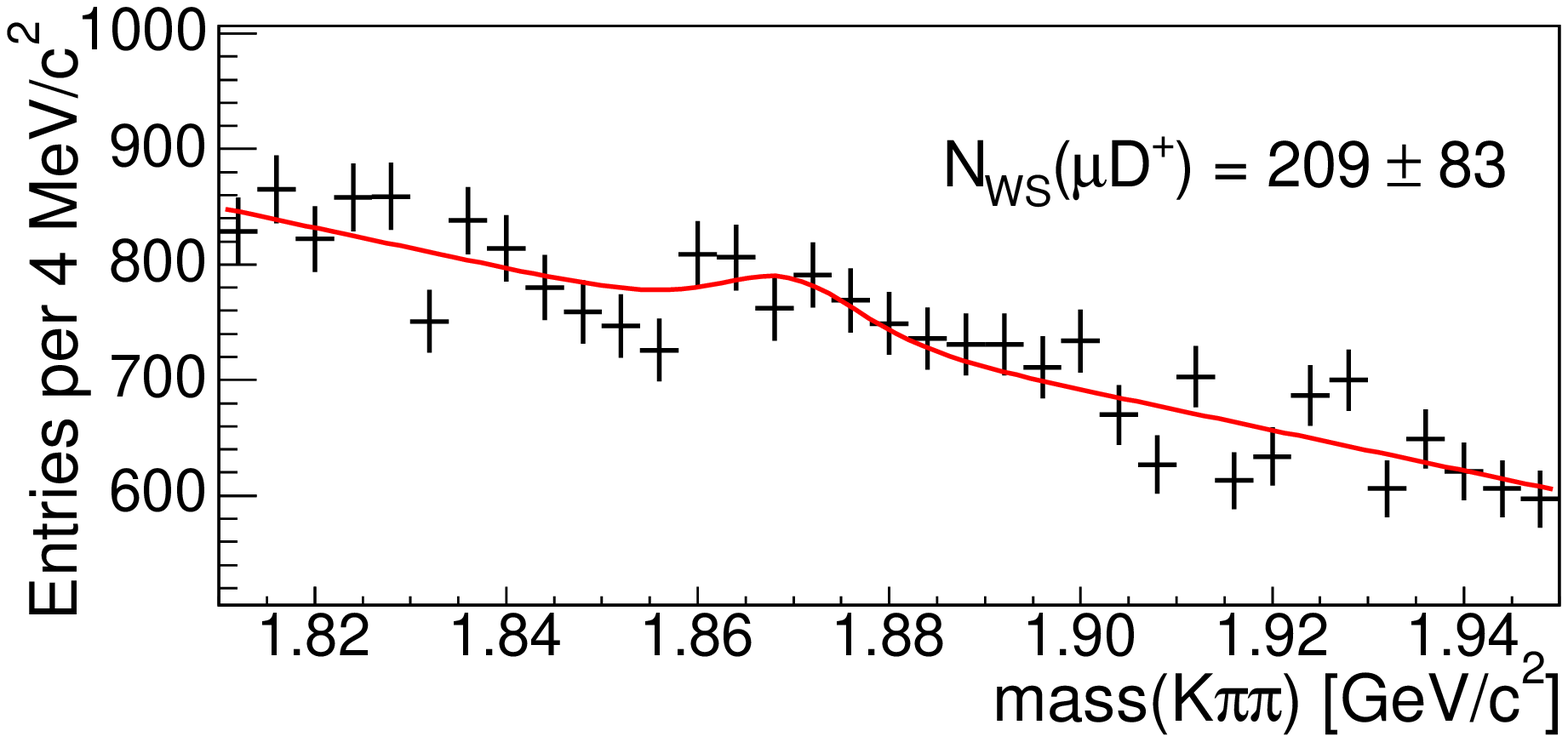}
\includegraphics[width=0.5\hsize]{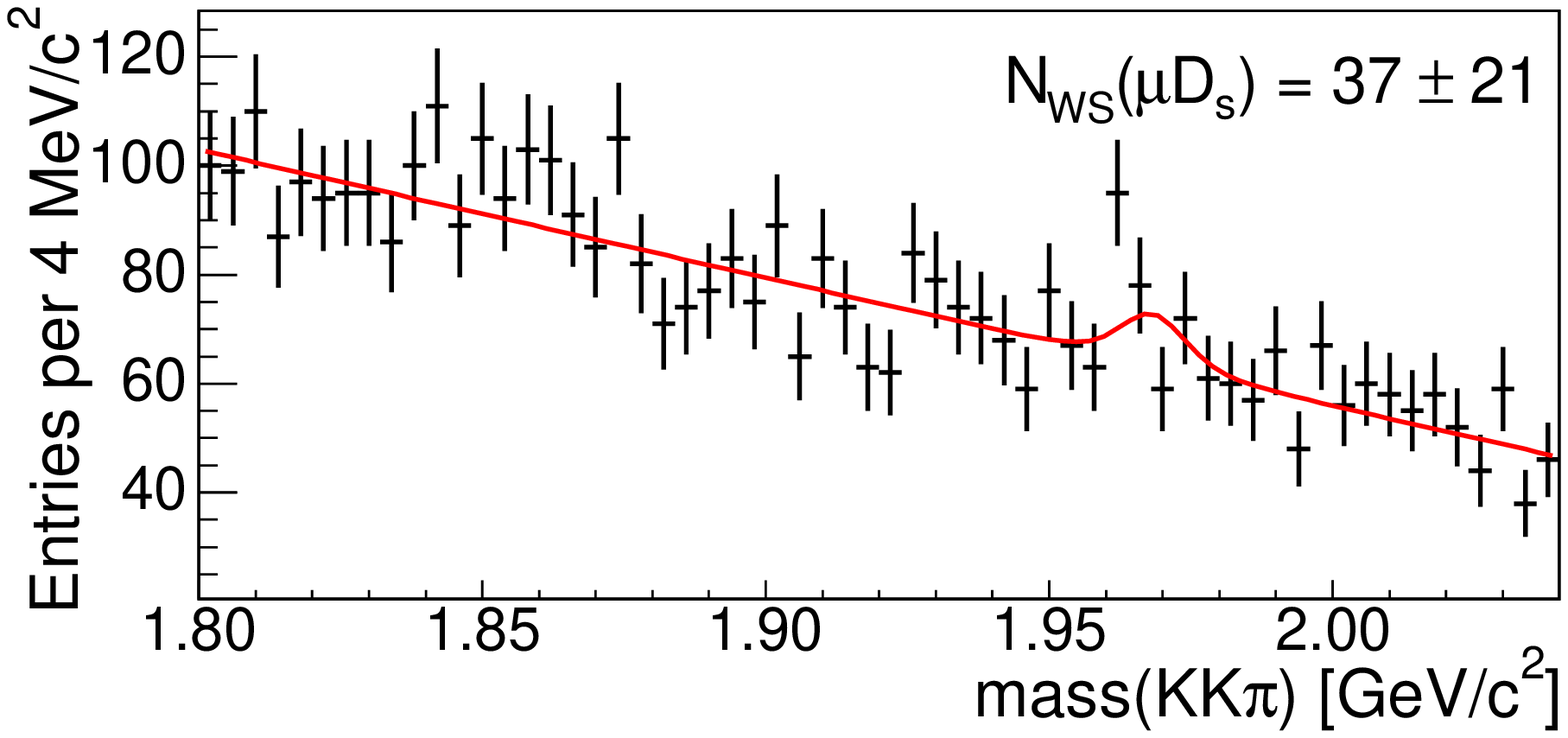}
\put(-425,32){\large\bf (c)}
\put(-185,32){\large\bf (d)}
}
\centerline{
\includegraphics[width=0.5\hsize]{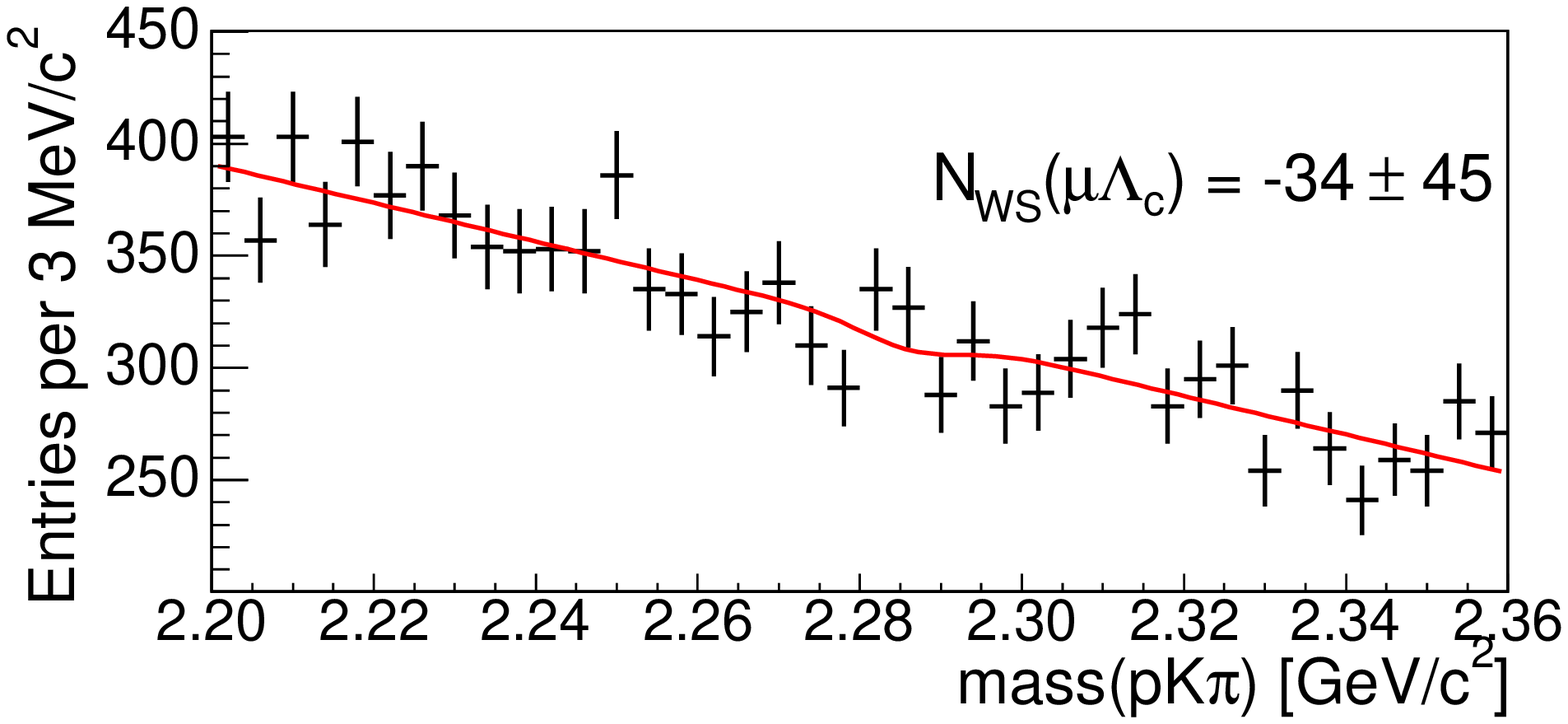}
\put(-190,32){\large\bf (e)}
}
\caption{\label{fig:ws_mu}
$\mu$+SVT wrong sign invariant mass distributions of (a) \Dz, (b) \Dst,
(c) \Dp, (d) \Ds, and (e) \Lc~after the $ct^*(\mu^-D)$ requirement.}
\end{figure*}

\begin{figure*}
\centerline{
\includegraphics[width=0.5\hsize]{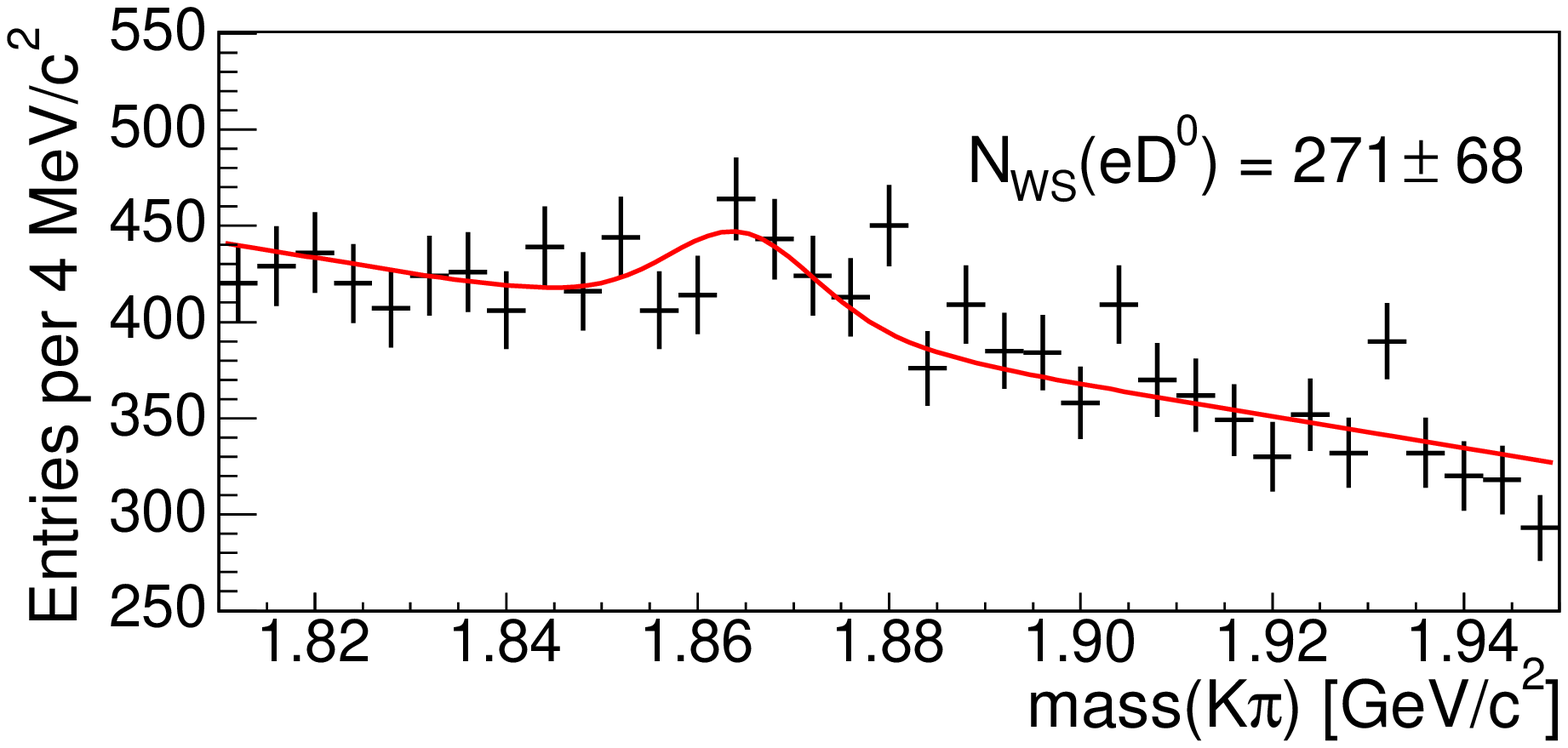}
\includegraphics[width=0.5\hsize]{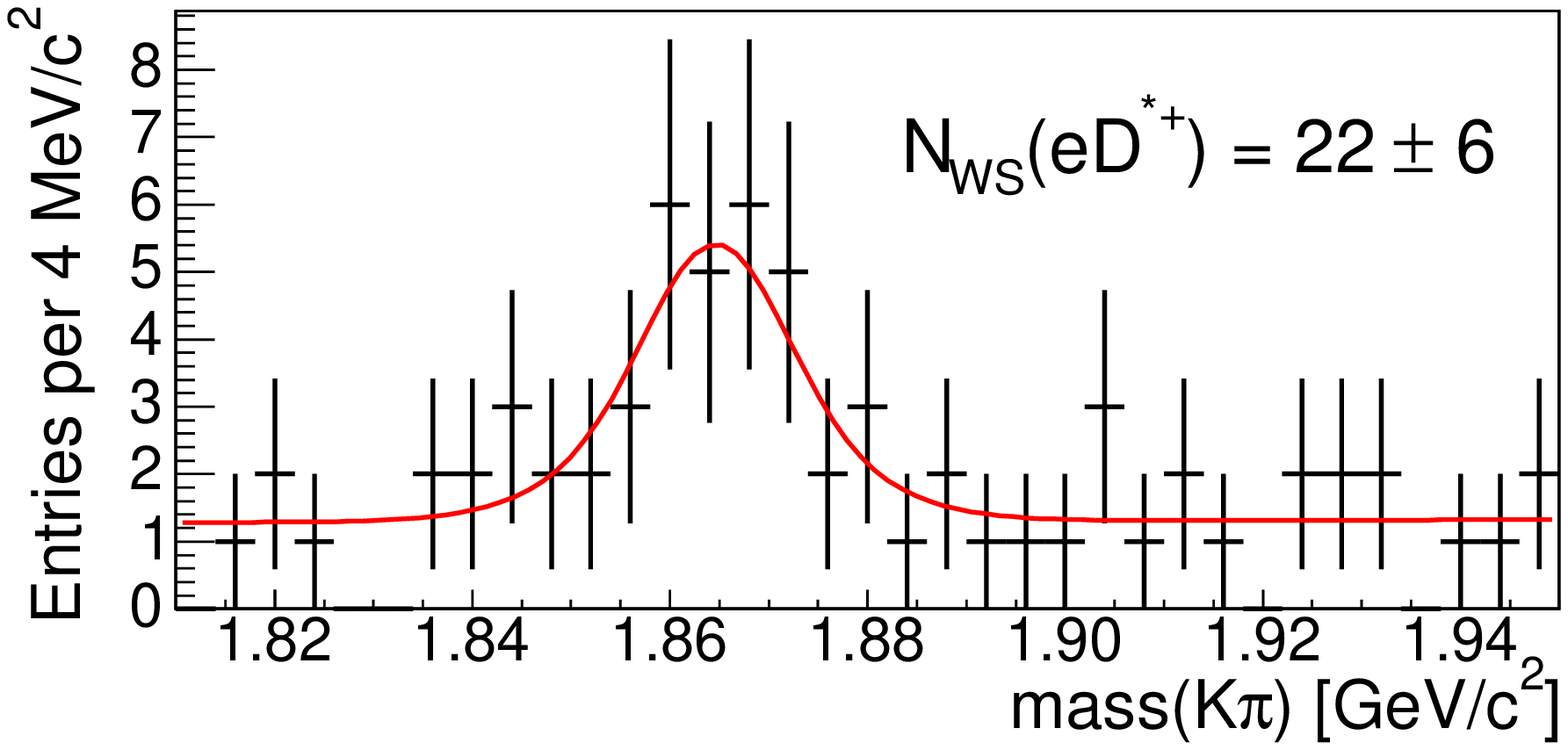}
\put(-425,32){\large\bf (a)}
\put(-185,82){\large\bf (b)}
}
\centerline{
\includegraphics[width=0.5\hsize]{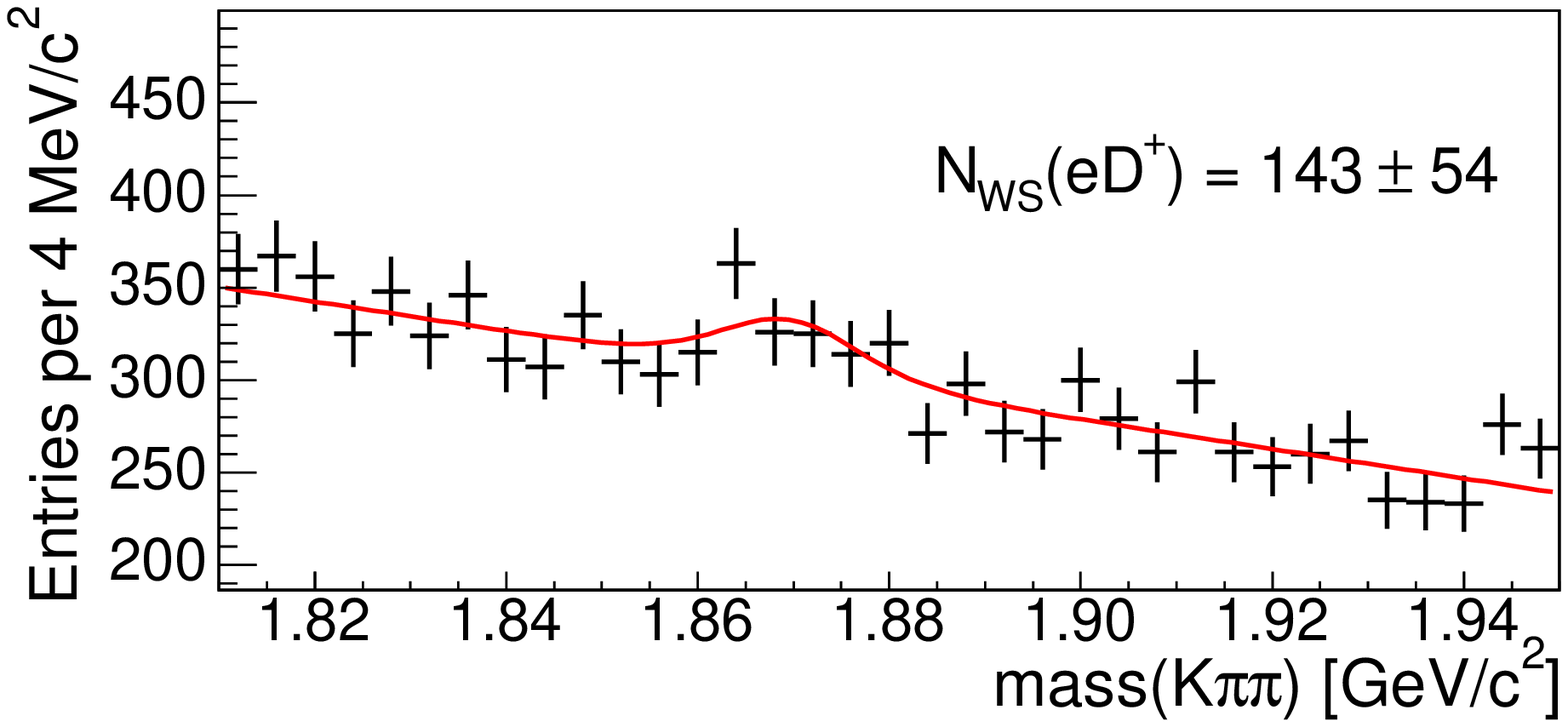}
\includegraphics[width=0.5\hsize]{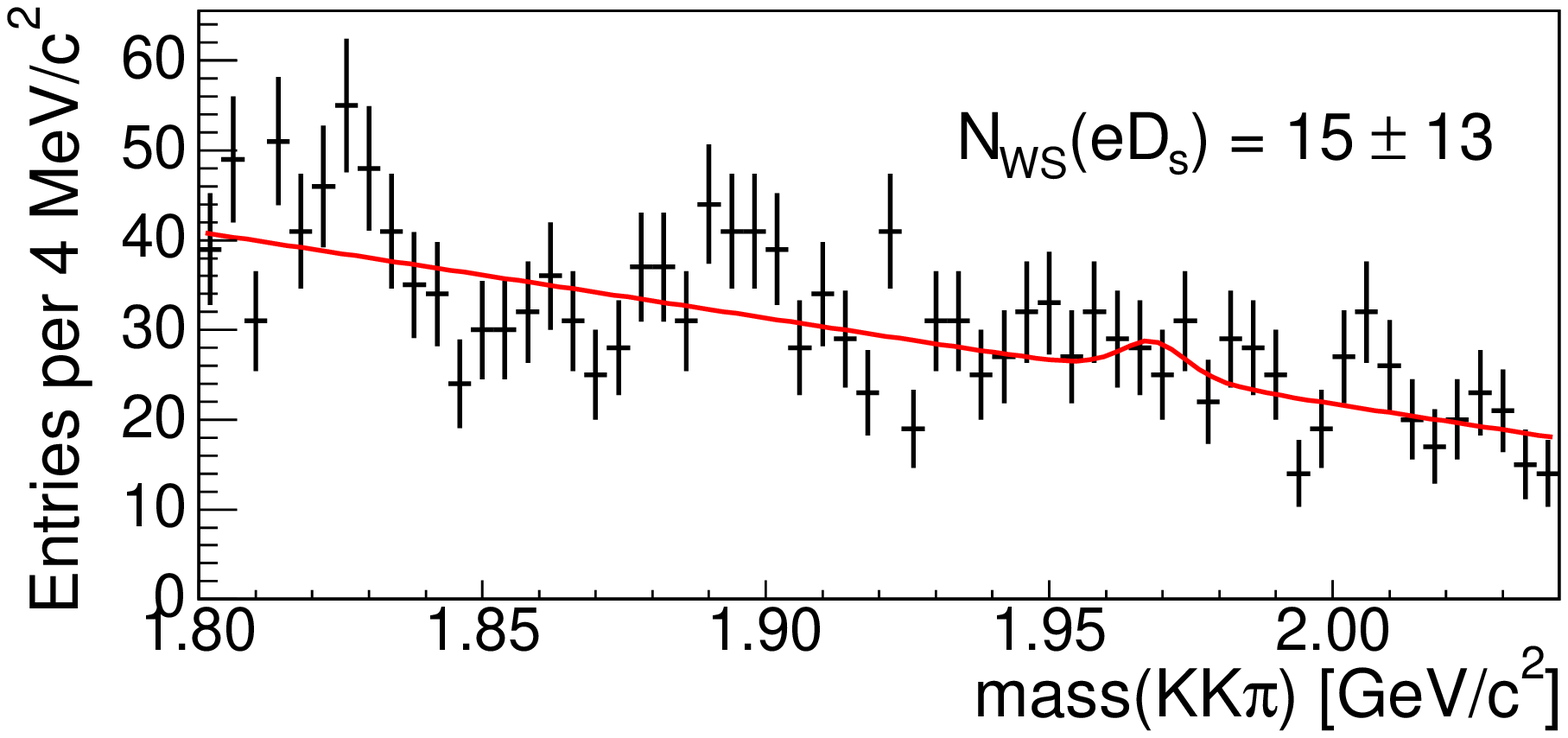}
\put(-425,32){\large\bf (c)}
\put(-185,32){\large\bf (d)}
}
\centerline{
\includegraphics[width=0.5\hsize]{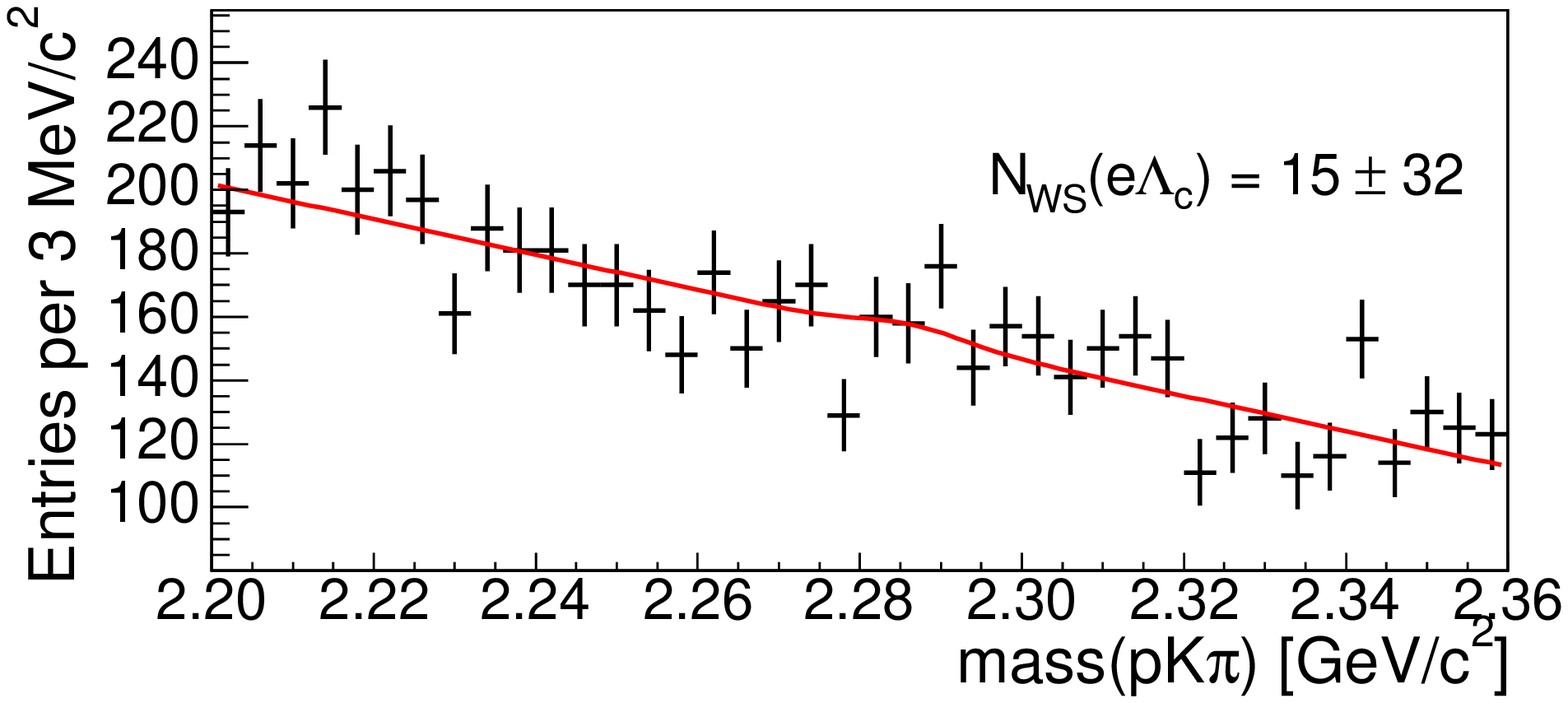}
\put(-190,32){\large\bf (e)}
}
\caption{\label{fig:ws_el}
$e$+SVT wrong sign invariant mass distributions of (a) \Dz, (b) \Dst,
(c) \Dp, (d) \Ds, and (e) \Lc~after the $ct^*(e^-D)$ requirement.}
\end{figure*}

The wrong sign lepton-charm combinations represent several possible
backgrounds that may be present in the right sign signals with a
significant contribution to the wrong sign combinations expected to
arise from false lepton candidates. Another contribution originates
from real leptons from non-bottom sources, such as electrons from
photon conversion $\gamma\ra e^+e^-$ or muons from kaon and pion
decay-in-flight. These sources are included in the discussion of false
lepton backgrounds.  The wrong sign signals are present in the data
even after the prompt region is removed by requiring $ct^*(\lD)>
200$~$\mu$m, as can be seen in Figs.~\ref{fig:ws_mu} and
\ref{fig:ws_el}.  Additionally, some discrepancy is still observed in
the $ct^*(\lD)$ comparisons between data and simulation, possibly
indicating a residual background from false leptons.  False leptons
that originate from a ``\Bb''-like hadron ({\it i.e.} a relatively
long-lived particle) are not necessarily represented equally between
right sign and wrong sign combinations, as is the case with prompt
false leptons.  Since the false leptons of concern most likely come
from a real bottom hadron in which a hadronic track has been
mis-identified as a lepton, they are enhanced in the right sign over
wrong sign lepton-charm combinations.

This systematic uncertainty is studied by utilizing the large false
lepton sample available from the TTT semileptonic \Bb~decays, which
has approximately five times more \lDp\ and \lDz~events than the
$\ell$+SVT trigger sample.  Since the statistics are much larger in
the TTT sample and the average lepton transverse momentum is lower, a
larger sample of false leptons is available for the study.  Lepton
candidates with a low probability of being true leptons, as measured
from a likelihood weighting of lepton identification
variables~\cite{Ref:Giurgiu, Ref:Tiwari}, are selected from
semileptonic TTT events in which one of the charm daughters is matched
to one SVT track and the lepton is matched to the other SVT trigger
track.  This sample of false leptons is then used to estimate the
factor required to scale the residual wrong sign ``signals'' in the
$\ell$+SVT data to the right sign signals, giving an estimate of the
false lepton contamination from bottom hadrons in the right sign
signals.  All selection requirements used in this analysis are applied
and the numbers of right sign and wrong sign events are compared.  The
scaling obtained for false leptons from long-lived ``bottom''-like
hadrons is $w_{e} = 2.93\pm 0.47$ and $w_{\mu} = 3.91\pm 0.73$,
assuming that all of the wrong sign events originate from long-lived
sources. To obtain the systematic uncertainty, the right sign
lepton-charm yields are decreased by the wrong sign yields scaled by
the appropriate factor for the $e$+SVT and $\mu$+SVT datasets.  The
fit for the fragmentation fractions is then repeated and the resulting
systematic uncertainties noted in Table~\ref{tab:systematics}.

\subsection{Variation of Selection Requirements}

The selection requirements have been chosen to be similar across the
five lepton-charm channels, in order to cancel as many systematic
uncertainties as possible while still respecting the different
kinematic features of the decays.  To check the dependence of the
final result on these selection criteria, the signal selection has
been varied in such a way that the $p_T(D) > 5~\gevc$ cut is applied
to all channels, while the $\chi^2_{xy}(D)$ and vertex probability
requirements are the same.  The varied cuts used to assign the
systematic uncertainty are listed in Table~\ref{tab:varied_cuts}.

\begin{table}
\caption{\label{tab:varied_cuts}
Alternative signal selection.}
\begin{ruledtabular}
\begin{tabular}{cccccc}
Cuts  & $\lDz$ & $\lDst$ & $\lDp$ & $\lDs$  & $\lLc$\\
    \hline
$ct(D)$~[cm] $\in$          & 
(-0.01,0.10) & (-0.01,0.10) & (-0.01,0.20) & (-0.01,0.10) & (-0.01,0.05) \\
$ct^*(\lD)$~[cm] $>$    &
0.02         & 0.02         & 0.02         & 0.02         & 0.02         \\
$\sigma_{ct^*}(\lD)$~[cm] $<$ &
0.04         & 0.04         & 0.04         & 0.04         & 0.04         \\
$m(\lD)$~[\gevcc] $\in$ &
(2.4,5.1)    & (2.4,5.1)    & (2.4,5.1)    & (2.4,5.1)    & (3.4,5.5)    \\
$p_T(D)$ [\gevc] $>$        &
5.0          & 5.0          & 5.0          & 5.0          & 5.0          \\
$p_T(p)$ [\gevc] $>$        & 
N/A          & N/A          & N/A          & N/A          & 2.0          \\
$p_T(K)$ [\gevc] $>$        & 
0.6          & 0.6          & 0.6          & 0.6          & 0.6          \\
$\chi^2_{2D}(D) <$          &
10           & 10           & 10           & 10           & 10            \\
vertex prob.$(\lD) >$   &
$10^{-7}$    & $10^{-7}$    & $10^{-7}$    & $10^{-7}$    & $10^{-7}$    \\
$\lxy/\sigma_{Lxy}(D) >$    &
4.5          & 4.5          & 4.5          & 4.5          & 4.5          \\
$\Delta m(\Dst,\Dz)$~[\gevcc] $\in$ &
N/A          & (0.1440,0.1475) & N/A       & N/A          & N/A          \\
$|m(\phi)-1.019|$~[\gevcc] $<$ & 
N/A          & N/A          & N/A          & 0.0095       & N/A          \\
\dedx~${\cal LR}(p)\ >$     &
N/A          & N/A          & N/A          & N/A          & 0.3          \\
\end{tabular}
\end{ruledtabular}
\end{table}

\subsection{\boldmath \Ds~Reflection in the \Dp~Signal}

In addition to residual wrong sign backgrounds, another source of
irreducible non-combinatoric background arises from the $\Ds\ra
K^+K^-\pi^+$ reflection into the $\Dp\ra K^-\pi^+\pi^+$ signal.  This
effect has been measured from the data, using the simulation to scale
the expected rates of generic $\Ds\ra K^-K^+\pi^+$ decays to the
$\Ds\ra\phi\pi^+$ decay (see Section~\ref{sec:reflec}).  By default,
the \Ds~reflection is included in the fit to the \Dp~signal by
constraining the normalization of the
\Ds~reflection within its uncertainty.  In order to assign a systematic
uncertainty on this method, the normalization of the reflection is
allowed to vary, both by fixing $N(\Ds)$ to the number measured from
data, and also by allowing $N(\Ds)$ to float in the fit to obtain the
\Dp~signal, listed for both scenarios in Table~\ref{tab:reflec}.  The
larger effect is observed when $N(\Ds)$ is a free fit parameter, while
fixing the normalization produces a more moderate shift.  Since the
normalization procedure of $N(\Ds\ra\phi\pi^+)$ relative to
$N(\Ds\ra K^+K^-\pi^+)$ is, in principle, well-understood from the
data and simulation, the variations obtained from fixing the
normalization are taken as the systematic uncertainty associated with
this method.

\begin{table}[tbp]
\caption{\label{tab:reflec}
$\lDp$ yields with different \Ds~reflection normalizations.}
\begin{ruledtabular}
\begin{tabular}{lcccc}
&\multicolumn{2}{c}{$e$+SVT}      &\multicolumn{2}{c}{$\mu$+SVT}     			\\
\Ds~Reflection &$N(\Ds)$        &$N(\Dp e^-)$         &$N(\Ds)$      &$N(\Dp\mu^-)$    \\
\hline		       	                                         		     
Constrained      &1,710$\pm$80    &10,780$\pm$150    &2,780$\pm$460   &20,240$\pm$220   \\
Fixed            &1,577           &10,780$\pm$150    &2,570           &20,270$\pm$250   \\
Floating         &3,270$\pm$1,100 &10,570$\pm$200    &5,050$\pm$1,670 &19,910$\pm$340   \\
None             &---             &11,020$\pm$160    &---             &20,640$\pm$260   \\  
\end{tabular}
\end{ruledtabular}
\end{table}

\subsection{XFT Efficiencies}

Knowledge of efficiencies that are different for the different
particle species is essential for the proper determination of the
relative efficiencies between lepton-charm channels.  One of these
sets of efficiencies is the XFT trigger efficiencies (described in
Section~\ref{sec:xft}), which cannot be accurately predicted by the
simulation and are not expected to readily cancel in the relative
efficiencies between the final state charm signals.  The systematic
uncertainty on this efficiency is determined by varying the default
XFT efficiencies by 1\,$\sigma$ of the fit parameters given in
Table~\ref{tab:XFT}.  To determine the systematic uncertainty of the
shift in the $K$- and $\pi$-XFT efficiencies, the $\pi$ efficiency,
which has the larger uncertainty of the two, is shifted up or down by
the uncertainties in the XFT parameterizations, while the $K$
efficiency is held constant.  The systematic uncertainty associated
with the proton efficiency is assigned by fitting the efficiency with
a constant line.  The proton parameterization was shifted by the full
uncertainty on the fit parameters (either all up or all down), which
are quite large as can be seen in Table~\ref{tab:XFT}.

\subsection{Single Track Efficiency}

The efficiency to add a single track, needed to adjust the two track
topology to the three track topology, is measured by reconstructing the
$\Dz\ra K^-\pi^+\pi^-\pi^+$ channel relative to the \DzKpi~decay mode
(see Section~\ref{sec:single_track}).  This method assumes that the two
additional pions in the $\Dz\ra K^-\pi^+\pi^-\pi^+$ decay are
uncorrelated.  Since the two tracks are identified and pass through
different parts of the detector, this assumption is reasonable.  The
only way the determination of the single track efficiency might be
biased by the correlation of the third and fourth tracks arises from
vertexing effects.  To assess the degree of the bias that might occur in
the vertexing of the $\Dz\ra K^-\pi^+\pi^-\pi^+$ mode due to the
correlation between the two additional pions, a three track vertex is
formed in the Monte Carlo simulation of $\mu^-\Dz$ decays with $\Dz\ra
K^-\pi^+\pi^-\pi^+$ and the impact parameter of the fourth track with
respect to the three track vertex is determined.  If the impact
parameter of the fourth track lies outside of 1\,$\sigma$ of the error
on the vertex and 1\,$\sigma$ of the error on the impact parameter, it
is assumed that the fourth track could bias the position of the vertex.
The impact parameter of the fourth track is found to be outside
1\,$\sigma$ of the three-track vertex $(5.3\pm 0.1)\%$ of the time. This
fraction is assumed to correspond, to good approximation, to the degree
of correlation in the efficiency of the third and fourth pions.  Another
source of uncertainty in the determination of the single track
efficiency is the error on the ratio of branching fractions of the two
reconstructed $\Dz$ decays, ${\cal B}(\Dz\ra K^-\pi^+\pi^-\pi^+)/{\cal
B}(\Dz\ra K^-\pi^+)= 2.10\pm 0.03\pm 0.06$~\cite{Ref:CLEO-c}. The
systematic uncertainty from the error on this ratio of branching
fractions is also included in the systematic uncertainty from the single
track efficiency listed in Table~\ref{tab:systematics}.

\subsection{Sample Composition Lifetimes}

\Bb~meson lifetimes relative to the \Bd~lifetime are included in the sample
composition procedure.  Consequently, the \Bb~lifetimes are needed to
determine the predicted number of lepton-charm mesons (see
Section~\ref{sec:fit}).  As there are uncertainties on the PDG
values~\cite{Ref:PDG_2004} of the lifetimes used in the fit, which are
listed in Table~\ref{tab:lifetimes}, the lifetimes and lifetime ratios
are varied in the process extracting the sample composition within
their PDG uncertainties.  The central value of the lifetime ratio
$\tau(\Bu)/\tau(\Bd) = 1.086\pm 0.017$ has changed several times in
several years and different values are used in the sundry measurements
of \fufd.  Although the lifetime ratio in the PDG is slightly higher
than that used in other measurements of
\fufd, the uncertainty on the PDG value covers the central value of
the other possible lifetime ratios.  No \Lb~lifetime is used in the
fit for the baryon sample composition, although $\fboud$ varies
slightly when the ratio $\tau(\Bu)/\tau(\Bd)$ is varied within the PDG
uncertainty, because the \Lb~fragmentation fraction is measured
relative to the \Bd\ and
\Bu~modes.

\subsection{Monte Carlo Simulation Statistics}

Since a finite number of Monte Carlo simulation events are generated
for each exclusive decay to be used in the process to extract the
sample composition (see Section~\ref{sec:sample_comp},) the statistics
of the generated simulation is checked to see whether the statistical
uncertainties on the yields, which are used to determine the
efficiencies, contribute a significant uncertainty to the
measurement. The simulation yields in each decay are shifted by
$\pm$\,1\,$\sigma$ around their central values and the efficiencies
are re-determined accordingly.  To assign the systematic uncertainty,
half of the yields are randomly shifted up, while the other half are
shifted down.  In all cases, the shift in all three relative
fragmentation fractions is small compared to the other uncertainties,
as can be seen in Table~\ref{tab:systematics}.

\subsection{Bottom Hadron Lifetimes}

Knowledge of the bottom hadron lifetimes is also needed for the
generation of the various \Bb~simulation samples.  While the \Bd\ and
\Bu~lifetimes are well-measured, there are large uncertainties on the \Bs\
and \Lb~lifetimes~\cite{Ref:PDG_2004}. To assign a
systematic error due to the uncertainty in the knowledge of the \Bb\
and \Lb~lifetimes, the simulation is re-generated with the \Bs\ and
\Lb~lifetimes shifted by one sigma uncertainty on their PDG
values: $\tau(\Bs)=(438\pm17)~\mu$m and $\tau(\Lb)=(368\pm24)~\mu$m.
The shift in \fboud\ is one of the larger uncertainties in
Table~\ref{tab:systematics}, but it is still small compared to the
uncertainties due to the imprecise knowledge of the baryon branching
fractions.

\subsection{\label{sec:Lbspectrum} \boldmath $p_T$ Spectra}

The bottom hadron $p_T$ spectra are one of the biggest uncertainties
on the knowledge of the relative efficiencies. Consequently, the
systematic uncertainties arising from the $p_T$ spectra are estimated
conservatively, as no definitive measurements for the \Bs~meson and
\Lb~baryon are available.  The systematic uncertainty assigned to the
$p_T$ spectrum for \lLc\ is taken from a $\pm$2\,$\sigma$ variation of
the tuned semileptonic \Lb~$p_T$ spectrum.  The spectrum is varied by
$\pm$2\,$\sigma$ in order to provide a conservative error, since the
$p_T$~spectrum measured from the semileptonic $\Lb$ decay is
incomplete.  Although the \lDs~MC generated with the $p_T$~spectrum
obtained from the inclusive $J/\psi$~cross-section
measurement~\cite{Ref:CDFdet} agrees well with the data, there is the
possibility that the \Bs~meson $p_T$ spectrum is different from the
\Bd\ and \Bu~spectra.  This possibility is accounted for by
measuring the ratio \fsoud\ with the default generator input spectrum,
while the \Bd\ and \Bu~decays are generated with the $p_T$ spectrum
inferred from Ref.~\cite{Ref:CDFdet}. Since no significant discrepancy
is observed between the \lDs~data and the simulation using the $p_T$
spectrum from Ref.~\cite{Ref:CDFdet}, 
this is a conservative assessment of the systematic error due to the
uncertainty on the \Bs~momentum spectrum.

%

\subsection{\boldmath Specific Ionization Efficiency}

Accurate knowledge of the requirement on the \dedx~based likelihood
ratio ${\cal LR}(p)$ on the proton in the \LcpKpi~decays is important
for an accurate determination of the \lLc~efficiency relative to the
semileptonic bottom hadron decay efficiencies.  In order to assign an
uncertainty to the knowledge of the \dedx~efficiency, the measurement
of the fragmentation fractions is performed without any \dedx~cut
applied to either the data or simulation.  The \fboud~fit result is
stable and the difference with the default fit is treated as a
systematic uncertainty.  Removing the \dedx~cut does not produce a
significant change in either \fufd\ or \fsoud.

\subsection{\boldmath \Lb~Polarization}

The polarization of the \Lb~baryon in hadronic collisions is not
known.  By default, the \Lb~baryon is unpolarized in the simulation
used in this measurement. In order to assign a systematic uncertainty
to the possible polarizations of the \Lb, the extreme cases of the
\Lb\ being fully polarized are tested to bound the effect.  A
systematic uncertainty is assigned when the \Lb~is produced either
with entirely positive helicity or entirely negative helicity.

\subsection{Bottom Hadron Branching Fractions}

Systematic uncertainties due to the knowledge of the bottom hadron
branching fractions arise in two places in the sample composition procedure.
First, the indirect semileptonic decays ({\it e.g.}  $\Bb\ra
D\bar{D}X,\ \bar D\ra\ell^- X$) contributing to the
lepton-charm signals, many of which are poorly determined experimentally, 
and second the uncertainty in the PDG semileptonic \Lb~branching fraction, 
${\cal B}(\LbLcln X)=(9.2\pm2.1)\%$~\cite{Ref:PDG_2004}.  Since many of
the measured indirect semileptonic decays are poorly determined, the branching
fractions predicted from symmetry principles for these decay modes are
used in the sample composition process to determine the systematic shift in
the fragmentation fractions. The contributions of indirect semileptonic
\Bs~decays to the \lDp, \lDz, and \lDst~signatures are small 
(see Tables~\ref{tab:Dp_eff}-\ref{tab:Dst_eff}), but the rate for
$\Bd/\Bu \ra D_s^{(*)+}D^{(*)}(\ell \Ds X)$ decays contributing to the
\lDs~final state (see Table~\ref{tab:Ds_eff}) is an order of magnitude
larger due to the more copious fragmentation of $b$~quarks into \Bd\
and \Bu~mesons versus \Bs~mesons. To give a sense of the maximal
possible effect on \fsoud\ if no contributions from \Bd\ and
\Bu~mesons to the \lDs~yield were accounted for, the fragmentation
fraction \fsoud\ would increase by about 10\% from $\sim$\,0.160 to
$\sim$\,0.176.  This estimate is presented for general interest, though it not
used as a systematic uncertainty on the measurement, as it is known to
be an incorrect assumption.

To determine the systematic uncertainty associated with the inclusive
semileptonic \Lb~branching fraction, the PDG value is varied within its
uncertainties.  This is one of the largest systematic uncertainties
associated with the measurement of the \Lb~fragmentation fraction.

\subsection{Charm Branching Fractions}

Another source of systematic uncertainty due to the branching
fractions used in the sample composition procedure arises from the often
poor knowledge of the ground state charm branching fractions, which are
taken from the PDG and listed in Table~\ref{tab:charm}.  To determine
the uncertainty in the fragmentation fractions, the central values of
the ground state charm branching fractions included in the sample
composition are varied, one by one, within $\pm$1\,$\sigma$ of the PDG
uncertainty.  The largest shift in $\fufd$ comes from ${\cal 
B}(\DpKpipi)$, while the single largest uncertainty in \fsoud\ is due
to the large error on ${\cal B}(\Dsphipi)$. A poor knowledge of
${\cal B}(\LcpKpi)$ contributes the largest single systematic
uncertainty to \fboud.

In addition to the poorly measured ground state charm branching fractions,
many of the excited charm decays also have large uncertainties.  To assess
a systematic uncertainty for the limited knowledge of the excited charm
decays, the excited charm branching fractions are varied by shifting half of
the $D^{**}$ branching fractions randomly up by 30\%, while the other half are
shifted down.  When quoting the final result on the fragmentation
fractions, a separate systematic uncertainty is quoted due to uncertainties
on external branching fractions as indicated in Table~\ref{tab:systematics}.

\subsection{\boldmath \Lb~Sample Composition}
 
A systematic uncertainty is assigned for the uncertain knowledge of
the \Lb~sample composition (see Section~\ref{sec:sample_comp}). The
\Lb~sample composition is considered without any of the non-resonant 
baryon modes included, while the total semileptonic branching fraction in
both cases is required to be ${\cal B}(\LbLcln X)=9.2\%$.  A
systematic effect for a potential mis-modeling of the decay is also
considered and found to be negligible.  This uncertainty is determined
by evaluating the width difference of the $m(\lLc)$ distribution in
both the data and the Monte Carlo simulation.  The RMS of the data
distribution is 451 \mevcc, while the RMS of the MC is 455 \mevcc.
The ratio of excited to ground state \LbLcln\ decays is changed in the
simulation such that the RMS of the simulated $m(\lLc)$ distribution
decreases by 4 \mevcc, producing a 0.017 shift in \fboud.  Both
uncertainties result in a total systematic uncertainty of 0.047 on
\fboud.

\subsection{Total Systematic Uncertainty}

The total systematic uncertainties due to the knowledge of the
relative efficiencies, obtained by adding the individual systematic
uncertainties related to the determination of the efficiencies used in
the parameterization of the sample composition in quadrature, are
$\left\lbrace ^{+0.025}_{-0.045}\right\rbrace$ for \fufd,
$\left\lbrace ^{+0.011}_{-0.010}\right\rbrace$ for \fsoud, and
$\left\lbrace ^{+0.058}_{-0.056}\right\rbrace$ for \fboud.  
When uncertainties arising from branching fractions are included, the
uncertainties increase to
$\left\lbrace ^{+0.062}_{-0.074}\right\rbrace$ for \fufd, 
$\left\lbrace ^{+0.058}_{-0.035}\right\rbrace$ for \fsoud, and 
$\left\lbrace ^{+0.141}_{-0.103}\right\rbrace$ for \fboud, as given in
Table~\ref{tab:systematics}.

\begin{table}[tbp]
\caption{\label{tab:systematics}
Compilation of systematic uncertainties assigned.}
\begin{ruledtabular}
\begin{tabular}{lccc}
Systematics                             &$\fufd$              &$\fsoud$              &$\fboud$                    \\
\hline		                        		         
False Leptons                           &-0.039               &-0.001                &+0.018	 		  \\
Variation of cuts                       &$\pm 0.011$          &$\pm 0.0003$          &$\pm 0.019$                 \\
\Ds~reflection in \Dp\                  &+0.001               &+0.00002              &+0.0001                     \\
XFT eff.                                &$\pm 0.003$          &$\pm 0.0004$          &$\pm 0.006$                 \\
Single track eff.                       &$\pm 0.014$          &$\pm 0.002$           &$\pm 0.002$                 \\
Sample comp. lifetimes                  &$^{+0.018}_{-0.014}$ &$\pm 0.006$           &$\pm 0.002$                 \\
MC statistics                           &$\pm 0.005$          &$\pm 0.0007$          &$\pm 0.0006$                \\
Bottom hadron lifetimes                 &-                    &$^{+0.005}_{-0.001}$  &$^{+0.0077}_{-0.0136}$      \\
$p_T$ spectra                           &-                    &$\pm 0.008$           &$\pm 0.049$                 \\
\dedx~eff.                              &-                    &-                     &$\pm 0.012$                 \\
\Lb~polarization                        &-                    &-                     &$\pm 0.007$                 \\
{\bf Total (eff.)}                      &$^{+0.025}_{-0.045}$ &$^{+0.011}_{-0.010}$  &$^{+0.058}_{-0.056}$        \\
\hline
Physics bkgs                            &$\pm 0.001$          &$\pm 0.002$           &$\pm 0.001$                 \\
${\cal B}(\LbLcln X)$			&-                    &-                     &$^{+0.076}_{-0.048}$        \\
${\cal B}(\LcpKpi)$			&-                    &-                     &$^{+0.091}_{-0.053}$        \\
${\cal B}(\DpKpipi$			&$\pm 0.054$          &$\pm 0.003$           &$\pm 0.010$                 \\
${\cal B}(\DzKpi)$			&$\pm 0.020$          &$\pm 0.003$           &$\pm 0.003$                 \\
${\cal B}(\Dsphipi)$			&$\pm0.0006$          &$^{+0.057}_{-0.034}$  &$\pm 0.001$                 \\
${\cal B}(D^{**})$                      &$\pm 0.010$          &$\pm 0.004$           &$\pm 0.011$                 \\
\Lb~sample composition                  &-                    &-                     &$\pm 0.047$                 \\
{\bf Total (${\cal B}$)}                &$\pm 0.058$          &$^{+0.057}_{-0.034}$  &$^{+0.128}_{-0.087}$        \\
\hline
{\bf Total}                             &$^{+0.062}_{-0.074}$ &$^{+0.058}_{-0.035}$  &$^{+0.141}_{-0.103}$\\
\end{tabular}
\end{ruledtabular}
\end{table}

\section{\label{sec:results} 
Final Results and Discussion}

The weighted average of the fragmentation fractions between the
electron plus displaced track and muon plus displaced track samples
yields:
\begin{eqnarray*}
\frac{\fu}{\fd}    &=&1.054\pm 0.018\,{\rm (stat)}\, 
                     ^{+0.025}_{-0.045}\,{\rm (sys)}\,
                     \pm 0.058\,({\cal B}),\\
\frac{\fs}{\fu+\fd}&=&0.160\pm0.005\,{\rm (stat)}\,
                     ^{+0.011}_{-0.010}\,{\rm (sys)}\,
                     ^{+0.057}_{-0.034}\,({\cal B}),\\
\frac{\fb}{\fu+\fd}&=&0.281\pm0.012\,{\rm (stat)}\,
                  ^{+0.058}_{-0.056}\,{\rm (sys)}\,
                  ^{+0.128}_{-0.087}\,({\cal B}). 
\end{eqnarray*}
Since this analysis potentially ignores the production of $b$~baryons
that might not decay into the \Lb~final state, no constraint is
applied requiring the fragmentation fractions \fu, \fd, \fs, and \fb\
to sum to unity.  The correlation matrix for the fit is shown in
Table~\ref{tab:correlations}.

This result is in agreement with the world average of the
fragmentation fraction of \Bu~relative to \Bd, which is expected to be
equal to unity~\cite{Ref:PDG_2004}. The result on the relative
fragmentation fraction \fsoud\ presented in this paper agrees with the
LEP average $\fsoud=0.135\pm0.011$~\cite{Ref:PDG_2004} within one
standard deviation.  Separating ${\cal B}(\Dsphipi)$ from the result
of \fsoud, for comparison with the world average, gives:
\begin{eqnarray*}
\frac{\fs}{\fu+\fd}\times{\cal B}(\Dsphipi)&=&
(5.76\pm 0.18\,{\rm (stat)}\,^{+0.45}_{-0.42}\,
{\rm (sys)}\,)\times 10^{-3}. 
\end{eqnarray*}
There is no significant indication of a higher rate of $b$-quark
fragmentation to \Bs~mesons at the Tevatron which would contribute to
the anomalous Run\,I values of
$\bar{\chi}$~\cite{Ref:Paolo1,Ref:Paolo2,Ref:Berger}.  The uncertainty
on \fs\ will significantly decrease with an improved measurement of
${\cal B}(\Dsphipi)$, which is in preparation by the CLEO-c
experiment~\cite{Ref:CLEO_BR_Ds}.

Separating the poorly known ${\cal B}(\LcpKpi)$ and ${\cal B}(\LbLcln)$
from the results of \fboud\ yields:
\begin{eqnarray*}
\frac{\fb}{\fu+\fd}\times{\cal B}(\LcpKpi)&=&
(14.1\pm 0.6\,{\rm (stat)}\,^{+5.3}_{-4.4}\,
{\rm (sys)}\,)\times 10^{-3}\ \ {\rm or} \\ 
\frac{\fb}{\fu+\fd}\times{\cal B}(\LbLcln)\,{\cal B}(\LcpKpi)
&=&(12.9\pm 0.6\,{\rm (stat)}\,\pm 3.4\,{\rm (sys)}\,)\times 10^{-4}. 
\end{eqnarray*}
This quantity can be compared more naturally with the LEP results,
which quote $\fb\times {\cal B}(\LbLcln)\times {\cal
BR}(\LcpKpi)$~\cite{Ref:LEP_fb1,Ref:LEP_fb2}.  When all branching
fractions with large uncertainties are factored out, $\fb$ is
$\sim$\,2.3\,$\sigma$ higher than the LEP results, assuming that
$\fu=\fd=39.7\%$ at LEP.  In addition, this measurement of \fboud\ is
approximately twice as large as the world average of
$\fboud=0.125\pm0.020$~\cite{Ref:PDG_2004}, which is dominated by the
LEP results.

\begin{table}[tbp]
\caption{\label{tab:correlations}
Correlation matrix of fit parameters.}
\begin{ruledtabular}
\begin{tabular}{lccccccc}
Parameter     &$\fufd$   &$\fsoud$   &$\fboud$   &$\Gamma$   &$\Gamma^*$   &$\Gamma^{**}$   &$N(\Bd)$   \\
\hline		                        		         
$\fufd$       &1.0       &-0.021     &-0.053     &-0.011     &-0.135       &0.162           &-0.249      \\
$\fsoud$      &          &1.0        &0.077      &-0.015     &-0.058       &0.150           &-0.116      \\
$\fboud$      &          &           &1.0        &0.425      &0.563        &0.239           &-0.575      \\
$\Gamma$      &          &           &           &1.0        &0.657        &-0.122          &-0.674      \\
$\Gamma^*$    &          &           &           &           &1.0          &0.134           &-0.853      \\
$\Gamma^{**}$ &          &           &           &           &             &1.0             &-0.436      \\
$N(\Bd)$      &          &           &           &           &             &                &1.0         \\
\end{tabular}
\end{ruledtabular}
\end{table}

\subsection{\label{sec:discussion} 
Discussion of Results}

The difference between the \fb~result presented in this paper and the
LEP results may be explained, at least in part, by the different
environment of hadro-production of bottom hadrons in $p\bar{p}$
collisions. In addition to this effect, the transverse momentum of the
bottom hadrons is significantly lower for the data used in this
measurement, $\langle\,p_T(b)\,\rangle$\,$\sim$\,15~\gevc, than the
$b\bar{b}$ data collected at the $Z$ pole used in the LEP
measurements, $\langle\,p_T(b)\,\rangle$\,$\sim$\,45~\gevc. To study a
potential momentum dependence of \fb, the behavior of the
fragmentation fractions in bins of the lepton-charm $p_T$ is
investigated. Note that the fragmentation fractions can depend on
momentum, and the fractions reported here are for momenta integrated
above the effective $p_T^{min}$ which is chosen to be 7~\gevc\ in this
analysis.

\begin{figure*}
\centerline{
\includegraphics[width=1.0\hsize]{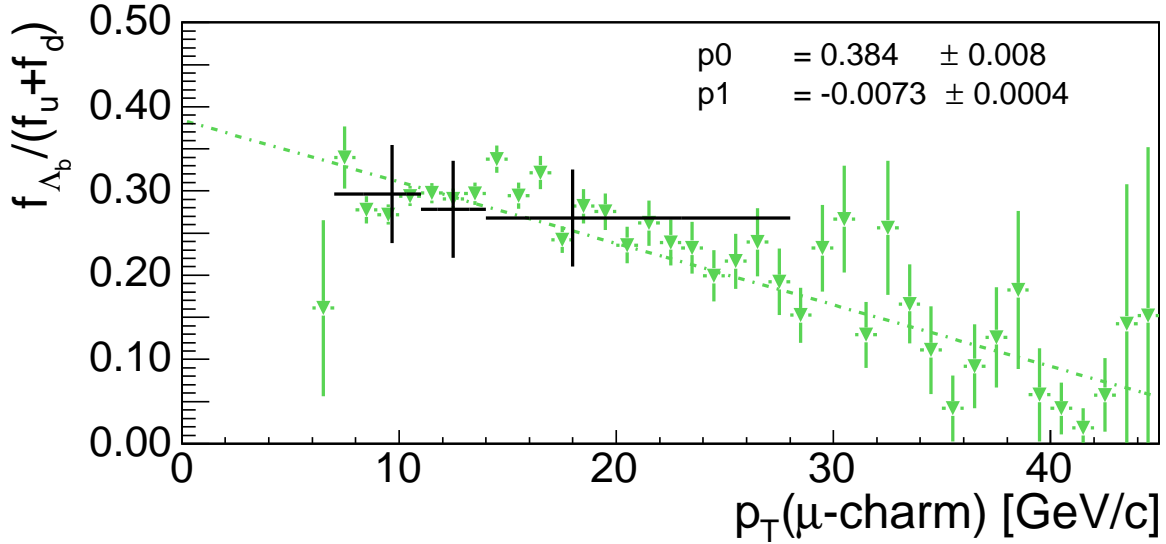}
}
\caption{\label{fig:frag_frac_pt}
Determination of \fboud\ for three momentum ranges (three points with
error bars) overlaid on Monte Carlo simulation scaling of \fboud\ as a
function of $p_T(\lLc)$ (triangles). The dashed line is a fit to the Monte
Carlo simulation.}
\end{figure*}

For this study the electron and muon datasets of the lepton-charm
candidates are divided into three momentum ranges with similar
statistics in each bin.  The chosen momentum bins are less than
11~\gevc, from 11 to 14~\gevc, and greater than 14~\gevc. The
lepton-charm yields and corresponding efficiencies are redetermined in
each momentum interval and the fit for the fragmentation fractions is
repeated.  The weighted average of \fboud\ obtained from the $e$+SVT
and $\mu$+SVT data in the three $p_T$~ranges is shown as three points
with error bars in Fig.~\ref{fig:frag_frac_pt}.  The uncertainties on
these points include the systematic uncertainties on the efficiencies,
but do not reflect the uncertainties from branching fractions.  The
data points are consistent with a decrease in the ratio \fboud\ with
increasing bottom hadron momentum.  However, in the near future,
larger CDF datasets of lepton-charm events will provide increased
statistics for a more adequate extrapolation of this suggested
momentum dependence of \fb\ as compared to $B$-hadron momenta at LEP.

To obtain a better extrapolation of the indicated momentum dependence
of $b$-quark fragmentation into \Lb~baryons, the Monte Carlo simulation
tuned on the data is used to estimate such a momentum dependence.  The
inclusive $\LbLcln X$ simulation generated with the tuned semileptonic
\Lb~$p_T$ spectrum is compared with the inclusive $\BdDpln X$ simulation
generated with the $p_T$~spectrum inferred from the inclusive $J/\psi$
cross-section measurement~\cite{Ref:CDFdet}.  Assuming that the
inclusive Monte Carlo samples provide a good description of the \lLc\
and \lDp~data, as demonstrated in Section~\ref{sec:data_mc}, the ratio
of the momentum dependence of both datasets provides an estimate of
the shape of the ratio of d$N/$d$p_T(\Lb)$ to d$N/$d$p_T(\Bd)$. The
ratio of both distributions is therefore proportional to $\fb/\fd$ and
thus to \fboud, assuming $\fu=\fd$. To obtain an absolute
normalization, the ratio of the two distributions is fixed at the mean
$p_T(\lLc)$ of the present measurement,
$\langle\,p_T(\lLc)\,\rangle\approx 14.1$~\gevc, to the central value
of $\fboud = 0.281$ as obtained in this analysis. The result of this
MC study is shown as triangles with error bars in
Fig.~\ref{fig:frag_frac_pt}. Fitting a straight line to these points
agrees well with the three data points of \fboud\ obtained in three
momentum bins as described above. Extrapolating the line to
$p_T(\lLc)\sim35~(40)~\gevc$ yields a value for \fboud\ of about 0.128
(0.092). This is close to the world average of
$\fboud=0.125\pm0.020$~\cite{Ref:PDG_2004}, which is dominated by the
LEP results.  This study indicates a possible momentum dependence of
the $b$-baryon fragmentation that would explain the difference between
the \fb~result presented in this paper and the LEP measurements.
 
Finally, knowledge of \fb\ will improve with better measurements of the
\LcpKpi~branching fraction and the semileptonic $\LbLcln X$ branching
fractions, in addition to better measurements of the \Lb~semileptonic
sample composition, particularly the measurement of the
$\Lb\ra\ell^-\bar{\nu}_{\ell}\Lambda_c(2593)^+$ and
$\Lb\ra\ell^-\bar{\nu}_{\ell}\Lambda_c(2625)^+$ branching fractions.
Additionally, a definitive measurement of the $p_T$ spectrum of
\Lb~baryons in $p\bar p$~collisions compared to the momentum spectrum
of \Bd~mesons measured with fully reconstructed \Lb\ and \Bd~decay
modes will shed light on expected differences in the momentum spectra
and significantly reduce the systematic uncertainty of a measurement
of the $b$-quark fragmentation fractions in the future.


\begin{acknowledgments}
We thank the Fermilab staff and the technical staffs of the
participating institutions for their vital contributions. This work
was supported by the U.S. Department of Energy and National Science
Foundation; the Italian Istituto Nazionale di Fisica Nucleare; the
Ministry of Education, Culture, Sports, Science and Technology of
Japan; the Natural Sciences and Engineering Research Council of
Canada; the National Science Council of the Republic of China; the
Swiss National Science Foundation; the A.P. Sloan Foundation; the
Bundesministerium f\"ur Bildung und Forschung, Germany; the Korean
Science and Engineering Foundation and the Korean Research Foundation;
the Science and Technology Facilities Council and the Royal Society,
UK; the Institut National de Physique Nucleaire et Physique des
Particules/CNRS; the Russian Foundation for Basic Research; the
Comisi\'on Interministerial de Ciencia y Tecnolog\'{\i}a, Spain; the
European Community's Human Potential Programme; the Slovak R\&D Agency;
and the Academy of Finland.
\end{acknowledgments}





\bibliography{fq_prd}    

\begin{thebibliography}{59}
\bibitem[{Ref()}]{Ref:pT}
\bibinfo{note}{The transverse momentum,~$p_T$, is the component of the track
  momentum,~$p$, transverse to the proton beam direction.}
\end{thebibliography}


\end{document}